\documentclass[aps,twocolumn,showpacs,groupedaddress,rmp]{revtex4}
\usepackage{amssymb,amsmath}
\usepackage{graphicx,array}
\usepackage{dcolumn}
\usepackage{bm}
\usepackage{float}
\usepackage{caption}
\usepackage{natbib}
\usepackage{setspace}
\usepackage{xkeyval}
\usepackage{rotating}

\begin{document}

\title{Simulation and understanding of quantum crystals}

\author{Claudio Cazorla}
\email{c.cazorla@unsw.edu.au}
\affiliation{School of Materials Science and Engineering, UNSW Australia, Sydney NSW 2052, Australia \\
Integrated Materials Design Centre, UNSW Australia, Sydney NSW 2052, Australia}

\author{Jordi Boronat}
\email{jordi.boronat@upc.edu}
\affiliation{Departament de F\'{i}sica, Universitat Polit\`{e}cnica de Catalunya, Campus Nord B4-B5, 
E-08034 Barcelona, Spain}

\begin{abstract}
Quantum crystals abound in the whole range of solid-state species. Below a certain threshold  
temperature the physical behavior of rare gases ($^{4}$He and Ne), molecular solids (H$_{2}$ 
and CH$_{4}$), and some ionic (LiH), covalent (graphite), and metallic (Li) crystals can be 
only explained in terms of quantum nuclear effects (QNE). A detailed comprehension of the nature 
of quantum solids is critical for achieving progress in a number of fundamental and applied 
scientific fields like, for instance, planetary sciences, hydrogen storage, nuclear energy, 
quantum computing, and nanoelectronics. This review describes the current physical understanding 
of quantum crystals formed by atoms and small molecules, as well as the wide palette of simulation 
techniques that are used to investigate them. Relevant aspects in these materials such as phase 
transformations, structural properties, elasticity, crystalline defects and the effects of reduced 
dimensionality, are discussed thoroughly. An introduction to quantum Monte Carlo techniques, 
which in the present context are the simulation methods of choice, and other quantum simulation 
approaches (e. g., path-integral molecular dynamics and quantum thermal baths) is provided. The 
overarching objective of this article is twofold. First, to clarify in which crystals and physical 
situations the disregard of QNE may incur in important bias and erroneous interpretations. And second, 
to promote the study and appreciation of QNE, a topic that traditionally has been treated in the 
context of condensed matter physics, within the broad and interdisciplinary areas of materials science.  
\end{abstract}

\pacs{67.80.-s, 64.70.-p, 31.15.A-, 82.20.Wt}

\maketitle

\tableofcontents

%===============================================================
%===============================================================
\section{Introduction}
\label{sec:intro}
%===============================================================
%===============================================================
\subsection{Quantum crystals: definition and interests}
\label{subsec:preliminary}
Quantum crystals are characterised by light-weight particles interacting through weak 
long-range forces. At low temperatures, the kinetic energy per particle in a quantum 
crystal, $E_{\rm k}$, is much larger than $k_{B}T$, where $k_{B}$ is the Boltzmann constant, 
and the spatial fluctuations about the equilibrium lattice sites are up to $30$\% of the 
distance to the neighboring lattice sites, that is, much larger than in any classical 
solid. These qualities can be understood only in terms of quantum mechanical arguments. 
Consider, for instance, the quantum expression of the atomic kinetic energy for a system of 
$N$ indistinguishable particles with mass $m$: 
\begin{equation}
E_{\rm k} = -\frac{\hbar^{2}}{2m} \left< \frac{{\bf \nabla}^{2} \Psi}{\Psi} \right>~,
\label{eq:kinetice}
\end{equation}
where $\hbar$ is the Planck constant, $\Psi$ the ground-state wave function of the system, 
and $\langle \cdots \rangle$ denotes expected value. If the light-weight particles were 
to rest immobile on the positions of the crystal arrangement, ${\bf R}_{0}$, that minimises 
their potential energy, $E_{\rm p}$, that is, $\Psi \propto \sqrt{\delta \left( {\bf R} 
-{\bf R}_{0} \right)}$, $E_{\rm k}$ would diverge. Rather, particles in a quantum crystal 
remain fluctuating around such equilibrium lattice sites in order to minimise their total 
energy $E = E_{\rm p} + E_{\rm k}$. The corresponding degree of spatial delocalisation is 
determined by a subtle balance between the accompanying gains in kinetic and potential 
energies. 
    
Examples of quantum solids include, Wigner crystals (Wigner, 1934; Ceperley and Alder, 1980;
Drummond \emph{et al.}, 2004; Militzer and Graham, 2006; Drummond and Needs, 2009), vortex lattices 
(Safar \emph{et al.}, 1992; Cooper, Wilkin, and Gunn, 2001; Abo-Shaeer \emph{et al.}, 2001), dipole 
systems (Astrakharchik \emph{et al.}, 2007; Matveeva and Giorgini, 2012; Boninsegni, 2013a; Moroni 
and Boninsegni, 2014), rare-gases, molecular solids, light metals, and many other similar systems
(see the next paragraphs). For the sake of focus, however, in this review we will concentrate 
on quantum crystals formed by atoms and small molecules. 
 
\begin{figure}
\centerline
        {\includegraphics[width=1.0\linewidth]{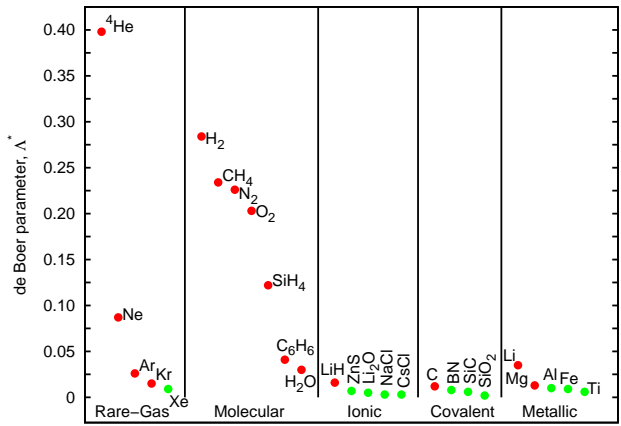}}
\caption{(Color online) De Boer parameter estimated in a series of crystals spanning over the
         whole range of solid-state species. The cases in which $\Lambda^{\ast}$ adopts
         a value larger than $0.012$ are indicated with red dots.}
\label{fig:deboer}
\end{figure}

A quantitative indicator of the degree of quantumness of a system is given by the de Boer 
parameter, $\Lambda^{\ast}$ (Sevryuk \emph{et al.}, 2010). This is defined as the ratio of 
the corresponding de Broglie wavelength, $\lambda (\epsilon)$, and a typical interatomic 
distance, $r_{0}$, namely: 
\begin{equation}
\Lambda^{\ast} =  \frac{\lambda (\epsilon)}{r_{0}} = \frac{\hbar}{r_{0}\sqrt{m \epsilon}}~, 
\label{eq:deboer}
\end{equation}
where $\epsilon$ is an energy scale characterising the interactions between particles. 
The smaller $m$ and $\epsilon$ are, the larger $\Lambda^{\ast}$ results. 
Figure~\ref{fig:deboer} shows the de Boer parameter estimated in a series of crystals 
that are representative of the broad spectrum of solid-state species.~\footnote{$\Lambda^{\ast}$ 
values were calculated using reported Lennard-Jones potential parameters for ``Rare-gas'' 
and ``Molecular'' solids. In the rest of cases, we used reported experimental 
cohesive energies as $\epsilon$'s and equilibrium lattice parameters as $r_{0}$'s.} 
The crystals in which quantum nuclear effects (QNE) are expected to be large, somehow arbitrarily 
defined here as $\Lambda^{\ast} \ge 0.012$ (which in the limiting case coincides with graphite), 
are indicated with red dots. As it is observed, most rare gases and light-weight molecular solids, 
among which we highlight helium, hydrogen, and methane, are quantum crystals. An important number 
of quantum specimens also are found in the remnant of solid-state categories like, for instance, 
metal hydride (ionic), carbon-based (covalent), and alkali metal (metallic) compounds.  

Quantum paraelectrics, although not included in Fig.~\ref{fig:deboer}, also conform to an intriguing class 
of quantum crystals. Quantum paraelectrics are materials in which the onset of ferroelectricity, that is,
the appearance of a spontaneous and externally switchable electrical polarisation, is suppressed by quantum
nuclear fluctuations (M\"{u}ller and Burkard, 1979; Rytz, H\"{o}chli, and Bilz, 1980; Conduit and Simons, 2010). 
Examples of quantum paraelectrics include SrTiO$_{3}$ and KTaO$_{3}$, which normally are classified as complex 
oxide perovskites (Ohtomo and Hwang, 2004; Cazorla and Stengel, 2012). Quantum paraelectrics do not follow the 
conventional definition of a quantum crystal since they contain heavy atomic species that interact through 
strong covalent and ionic forces. Actually, the size of QNE in these materials should be rather small 
(i. e., $\Lambda^{\ast} \ll 0.01$). At low temperatures, however, quantum paraelectrics are on the verge of 
a phase transition involving crystal structures with very similar energies and thus the impact of QNE 
in these and other related materials is large (see Sec.~\ref{sec:matscience}).  

The study of quantum solids is very important to understand nature. Hydrogen and helium, for instance, are 
the most abundant elements in the universe; they represent the $\sim 70-95$~\%~ of the mass of giant planets 
in our Solar System such as Jupiter and Saturn (Fortney, 2004; Baraffe, Chabrier, and Barman, 2010). An 
exhaustive knowledge of their condensed matter phases at extreme thermodynamic conditions is then crucial for 
understanding the chemical composition and past and future evolution of planetary bodies. Quantum solids are 
also sought after for technological applications. Rare-gases alloys, for instance, are intensively employed as 
pressure-transmitting media in high-load compression experiments and synthesis processes, due to their intriguing 
elastic properties (Errandonea \emph{et al.}, 2006; Dewaele \emph{et al.}, 2008; Cazorla, Errandonea, and Sola 2009). 

Other examples of scientific fields in which quantum crystals are important include nuclear energy, gas 
storage, quantum computing, and nanoelectronics. For instance, lithium hydride (LiH) and deuteride (LiD) are 
thoroughly used in the nuclear industry either as shielding agents or fuel in energy reactors (Welch, 1974; 
Veleckis, 1977). Metal hydrides are also promising for hydrogen storage applications (Grochala and Edwards, 2004; 
Shevlin and Guo, 2009) since they can supply large amounts of gas upon thermodynamic destabilisation. Likewise, 
carbon-based nanostructures (e. g., graphene, nanotubes, and fullerenes) exhibit large gas uptake capacities 
(Cazorla, Shevlin, and Guo, 2011; Gadipelli, 2015; Cazorla, 2015) as a consequence of their large surface-to-volume 
ratio, light atomic weight, and great thermodynamic stability. Diamonds with negatively charged nitrogen-vacancy 
centers, another type of carbonaceous nanomaterial, are playing a crucial role on the development of scalable 
quantum computing components (Fuchs \emph{et al.}, 2011; Nemoto \emph{et al.}, 2014). This class of 
crystals also can be employed as tunable \emph{quantum simulators} that, in analogy to ultracold atom gases 
trapped in optical lattices (Lewestein \emph{et al.}, 2007), can be used to answer fundamental questions  
in the fields of condensed matter, biology, and high energy physics (Georgescu \emph{et al.}, 2014; Wang 
\emph{et al.}, 2015). Finally, quantum paraelectrics find numerous applications in nanoelectronics as varistors, 
supercapacitor electrodes, and substrates on which to grow epitaxial films of other perovskite compounds (Lawless, 
1974; Schlom \emph{et al.}, 2007; Cazorla and Stengel, 2012).      

Besides of their fundamental and applied interests, quantum crystals are also very important in the 
framework of development and testing of new theories. The interactions between particles in quantum 
crystals typically are of dispersion, hydrogen bond, and multipole-multipole types, which are long-ranged 
and weak. The cohesion between atoms in solid helium, for instance, is so weak that to a first approximation
this crystal can be described by a system of hard spheres (Kalos, Levesque, and Verlet, 1974). 
Nevertheless, the description of long-ranged and very weak interactions poses a serious challenge to some 
families of first-principles methods (also known as \emph{ab initio} because do not rely on any predetermined 
knowledge of the atomic forces) as the analytical expression of the corresponding electronic exchange and 
correlation energies are intricate and difficult to approximate for computational purposes (Klime\v{s} and 
Michaelides, 2012; Cazorla, 2015). This circumstance converts quantum solids into an ideal playground in 
which to perform benchmark calculations for assessing the performance of standard and advanced electronic 
band-structure first-principles methods like, for instance, density functional theory (DFT) and electronic 
quantum Monte Carlo (eQMC) [Driver \emph{et al.}, 2010; Henning \emph{et al.}, 2010; Clay III \emph{et al.}, 
2014; Clay III \emph{et al.}, 2016] (see Sec.~\ref{subsec:firstprinciples}). 

Likewise, quantum nuclear effects (QNE) must be fully accounted for in any study dealing with quantum 
solids since they may affect noticeably the most fundamental properties of crystals like, for instance, 
atomic structure, vibrational phonon excitations, magnetic spin order, and electronic energy band gap. 
This fact leads to the situation in which approaches describing QNE only at a qualitative or approximate 
level (e. g., the Debye model and quasi-harmonic approximation) normally are inadequate for investigating 
genuine quantum crystals (see, for instance, Morales \emph{et al.}, 2013; Monserrat \emph{et al.}, 2014; 
Cazorla and Boronat, 2015); instead, full quantum approaches based on the solution to the Schr\"odinger 
equation or path-integral formulation of quantum mechanics due to Feynman (1948) must be employed (the 
fundamentals of these and others quantum simulation techniques will be reviewed in Sec.~\ref{sec:simulation}).  

\subsection{A bit of history and theory}
\label{subsec:historytheory}
The experimental study of quantum solids was initiated with the solidification of $^{4}$He at the 
Kamerlingh Onnes Laboratories in Leiden, by W. H. Keesom on June 1926 (Keesom, 1942; Domb and Dugdale, 
1957). It was not until the late 1960's and early 1970's, however, that with the establishment of 
neutron inelastic scattering techniques solid helium started being investigated thoroughly (Klein and 
Venables, 1974). The aim of those early neutron inelastic scattering experiments (Lipschultz \emph{et al.}, 
1967; Minkiewicz \emph{et al.}, 1968) was to understand the phonon dynamics in such a highly anharmonic 
solid. Actually, harmonic calculations render a mechanically unstable solid (that is, with imaginary lattice 
phonon frequencies) at low densities (Wette and Nijboer, 1965), hence it was very appealing to rationalise the 
real dynamics of the crystal. An interest on understanding how hydrostatic pressure could modify the physical 
properties of quantum solids started to develop also at that time (Eckert \emph{et al.}, 1977; Stassis 
\emph{et al.}, 1978). 

Likewise, the initial theoretical efforts were concentrated in finding a theory that could describe 
correctly the dynamical stability observed in highly anharmonic crystals. This was accomplished with 
the development of the self-consistent phonon (SCP) theory (Koehler, 1966; Glyde, 1994). In the SCP 
approach, one essentially assumes an harmonic solid with force constants that best represent the real 
anharmonic crystal, which are determined on the basis of a variational principle. We note that in 
recent years variants of the SCP approach have been successfully applied to the study of highly anharmonic 
metallic, molecular, and superconductor materials, in the context of electronic first-principles calculations 
(Errea, Rousseau, and Bergara, 2011; Errea, Calandra, and Mauri, 2013; Errea, Calandra, and Mauri, 2014; 
Monserrat \emph{et al.}, 2014; Engel, Monserrat, and Needs, 2015; Errea, \emph{et al.}, 2015).   

In quantum crystals, due to the large excursions of the atoms around the equilibrium positions, 
a good treatment of the short-range correlations is necessary. The need for considering such 
microscopic effects, which are beyond the extent of harmonic and quasi-harmonic approaches, 
led to the development of the variational theory of quantum solids (Nosanow, 1966; Koehler, 
1967). Nosanow proposed a general wave function model for a quantum solid of the form:
\begin{equation}
\Psi \left( {\bf r}_{1}, \cdots , {\bf r}_{N} ; \lbrace {\bf R}_{I} \rbrace \right) =
\prod_{j < k} f(r_{jk}) \prod_{i = 1}^{N} g\left( |{\bf r}_{i} - {\bf R}_{i}| \right)~,
\label{eq:nosanow}
\end{equation}
where ${\bf R}_{I}$ are the position vectors defining the equilibrium crystal lattice, ${\bf r}_{i}$ 
the position vectors of the particles, $r_{jk} \equiv |{\bf r}_{j} - {\bf r}_{k}|$, and $g(r)$ and 
$f(r)$ Gaussian and two-body correlation factors, respectively. The second factor in Eq.~(\ref{eq:nosanow}) 
localises each particle around a particular equilibrium lattice site while the first 
accounts for the interparticle correlations introduced by the atomic interactions.

After McMillan's and other authors' works on liquid $^{4}$He (McMillan, 1965; Schiff 
and Verlet, 1967), the two-body correlation factors in $\Psi$ are frequently expressed 
as $f(r) = \exp{\left[ -\frac{1}{2}\left(\frac{b}{r}\right)^{5} \right]}$. This function 
corresponds to the asymptotic solution of the Schr\"odinger equation in the $r \to 0$ limit 
of a two-body problem in which the interparticle interaction is of the Lennard-Jones type. 
With such a relatively simple analytical model of $\Psi$ and by employing Monte Carlo 
multidimensional integration techniques (Metropolis \emph{et al.}, 1953; Wood and Parker, 
1957), it was then possible to perform variational calculations of the ground state of solid 
helium (Hansen and Levesque, 1968) and other quantum solids (Hansen, 1968; Bruce, 1972). 
These advancements set the foundations of the variational Monte Carlo (VMC) method as 
applied to the study of quantum solids (see Sec.~\ref{subsub:approxzeroT}).    

Despite that variational approaches may be very insightful from a physical point of view, they 
rarely provide the exact quantitative answer in realistic problems. In order to obtain the precise 
solution to a quantum many-body problem one, for instance, may deal explicitly with the corresponding 
Schr\"odinger equation. To this end, more sophisticated techniques than VMC, albeit related, 
were developed during the 1970s, among which we highlight the Green's function Monte Carlo (GFMC) 
method due to Kalos and co-workers (Kalos, 1962; Kalos, Levesque, and Verlet, 1974; Ceperley 
\emph{et al.}, 1976; Whitlock and Kalos, 1979; Whitlock \emph{et al.}, 1979; Schmidt and Kalos, 1984). 
The basic idea behind GFMC is to employ Monte Carlo sampling techniques to solve the time-independent 
Schr\"odinger equation of a many-body system, when that is expressed as an integral equation 
containing a Green's function. Although the exact form of the Green's function normally 
is not known, this can be reproduced with stochastic sampling techniques involving probability 
distribution functions that are generated with the help of Trotter's product formula (Trotter, 
1959) [see Sec.~\ref{subsub:dmc}]. 

An intimately related method to GFMC is diffusion Monte Carlo (DMC), in  which the imaginary time-dependent 
Schr\"odinger equation, rather than the time-independent, is integrated by using a analytical short-time 
approximation to the Green's function (Ceperley and Alder, 1980; Reynolds \emph{et al.}, 1982; Guardiola, 
1986; Hammond \emph{et al.}, 1994). Both GFMC and DMC are \emph{exact} ground-state methods, in the sense 
that provide results for the energy that in principle are affected only by statistical errors. These two 
methods belong to the family known as ``projection techniques'', in which a projector operator is iteratively 
applied in order to cast out the ground state of the targeted quantum many-body system [in this latter category 
we also find, for instance, the reptation Monte Carlo method due to Baroni and Moroni (1999)]. Nonetheless, 
the DMC method is more efficient in dealing with arbitrary boundary conditions and potential energy functions
(Anderson, 2002), hence the use of GFMC is very infrequent nowadays. In Sec.~\ref{subsub:dmc}, we will review 
the fundamentals of the DMC method as applied to the study of quantum bosonic crystals. 

Quantum nuclear effects are also crucial to understand quantum solids at finite temperature (i. e., $T 
\neq 0$). The threshold temperature below which QNE are important can be considered to be equal to the 
Debye temperature $\Theta_{D}$ (Born and Huang, 1954). $\Theta_{D}$ is defined as $\hbar \omega_{m} / 
k_{B}$, where $\omega_{m}$ is the largest vibrational frequency in the crystal (that is, at 
$\Theta_{D} \le T$ all phonon modes in the solid are excited). This threshold temperature can be 
obtained directly from neutron inelastic scattering or specific heat measurements, and in the 
particular case of rare-gases $\Theta_{D}$ ranges from $25$ to $85$~K. It is important to note 
that $\Theta_{D}$ can increase dramatically under the application of external pressure, hence making 
unavoidable the consideration of QNE in the study of highly compressed quantum crystals. In molecular 
hydrogen, for instance, the Debye temperature at normal pressure conditions amounts to $\sim 100$~K 
whereas at $P = 20$~GPa turns out to be $\sim 1,000$~K (Diatschenko \emph{et al.}, 1985).     

The theoretical method of choice for simulation of quantum solids at $T \neq 0$ is path-integral 
Monte Carlo (PIMC). The PIMC method is based on Feynman's formulation of non-relativistic quantum mechanics,
which can be thought of as a generalisation of the action principle of classical mechanics (Feynman, 1948; 
Feynman and Hibbs, 1965). In Feynman's path integral theory, however, a functional integral over an infinity 
of possible trajectories (that is, a path integral) replaces the notion of probability amplitude. From a 
computational perspective, Feynman's formalism allows one to map out the atomic quantum system of interest 
onto a classical model of interacting polymers that evolve in imaginary time. This idea, which is known as 
the ``classical isomorphism'' (Feynman, 1972; Barker, 1979; Chandler and Wolynes, 1981; Ceperley, 1995), makes 
it possible to sample the corresponding space of possible configurations with stochastic techniques, laying the 
foundations of the PIMC method. PIMC relies exclusively on the knowledge of the many-body Hamiltonian and, in 
contrast to other simulation techniques like for instance GFMC and DMC, does not comprise the use of projector 
operators (this method will be explained in detail in Sec.~\ref{subsub:pimc}). Interestingly, the PIMC approach 
can be generalised to zero-temperature calculations by exploiting the formal similarities between imaginary time 
propagators and thermal density matrices (Sarsa \emph{et al.}, 2000). This methodological extension is named 
path-integral ground state (PIGS) and will be reviewed in Sec.~\ref{subsub:pigs}.

The isomorphism between classical and quantum systems also allows to employ molecular dynamics 
simulation techniques for sampling of path integrals (Chakravarty, 1997; Tuckerman and Hughes, 1998). 
In this last framework, generally known as path-integral molecular dynamics (PIMD), the atoms 
are treated as distinguisable particles. Consequently, genuine quantum statistical effects, that in 
liquids and disordered systems at low temperatures may give rise to intriguing quantum phenomena like 
Bose-Einstein condensation and superfluidity, are neglected. Nonetheless, in situations in which 
atomic quantum exchanges are not relevant (i. e., high temperatures) PIMD becomes a very powerful method 
that can be used, for instance, to compute quantum time-correlation functions and transition state rates 
very efficiently (Gillan, 1990; Habershon \emph{et al.}, 2013; Herrero and Ram\'irez, 2014). 

Recently, an alternative to path-integral quantum simulation approaches has been proposed that relies on the 
combined action of ``quantum thermal baths'' (QTB) and molecular dynamics (Wang, 2007; Buyukdagli \emph{et al.}, 
2008; Dammak \emph{et al.}, 2009; Hern\'andez-Rojas, Calvo, and Gonz\'alez-Noya, 2015). In the QTB formalism, the 
dynamics of the system is governed by a Langevin-type equation including dissipative and Gaussian random forces 
that mimics the power spectral density given by the quantum fluctuation-dissipation theorem (Callen and Welton, 1951). 
Although quantum statistical effects are also neglected in QTB approaches, these methods are becoming increasingly 
more popular in the last years due to their reduced computational expense as compared to path-integral based 
techniques. Meanwhile, hybrid PIMD and QTB schemes have been developed recently that exhibit improved convergence 
and scalability as compared to PIMD (Ceriotti, Bussi, and Parrinello, 2009; Ceriotti, Manolopoulos, and Parrinello, 
2011; Ceriotti and Manolopoulos, 2012; Brieuc, Dammak, and Hayoun, 2016). In Sec.~\ref{subsec:finiteT}, we shall 
provide a brief introduction to these emergent quantum simulation techniques.     

\subsection{Quantum vs. classical solids}
\label{subsec:quantumvsclassical}
\begin{figure}
\centerline
        {\includegraphics[width=1.0\linewidth]{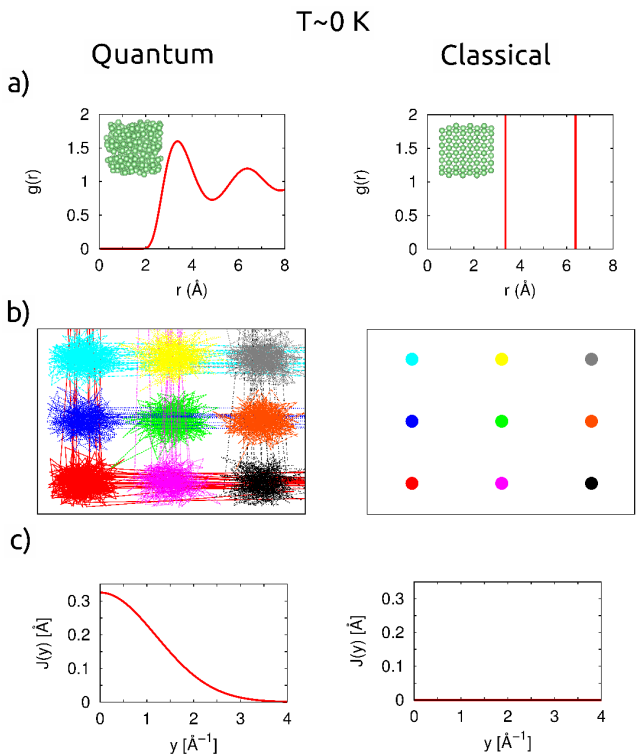}}
\vspace{-0.50cm}
\caption{(Color online) Sketch of the main differences between quantum and classical solids in the zero-temperature 
         limit. (a)~Radial pair distribution function, $g(r)$. (b)~Localisation of the atoms around their equilibrium 
         lattice positions; in the quantum case, lines represent the evolution of the atomic positions along time and 
         show the occurrence of atomic quantum exchanges. (c)~Compton profile of the longitudinal momentum distribution, $J(y)$.}
\label{fig:qvsc}
\end{figure}

Let us take a deeper look into the main differences between classical and quantum solids 
in the zero-temperature limit (see Fig.~\ref{fig:qvsc}). Atoms in a classical solid remain 
practically immobile on the positions of the periodic arrangement that minimises their 
potential energy (i. e., $E_{\rm k} \ll k_{B}T$), whereas in a quantum solid particles remain 
loosely localised around those sites (i. e.,  $E_{\rm k} \gg k_{B}T$). 
As a consequence, large Lindemann ratios (i. e., $\gamma \equiv \sqrt{\langle u^{2} \rangle}/a$, 
where $\langle u^{2} \rangle$ represents the atomic mean squared displacement and $a$ the lattice 
parameter) of the order of $\sim 0.1$ are observed in the latter case. Also, the radial pair 
distribution function, $g(r)$ (i. e., the average number density at a distance $r$ from an atom 
divided by the overall particle density), presents different features in the two types of crystals. 
In the zero-temperature limit, the $g(r)$ of a classical solid exhibits a series of sharp peaks 
signaling the radial distances between crystal lattice sites. By contrast, in a quantum solid $g(r)$ 
is continuous and displays a pattern of peaks-and-valleys that oscillates around unity at large distances 
(see Fig.~\ref{fig:qvsc}a). Likewise, the structure factor in a quantum solid, which is related to the 
Fourier transform of $g(r)$, presents some broadening and depletion of the main scattering amplitudes 
as compared to that in a classical solid (Whitlock \emph{et al.}, 1979; Draeger and Ceperley, 2000). 

A further difference between quantum and classical crystals is provided by the momentum distribution, 
$n({\bf k})$. In classical solids, $n({\bf k})$ is always (that is, independently of the interactions 
between the atoms) equal to the Maxwell-Boltzmann distribution:
\begin{equation}
n({\bf k})^{\rm class} = \left( \frac{1}{2 \pi \hat{\alpha_{2}}} \right)^{3/2} \exp{\Big[-\frac{{\bf k}^{2}}{2\hat{\alpha_{2}}}\Big]}~, 
\label{eq:maxboltz}
\end{equation}  
where by the equipartition theorem $\hat{\alpha_{2}} \equiv \frac{m k_{B} T}{\hbar^{2}}$. In quantum solids, 
however, the momenta and positions of the atoms are not independent and consequently $n({\bf k})$ may depart 
significantly from $n({\bf k})^{\rm class}$. In solid $^{4}$He, for instance, $n({\bf k})$ is non-Gaussian as 
it has a larger occupation of low momentum states as compared to a Maxwell-Boltzmann distribution (Diallo 
\emph{et al.}, 2004; Rota and Boronat, 2011). 

The atomic momentum distribution of condensed matter systems can be measured by inelastic neutron scattering performed 
at high momentum transfer (Glyde, 1994; Diallo \emph{et al.}, 2007). In this case, the Compton profile of the longitudinal 
momentum distribution, $J(y)$, is the quantity that is directly measured, that in the impulse approximation is related to 
$n({\bf k})$ through the expression (Withers and Glyde, 2007):
\begin{equation}
J(y) = 2 \pi \int_{|y|}^{\infty}  dk \, k n({\bf k})~,
\label{eq:compton}
\end{equation}  
where $y$ is a scaling variable. Compton profile experiments can provide a wealth of information about the nature of 
quantum solids (Glyde, 1994). For instance, recent neutron scattering measurements have found that the atomic kinetic 
energy in solid helium at $T \approx 0$ amounts to $24.25(0.30)$~K (the number within parentheses represents the 
accompanying uncertainty) [Diallo \emph{et al.}, 2007], which is in very good agreement with quantum Monte Carlo 
estimations (Ceperley \emph{et al.}, 1996; Cazorla and Boronat, 2008a; Vitiello, 2011).  
       
Kinetic isotopic effects, which attribute different kinetic energies to the isotopes of a same chemical element 
[see Eq.~(\ref{eq:kinetice})], are indicators of the existence of QNE. The magnitude of these effects can be 
inferred from inspection of functions $g(r)$ and $n({\bf k})$ (Mao and Hemley, 1994; Boninsegni \emph{et al.}, 
1994; Cazorla and Boronat, 2005). For instance, narrowing~(widening) of the peaks in $g(r)$ may be caused by the 
presence of heavier~(lighter) species. Kinetic isotopic effects can also manifest in the thermal expansion  
of quantum solids (Pamuk \emph{et al.}, 2012; Herrero and Ram\'irez, 2011a) and corresponding $P-T$ boundaries 
delimiting the thermodynamic stability regions of different phases (Lorenzana, Silvera, and Goettel, 1990; 
Goncharov, Hemley, and Mao, 2011).

In quantum mechanics, atoms of a same species are indistinguishable, that is, they can exchange positions while 
leaving the configuration of the system (namely, the square of the wave function) invariant. These atomic exchanges 
can occur as pairwise interchanges, three-particle, four-particle, and so on cyclic permutations. When the particles 
involved in such permutations are bosons and their number grows to infinity, the system becomes superfluid and the 
atoms on it can flow coherently without any resistance (Feynman, 1972; Ceperley, 1995). In quantum solids, as 
opposed to classical crystals, atoms can swap their positions and further delocalise in configurational space 
(see Fig.~\ref{fig:qvsc}b for a schematic representation). An illustrative example of this class of QNE is given 
by solid $^{3}$He. $^{3}$He atoms, which are fermions, have a non-zero magnetic moment and at low pressure the stable 
phase is a cubic bcc crystal. At temperatures below $1.5$~mK, this system adopts an exotic magnetic order that consists 
of two planes of up spins followed by two planes of down spins (Roger, Hetherington, and Delrieu, 1983). In terms of 
classical interaction arguments, that is, if only nearest-neighbor pair exchanges were important, the magnetic order 
in this crystal should be antiferromagnetic. However, quantum exchanges between more than two $^{3}$He atoms are very 
frequent and as a result a strong competition between ferromagnetism and antiferromagnetism appears in the crystal
that leads to the observed magnetic order (Ceperley, 1995). 

It has been theoretically shown that in \emph{commensurate} $^{4}$He crystals (i. e., crystals with exactly two 
atoms per hcp unit cell, without any point or line defects such as vacancies, dislocations, or grain boundaries) 
typical cyclic permutations occurring at few tenths of K only involve a small number of atoms. Consequently, the 
superfluid density in perfect quantum solids is null (Ceperley and Bernu, 2004; Bernu and Ceperley, 2005; 
Boninsegni, Prokof'ev, and Svistunov, 2006b). This conclusion appears to be consistent with the results of most 
recent and conclusive torsional oscillator experiments performed by Kim and Chan (2012, 2014). Meanwhile, in the 
presence of crystalline defects or atomic disorder quantum Monte Carlo calculations agree in predicting that the 
length of ring quantum exchanges increases noticeably, and thus the possibility of realising superfluidity starts 
to depart from zero (Boninsegni, Prokof'ev, and Svistunov, 2006b; Boninsegni \emph{et al.}, 2007; Rota and Boronat, 
2012). We must note, however, that convincing experimental evidence of superfluid-like manifestations in quantum 
crystals are yet elusive (Chan \emph{et al.}, 2013; Hallock, 2015). We will discuss these topics in more detail in 
Sec.~\ref{sec:defects}. 

\subsection{Incomplete understanding of quantum crystals}
\label{subsec:challengesquantumsolids}
\begin{figure}
\centerline
        {\includegraphics[width=1.0\linewidth]{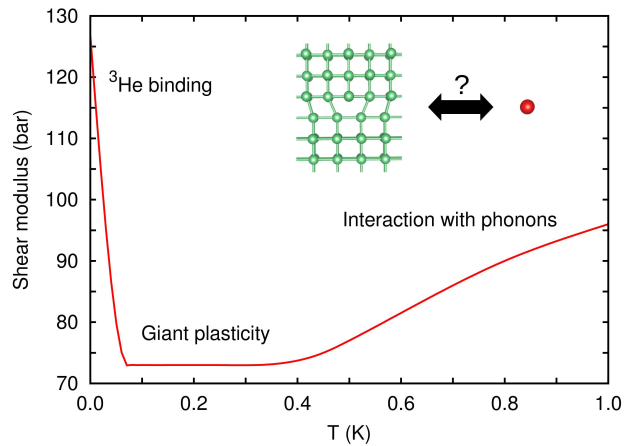}}
\vspace{-0.20cm}
\caption{(Color online) Shear modulus in an extremely pure $^{4}$He crystal expressed as a function 
         of temperature (adapted from Haziot \emph{et al.}, 2013c). Three different regimes are
         observed that can be explained in terms of dislocation dynamics (see text). The unsolved 
         problem about which are the interactions between dislocations and isotopic impurities is 
         noted schematically.}
\label{fig:gplast}
\end{figure}

Although a lot is already known on the physics of quantum crystals, there are still few puzzling and controversial 
aspects that urge for an improved understanding. One of these aspects is related to the interactions between different 
types of crystalline defects, their formation energy, and transport properties. In a seminal work, Day and Beamish 
(2007) reported the experimental dependence of the shear modulus, $\mu$, in solid $^{4}$He as a function of temperature. 
They found that $\mu$ increased with decreasing $T$ below a certain temperature of $0.15$~K. The observed 
increase in stiffness was rationalised in terms of line defects mobility: below a particular temperature threshold
the dislocations present in the crystal could be pinned by $^{3}$He impurities, in spite of the incredibly small 
concentration of the latter (i. e., just $200$ parts per billion of $^{4}$He atoms). This argument has been subsequently 
ratified by a number of compelling experimental works carried out by the groups of Beamish, in the University of Alberta, 
and Balibar, in the Ecole Normale Sup\'erieure de Paris (see, for instance, Haziot \emph{et al.}, 2013a; Haziot 
\emph{et al.}, 2013b; Fefferman \emph{et al.}, 2014; Souris \emph{et al.}, 2014a). Remarkably, Haziot \emph{et al.} (2013c) 
have recently shown that in ultra-pure single crystals of $^{4}$He the resistance to shear along one particular direction 
nearly vanishes at around $T = 0.1$~K, whereas normal elastic behavior is observed in the others. The exact origins of 
this intriguing effect, which has been termed as ``giant plasticity'', however, are still under debate (Zhou \emph{et al.}, 
2013; Haziot \emph{et al.}, 2013d), and the exact ways in which dislocations and isotopic impurities interact remain  
not fully understood (see Fig.~\ref{fig:gplast}).         

Recent theoretical arguments put forward by Kuklov \emph{et al.} (2014) suggest that quantum crystals might constitute 
a unique kind of materials in which topological lattice defects, that is, dislocations, could display quantum behavior 
like, for instance, quantum tunneling of kinks and jogs. Kuklov's hypotheses appear to be sustained by recent experimental 
observations in $^{4}$He and $^{3}$He crystals (see, for instance, Ray and Hallock, 2008; Lisunov \emph{et al.}, 2015). 
Verifying such a possible quantum scenario, however, turns out to be very challenging in practice due to the difficulties 
encountered both in the experiments and atomistic simulations. For instance, according to recent reports it appears 
to be extremely challenging to grow perfect helium crystals totally free of dislocations (Souris \emph{et al.}, 2014b). 
Concerning the calculations, a detailed and reliable simulation of line defects entails the use of large systems containing 
up to several thousands of atoms (Bulatov and Cai, 2006; Proville, Rodney, and Marinica, 2012), which currently is in the 
edge of quantum simulations. Due to these issues, many fundamental questions remain yet unanswered like for instance: 
What is the magnitude of the formation energy of dislocations in quantum solids? Can dislocations really behave as quantum 
entities so that they delocalise in space? Through which exact mechanisms quantum impurities like $^{3}$He atoms, which are 
extremely mobile, interact with dislocations? Solving these and other similar puzzles is crucial for advancing the field of 
quantum solids; this knowledge could have also an impact on the areas of materials science in which plasticity has a central 
role (e. g., fatigue in crystals and amorphous and martensitic transformations) [Proville, Rodney, and Marinica, 2012]. We 
will comment further on these points in Sec.~\ref{sec:defects}. 

Another source of unawareness in quantum crystals is posed by their behavior at extreme thermodynamic conditions.   
When a crystal is compressed the bonds between atoms normally are shortened so that particles become more localised
in order to avoid increasing their (highly repulsive) potential energy. At the same time, the kinetic energy of 
the solid increases due to Heisenberg's uncertainty principle. In quantum crystals, such a pressure-induced energy gain 
may be compensated in part by quantum atomic exchanges and quantum tunneling, which tend to favor the delocalisation 
of particles (Kosevich, 2005). The existence of proton quantum tunneling, for instance, has been demonstrated in solid 
hydrogen and ice under pressure, a QNE that is key to understand their corresponding phase diagrams and vibrational 
properties (Benoit, Marx, and Parrinello, 1998; Hemley, 2000; Howie \emph{et al.}, 2012a; Drechsel-Grau and Marx, 2014). 
On the other hand, quantum fluctuations in highly compressed solid hydrogen are known to hinder molecular rotation, 
which counterintuitively leads to some kind of atomic localisation (Kitamura \emph{et al.}, 2000; Li \emph{et al.}, 
2013). The ways in which QNE manifest and affect the physical properties of highly compressed quantum crystals 
actually seem to be quite unpredictable. 

Simulation of QNE phenomena at high pressures is technically difficult and demands intensive computational resources. 
The main reason for this is that the interactions between atoms cannot longer be described correctly with semi-empirical 
approaches like, for instance, pairwise potentials, and thereby the treatment of both the electronic and ionic degrees 
of freedom needs to be done quantum mechanically (see Sec.~\ref{sec:modeling}). Likewise, carrying out high-$P$ high-$T$ 
experiments in the laboratory is extremely challenging due to the occurrence of unwanted chemical reactions between the 
samples and containers (Dewaele \emph{et al.}, 2010). In addition to this, it is complicated to determine the exact atomic 
structure in highly compressed solids with low $Z$ numbers because their x-ray scattering cross sections are very small 
(Goncharov, Howie, and Gregoryanz, 2013; Dzyabura \emph{et al.}, 2014). Due to all these difficulties, the $P-T$ phase 
diagram of many quantum solids remain contentious and a complete understanding of the accompanying QNE features (e. g., 
quantum atomic exchanges and kinetic energy) is still pending. Further progress in this field is essential for advancing 
our knowledge in condensed matter physics and earth and planetary sciences (see Sec.~\ref{sec:molsol} for more details).  

\subsection{Aims and organisation of this review}
\label{subsec:organisation}
This review is concerned with the simulation and understanding of quantum solids formed by atoms and small molecules
under broad $P-T$ conditions. Important aspects in these systems like, for instance, their energetic and structural 
properties, phase transitions and elasticity, are discussed in detail. The effects that crystalline defects and reduced 
dimensionality have on the physical properties of archetypal quantum solids (i. e., $^{4}$He and H$_{2}$), are also 
reviewed. Special emphasis is put on identifying those systems and physical situations in which QNE must be considered 
in order to avoid likely misconceptions. In fact, QNE have been traditionally analysed in the field of condensed matter 
physics, however, comprehension of this class of effects is crucial for advancing in many other research areas such as 
planetary and materials sciences.   

We start by explaining the basics of the simulation methods that are used most frequently in the study of quantum 
crystals (Secs.~\ref{sec:simulation} and~\ref{sec:modeling}). In Secs.~\ref{sec:archetypal}-\ref{sec:matscience}, we 
describe the phenomenology and current understanding of quantum solids by surveying a large number of experimental 
and theoretical studies on archetypal and other less popular quantum crystals (e. g., H$_{2}$O, N$_{2}$, CH$_{4}$, 
LiH, and BaTiO$_{3}$). Finally, we comment on promising research directions involving quantum solids and summarise our 
general conclusions in Sec.~\ref{sec:conclusions}.

%===============================================================
%===============================================================
\section{Quantum simulation methods}
\label{sec:simulation}
%===============================================================
%===============================================================
We review the basics of customary quantum simulations methods that are employed for the investigation 
of quantum crystals. We classify them into two major categories, namely, ground-state 
($T = 0$) and finite-temperature ($T \neq 0$) methods. In the zero-temperature case, we differentiate 
between ``approximate'' and ``exact'' techniques. Depending on the nature of the problem that is 
going to be investigated and the amount of computational resources that are available, one may opt 
for using one or another.  

%===============================================================
\subsection{Ground-state approaches}
\label{subsec:zeroT}
%===============================================================

\subsubsection{Approximate methods}
\label{subsub:approxzeroT}

\paragraph{Quasi-harmonic approximation.}
In the quasi-harmonic approach (QHA) one assumes that the potential energy
of a crystal can be approximated with a quadratic expansion around the
equilibrium atomic configuration of the form (Born and Huang, 1954; Kittel, 
2005): 
\begin{equation} 
E_{\rm qh} = E_{\rm eq} + \frac{1}{2}
\sum_{l\kappa\alpha,l'\kappa'\alpha'}
\Phi_{l\kappa\alpha,l'\kappa'\alpha'} u_{l\kappa\alpha}
u_{l'\kappa'\alpha'}~,
\label{eq:eqh}
\end{equation}
where $E_{\rm eq}$ is the total energy of the perfect lattice,
$\boldsymbol{\Phi}$ the corresponding force-constant matrix, and $u_{l\kappa\alpha}$
is the displacement along Cartesian direction $\alpha$ of atom
$\kappa$ at lattice site $l$. Normally, this dynamical problem 
is solved by introducing:
\begin{equation}
u_{l\kappa\alpha}(t) = \sum_{q} u_{q\kappa\alpha} \exp{ \left[ i
    \left(\omega t - \boldsymbol{q} \cdot (\boldsymbol{l}+
    \boldsymbol{\tau}_{\kappa} \right) \right] }~,
\end{equation}
where $\boldsymbol{q}$ is a wave vector in the first Brillouin zone (BZ)
that is defined by the equilibrium unit cell; $\boldsymbol{l}+\boldsymbol{\tau}_{\kappa}$ 
is the vector that locates atom $\kappa$ at cell $l$ in the equilibrium 
structure. The normal modes are then found by diagonalizing the dynamical matrix:
\begin{equation}
\begin{split}
& D_{\boldsymbol{q};\kappa\alpha,\kappa'\alpha'} =\\ &
  \frac{1}{\sqrt{m_{\kappa}m_{\kappa'}}} \sum_{l'}
  \Phi_{0\kappa\alpha,l'\kappa'\alpha'} \exp{\left[
      i\boldsymbol{q}\cdot(\boldsymbol{\tau}_{\kappa}-\boldsymbol{l'}-\boldsymbol{\tau}_{\kappa'})
      \right]}~,
\end{split}
\end{equation}
and thus the crystal can be treated as a collection of non-interacting
harmonic oscillators with frequencies $\omega_{\boldsymbol{q}s}$
(positively defined and non-zero) and energy levels:
\begin{equation}
E^{n}_{\boldsymbol{q}s} = \left( \frac{1}{2} + n \right)
\omega_{\boldsymbol{q}s}~,
\end{equation}
where $0 \le n < \infty$. In this approximation, the Helmholtz
free energy of a crystal with volume $V$ at temperature $T$ is 
given by: 
\begin{equation}
F_{\rm qh} (V,T) = \frac{1}{N_{q}}~k_{B} T \sum_{\boldsymbol{q}s}\ln\left[ 2\sinh \left( 
    \frac{\hbar\omega_{\boldsymbol{q}s}(V)}{2k_{\rm B}T} \right) \right]~,
\label{eq:fharm}
\end{equation}
where $N_{q}$ is the total number of wave vectors used for integration 
over the BZ, and the $V$-dependence of the vibrational frequencies has 
been noted explicitly. In the zero-temperature limit, Eq.~(\ref{eq:fharm}) 
transforms into:
\begin{equation}
F_{\rm qh} (V,0) = \frac{1}{N_{\rm q}} \sum_{\boldsymbol{q}s}
\frac{1}{2}\hbar\omega_{\boldsymbol{q}s}(V)~,
\label{eq:zpe}
\end{equation}
which is usually referred to as the ``zero-point energy'' (ZPE).
We note that despite quasi-harmonic approaches may not be adequate for the
study of archetypal quantum solids (Morales \emph{et al.}, 2013; Monserrat 
\emph{et al.}, 2014; Cazorla and Boronat, 2015), QHA ZPE corrections normally 
are decisive in predicting accurate phase transitions in other materials 
(Cazorla, Alf\`e, and Gillan, 2008; Shevlin, Cazorla, and Guo, 2012; Cazorla 
and ${\rm \acute{I}}$${\rm \tilde{n}}$iguez, 2013). 

\paragraph{Variational Monte Carlo.}
Variational theory has been one of the most fruitful computational approaches
to study quantum fluids and solids. The strong repulsive interaction at short 
distances between particles produce a failure of conventional perturbation methods. 
The variational principle of quantum mechanics states that the expectation value 
of a Hamiltonian, $\hat{H}$, obtained with a model wave function, $|\Psi\rangle$, provides 
an upper bound to the true ground-state energy of the system, $E_0$, namely:
\begin{equation}
E=\frac{\langle \Psi | \hat{H} | \Psi \rangle}{\langle \Psi | \Psi\rangle} \ge E_0~.
\label{varmeth}
\end{equation}

In a many-body system, the evaluation of $E$ is not an easy task because one has to 
deal with a $3N$-dimensional integral. In this context, Monte Carlo integration techniques 
emerge as one of the most efficient computational methods. In variational Monte Carlo 
(VMC, named so because of its variational nature), one defines the multivariate 
probability density function (p.d.f.):
\begin{equation}
f(\bm{R})=\frac{|\Psi(\bm{R})|^2}{\int d \bm{R} ~ |\Psi(\bm{R})|^2}~,
\label{pdf}
\end{equation}   
which is normalised and positively defined. Meanwhile, the local energy, which adopts 
the form: 
\begin{equation}
E_{\text{L}}(\bm{R})= \frac{1}{\Psi(\bm{R})} \, \hat{H} \Psi(\bm{R})~,
\label{lenergy}
\end{equation}
allows one to express the expectation value of the Hamiltonian in the integral form:
\begin{equation}
\langle \hat{H} \rangle_\Psi = \int d \bm{R}~ E_{\text{L}}(\bm{R}) f(\bm{R})~.
\label{expectedenergy}
\end{equation}
In the two expressions above $\bm{R}$ stands for a multidimensional point
(also called ``walker''), $\bm{R} \equiv \{\bm{r}_1,\ldots,\bm{r}_N\}$. The 
expected value of the Hamiltonian then is calculated as the mean value of 
$E_{\text{L}}(\bm{R})$, evaluated in a series of points, $n_s$, that are 
generated according to the p.d.f. $f(\bm{R})$, namely:
\begin{equation}
\langle \hat{H} \rangle_\Psi = \frac{1}{n_s} \sum_{i=1}^{n_s}
E_{\text{L}}(\bm{R}_i)~.
\label{mcenergy}
\end{equation}
Effective sampling of the multidimensional p.d.f. $f(\bm{R})$ can be done 
with the Metropolis method (Metropolis \emph{et al.}, 1953; Wood and Parker, 
1957). Given a trial wave function, $\Psi(\bm{R})$, the VMC method provides the 
exact value of $\langle \hat{H} \rangle_\Psi$ to within statistical errors. Trial 
wave functions normally contain a set of parameters that are optimised in order 
to find the absolute minimum of $\langle \hat{H} \rangle_\Psi$. Alternatively, one can 
search for the parameter values that minimise the variance of the energy, whose 
lower bound \emph{a priori} is known to be zero (Hammond, Lester, and Reynolds, 
1994). 

With regard to Bose crystals (that is, formed by boson particles), the most widely 
used wave function is the Nosanow-Jastrow (NJ) model:
\begin{equation}
\Psi_{\rm{NJ}}({\bm r}_1,\ldots,{\bm r}_N) = \prod_{i<j}^{N} f(r_{ij}) \,
\prod_{i,I=1}^{N} g(r_{iI})~,
\label{njtrial}
\end{equation}
where $N$ is the number of particles and lattice sites, $f(r)$  a two-body Jastrow 
correlation function, and $g(r)$ a one-body localization factor that links particle 
$i$ to site $I$ (see Sec.~\ref{subsec:historytheory}). The Jastrow factor takes into 
account, at the lowest order, the dynamical correlations between particles induced by 
the interatomic potential, whereas the one-body term introduces the symmetry of the 
crystal.

Wave function $ \Psi_{\rm{NJ}}$ leads to an excellent description of the equation of 
state and structural properties of atomic quantum solids. However, it cannot be used 
to calculate properties that depend directly on the Bose-Einstein statistics (e. g., 
superfluidity and off-diagonal long-range order) because it is not symmetric under the 
exchange of particles. The latter symmetry requirement can be formally written as:
\begin{equation}
\Psi_{\rm{PNJ}}({\bm r}_1,\ldots,{\bm r}_N) = \prod_{i<j}^{N} f(r_{ij}) \,
\left(  \sum_{P(J)} \prod_{i=1}^{N} g(r_{iJ}) \right)~,
\label{pnjtrial}
\end{equation}
where $P(J)$ indicates a sum over all possible particle permutations involving the 
lattice sites. This wave function model, however, presents some practical issues since 
the number of configurations that needs to be sampled in order to reach convergence grows 
exponentially with the number of particles. 

Effective calculations involving a symmetric NJ wave function can be performed with 
the model:
\begin{equation}
\Psi_{\rm{SNJ}}({\bm r}_1,\ldots,{\bm r}_N) = \prod_{i<j}^{N} f(r_{ij}) \,
\prod_{J=1}^{N} \left( \sum_{i=1}^{N} g(r_{iJ}) \right)~,
\label{snjtrial}
\end{equation}
which has been introduced recently by Cazorla \emph{et al.} (2009). This symmetric wave 
function possesses a localization factor that suppresses lattice voids arising from double 
site occupancy, a desirable feature that also is reproduced by wave function $\Psi_{\rm{PNJ}}$. 

Other symmetric wave functions have been proposed in the context of quantum 
solids that do not rely on the symmetrization of $\Psi_{\rm{NJ}}$. 
These include a Bloch-like function (Ceperley, Chester, and Kalos, 1978), 
inspired in the band theory of electrons, and the shadow wave function (Galli, 
Rossi, and Reatto, 2005). The first model was introduced in a VMC study of the 
Yukawa system (Ceperley, Chester, and Kalos, 1978); the resulting variational 
energies, however, were significantly higher than those estimated with the 
non-symmetric NJ wave function, and the creation of vacancies or double occupancy 
of a same lattice site in the crystal could not be prevented. Consequently, this 
model has been overlooked in posterior studies. A more realistic symmetric model 
is provided by the shadow wave function (Vitiello, Runge, and Kalos, 1988; MacFarland 
\emph{et al.}, 1994), which is defined as: 
\begin{equation}
\Psi_{\rm{sh}}({\bm r}_1,\ldots,{\bm r}_N) = \Phi_p({\bm R}) \, \int d {\bm
S} \Theta({\bm R},{\bm S}) \Phi_s({\bm S})~,
\label{shadow}
\end{equation}
in which auxiliary variables ${\bm S}$ (also called ``shadows'') are introduced  
in order to avoid the explicit definition of any particular atomic arrangement. 
In Eq.~\ref{shadow}, $\Phi_p({\bm R})$ and $\Phi_s({\bm S})$ are Jastrow factors that 
correlate particles and shadows separately; function $\Theta({\bm R},{\bm S})$, on the 
other hand, introduces a coupling between particles and shadows. The shadow variables 
finally are integrated out of $\Psi_{\rm{sh}}$ in order to remove any explicit dependence
on them.

\subsubsection{Diffusion Monte Carlo}
\label{subsub:dmc}
Despite that variational methods may provide qualitatively correct results, it is not 
possible to determine their accuracy in absolute terms. Green's function Monte Carlo 
(GFMC) methods eliminate any variational constraints by solving directly the Schr\"odinger 
equation for a $N$-body problem. The most advanced of these methods is domain GFMC, 
in which the corresponding Green's function is time-independent (Kalos, 1962; Kalos, 
Levesque, and Verlet, 1974; Ceperley \emph{et al.}, 1976; Whitlock and Kalos, 1979; 
Whitlock \emph{et al.}, 1979; Schmidt and Kalos, 1984). A related method is diffusion 
Monte Carlo (DMC), which is time-dependent and nowadays widely used (Ceperley and Alder, 
1980; Reynolds \emph{et al.}, 1982; Hammond, Lester, and Reynolds, 1994; Anderson, 2002).

DMC is a projector method that, by working in imaginary time, is able to retrieve 
exact energy results for the ground state of a many-particle system. In imaginary
time, $\tau$, the Schr\"odinger equation becomes:
\begin{equation}
- \frac{\partial \Psi({\bm R},\tau)}{\partial \tau} = (\hat{H}-E_{0}) \Psi({\bm
R},\tau)~,
\label{schrodingertau}
\end{equation}      
where ${\bm R}=\{ {\bm r}_1, \ldots,{\bm r}_N \}$ and time is expressed in units
of $\hbar$. The time-dependent wave function of the system, $\Psi({\bm R},\tau)$, 
can be expanded in terms of the complete set of eigenfunctions of the Hamiltonian,
$\phi_i({\bm R})$, namely:
\begin{equation}
\Psi({\bm R},\tau)=\sum_{n}c_n \, \exp \left[\, -(E_i-E_{0}) \tau \, \right]\,
\phi_i({\bf R})\ ,
\label{dmc.eq1b}
\end{equation}
where $E_i$ is the eigenvalue associated to $\phi_i({\bm R})$.  The asymptotic 
solution of Eq.~(\ref{schrodingertau}) in the $\tau \rightarrow \infty$ 
limit then corresponds to $\phi_0({\bm R})$, provided that there is a non-zero 
overlap between $\Psi({\bm R},\tau=0)$ and the true ground-state wave function, 
$\phi_0({\bm R})$.

Direct application of Eq.~(\ref{schrodingertau}) to condensed matter problems 
is hindered by the repulsive interactions that atoms experience at short distances, 
which translates into large energy variances. To overcome this problem, 
one introduces importance sampling, a technique that is widely used in MC 
calculation of integrals. Importance sampling as applied to Eq.~(\ref{schrodingertau}) 
consists in rewriting the Schr\"{o}dinger equation in terms of the p.d.f.: 
\begin{equation}
f({\bm R},\tau)\equiv \psi({\bm R})\,\Psi({\bm R},\tau)~,
\label{dmc.eq2}
\end{equation}
where $\psi({\bm R})$ is a time-independent trial wave function that at the variational
level describes the ground state of the crystal correctly. 
By considering a Hamiltonian of the form:
\begin{equation}
\hat{H} = -\frac{\hbar^2}{2\,m} \, {\mbox{\boldmath $\nabla$}}^2_{{\bm R}} 
+ \hat{V} ({\bm R})~,
\label{dmc.eq3}
\end{equation}
Eq.~(\ref{schrodingertau}) turns into:
\begin{eqnarray}
-\frac{\partial f({\bm R},t)}{\partial \tau} & = &  -D\, 
{\mbox{\boldmath $\nabla$}}^2 f({\bm R},\tau) + \nonumber \\
& & D\,{\mbox{\boldmath $\nabla$}} \left[{\bm F}({\bm R}) \,f({\bm R},\tau)\right] + \nonumber   \\
  & & \left[ E_L({\bm R})-E_{0} \right] \cdot f({\bm R},\tau)~, 
\label{dmc.eq4}
\end{eqnarray}
where $D \equiv \hbar^2 /(2m)$, $E_L({\bf R}) \equiv \psi({\bm R})^{-1} H \psi({\bm R})$
is the local energy, and
\begin{equation}
{\bm F}({\bm R}) \equiv 2\, \psi({\bm R})^{-1} 
{\mbox{\boldmath $\nabla$}} \psi({\bm R})
\label{dmc.eq5}
\end{equation}
is the so-called drift or quantum force. ${\bm F}({\bm R})$ acts as an external 
force that guides the diffusion process rendered by the first term in the right-hand
side of Eq.~(\ref{dmc.eq4}). In particular, walkers are attracted towards regions 
in which the value of $\psi({\bm R})$ is large, thus avoiding the repulsive core 
of the interaction that produces large fluctuations in the energy. 

The right-hand side of Eq.~(\ref{dmc.eq4}) can be written as the action of three operators,
 $\hat{A_i}$, acting on the p.d.f. $f({\bm R},\tau)$, namely:
\begin{equation}
-\frac{\partial f({\bm R},\tau)}{\partial \tau} = (\hat{A}_1+ \hat{A}_2+ \hat{A}_3)\, 
f({\bm R},\tau) \equiv \hat{A}\, f({\bm R},\tau)~.
\label{dmc.eq4p}
\end{equation}
Operator $\hat{A}_1$ corresponds to a free diffusion process with coefficient $D$, $\hat{A}_2$ 
to a driving force produced by an external potential, and $\hat{A}_3$ to a birth/death branching 
term. In quantum Monte Carlo, the Schr\"odinger equation is most manageable when expressed 
in a integral form. This is achieved by introducing a Green's function, $G({\bm R}^{\prime},
{\bm R},\tau)$, that describes the transition probability to move from an initial state, 
${\bm R}$, to a final state, ${\bm R}^{\prime}$, during the time interval $\Delta \tau$, 
that is:
\begin{equation}
     f({\bm R}^{\prime},\tau+\Delta \tau) =\int G({\bm R}^{\prime},{\bm R},
\Delta \tau)\, f({\bm R},\tau)\, d{\bm R}~.
\label{dmc.eq6}
\end{equation}
More explicitly, the Green's function can be expressed in terms of the $\hat{A}$ operator 
as:
\begin{equation}
    G({\bm R}^{\prime},{\bm R}, \Delta \tau) =  
    \left \langle\,
{\bm R}^{\prime}\, | \, \exp(-\Delta \tau \hat{A})\, |\, {\bm R}\, \right \rangle~,
\label{dmc.eq7}
\end{equation}
and be approximated in practice with Trotter's product formula (Trotter, 1959):
\begin{equation}
e^{- \tau \left( \hat{A}_{1} + \hat{A}_{2} \right)} = \lim_{n \to \infty} \left( e^{-\frac{\tau}{n} \hat{A}_{1}} e^{-\frac{\tau}{n} \hat{A}_{2}} \right)^{n}~. 
\label{trotter1}
\end{equation}

DMC algorithms rely on reasonable approximations to the propagator $G({\bm R}^{\prime},{\bm
R},\Delta \tau)$ in the $\Delta \tau \to 0$ limit, which are iterated repeatedly until 
reaching the asymptotic regime $f({\bm R},\tau \rightarrow \infty)$ (that is, when the 
ground state is effectively sampled). The order of the employed $G({\bm R}^{\prime},{\bm R},\Delta \tau)$ 
approximation introduces a certain time-step bias on the results, that needs to be removed 
in order to provide perfectly converged solutions (Boronat and Casulleras, 1994).
In DMC, the sampling of an operator, $\hat{O}$, is performed according to the mixed 
distribution $f({\bm R},\tau)$ [see Eq.~(\ref{dmc.eq2})], rather than to $\phi_0({\bm R})$. 
Therefore, the standard DMC output corresponds to the so-called ``mixed'' estimator. Mixed 
estimators in general are biased by the trial wave function that is used for importance 
sampling. Only when $\hat{O}$ is the Hamiltonian of the system or an operator that commutes 
with it, the mixed estimator and exact result coincide. A simple scheme that is employed 
to remove partially the bias introduced by $\psi({\bm R})$ is:  
\begin{equation}
\langle \hat{O}({\bm R}) \rangle_e=2 \, \langle \hat{O}({\bm R})
\rangle_m - \langle \hat{O}({\bm R}) \rangle _v~,
\label{dmc.extrap}
\end{equation}
which is built from the mixed~($m$) and variational~($v$) estimators, and is known as
the ``extrapolated'' estimator~($e$) [Ceperley and Kalos, 1979]. Nevertheless, expectation 
values obtained with the extrapolation approach never are totally free of bias, and it is 
difficult to estimate \emph{a priori} the size of the accompanying errors. In order to overcome 
such limitations, one can calculate ``pure'' expectation values (that is, exact to within the 
statistical errors) by using the forward walking technique (Casulleras and Boronat, 1995).

\subsubsection{Path-integral ground-state Monte Carlo}
\label{subsub:pigs}
An interesting alternative to the DMC method has been put forward by Sarsa \emph{et al.} (2000), 
based on a previous proposal by Ceperley (1975). This method is termed path integral ground-state 
method (PIGS) and it is directly related to the path integral Monte Carlo (PIMC) method used at 
finite temperature (see next Sec.~\ref{subsub:pimc}). The integral version of the Schr\"odinger 
equation can be written in terms of the Green's function as:
\begin{equation}
\Psi({\bm R},\tau)=\int d {\bm R}^\prime \ G({\bm R},{\bm R}^\prime; \tau-\tau_0)
\Psi({\bm R}^\prime,\tau_0)~.
\label{green1}
\end{equation}
In the PIGS method one exploits the formal identity between the Green's function at imaginary 
time $\tau$, $G({\bm R},{\bm R}^\prime; \tau-\tau_0)$, and the statistical density matrix 
operator at an inverse temperature $\beta \equiv 1/T$, $\rho({\bm R},{\bm R}^\prime; \beta)$. The
convolution property of the density matrix permits to estimate $\rho({\bm R},{\bm R}^\prime; \beta)$ 
through a convolution of density matrices calculated at smaller values, $\beta/N_b$, namely:
\begin{eqnarray}
&\rho({\bm R},{\bm R}^\prime; \beta) & =   \int d{\bm R}_1 \ldots d{\bm
R}_{N_b-1}  \nonumber \\ 
 \times & \rho({\bm R},{\bm R}_1; \beta/N_b) & \ldots \rho({\bm R}_{N_b-1},{\bm
R}^\prime; \beta/N_b)~.
\label{convolut}
\end{eqnarray}

In PIMC ($T \neq 0$), one has to deal with the trace of the density matrix operator and 
hence the boundary condition ${\bm R} = {\bm R}^\prime$ is imposed; this makes a closed 
path. In the context of the ``classical isomorphism'' (Feynman, 1972; Barker, 1979; 
Chandler and Wolynes, 1981; Ceperley, 1995), a path is interpreted as a polymer in 
which first neighbors are connected with springs; moving a quantum particle is equivalent 
to evolve such a polymer. In PIGS ($T = 0$), at difference with PIMC, one truncates the path 
by imposing that the end points, ${\bm R}^\prime$, terminate in a trial wave function, 
$\psi$; the path then is open. In this case, the expectation value of an operator, $\hat{O}$, 
is determined by:
\begin{equation}
\hat{O}=\frac{\langle \psi | G(\tau/2) \hat{O} G(\tau/2) | \psi \rangle}{\langle
\psi | G(\tau) | \psi \rangle}~,
\label{middlechain}
\end{equation} 
where $\tau$ is the total imaginary time that the system takes to move from the initial 
to the end point. The most remarkable aspect of this method is that in the middle of 
the path, $\tau/2$, the sampling of any operator is exact, independently of whether 
$\hat{O}$ commutes or not with the Hamiltonian of the system. In other words, calculation 
of ``pure'' estimators is the standard output in PIGS, contrarily to what occurs in 
DMC (although for operators that commute with the Hamiltonian both methods shall provide 
equivalent results). Actually, for non-diagonal operators, like for instance the one-body 
density matrix, only PIGS is able to provide unbiased zero-temperature results in an efficient 
manner.

In order to perform PIGS calculations as efficiently as possible in practice, it is necessary 
to develop approximations for the propagator operator that are accurate to within a certain 
order in the time step. To this regard, significant progress has been achieved in recent 
years by developing splitting schemes for the exponential of the Hamiltonian operator, 
$\hat{H} = \hat{K} + \hat{V}$ (where $\hat{K}$ and $\hat{V}$ are the kinetic and potential 
energy operators, respectively), of the form:
\begin{equation}
\exp[\varepsilon (\hat{T}+\hat{V})] = \prod_{i=1}^{N} \exp(t_i\varepsilon
\hat{T}) \exp(v_i \varepsilon \hat{V})~,
\label{split1} 
\end{equation}
where the value of the parameters $\{t_i\}$ and $\{v_i\}$ are selected in a way that  
satisfies forward propagation (Chin and Chen, 2002). Under this constraint, one can 
write algorithms that are accurate up to fourth-order (Rota \emph{et al.}, 2010) and 
which produce very consistent convergence towards the ground state (see next Sec.~\ref{subsub:pimc}). 
Recent applications of the PIGS method involving high-order decomposition methods 
have shown that it is actually possible to obtain results that are completely independent 
of the trial wave function that is used as boundary condition (Rossi \emph{et al.}, 2009;
Rota \emph{et al.}, 2010). Even in the limiting case of considering only the symmetry 
requirement of the system (e. g., $\psi({\bm R})=1$ in the bosonic case) the PIGS 
method works reliably, with the only penalty of producing slightly larger variances. 
These methodological advancements permit to achieve accurate zero-temperature results 
in systems for which is difficult to find a good trial wave function.

A related method to PIGS is the reptation quantum Monte Carlo method (RQMC) 
due to Baroni and Moroni (1999); the starting point in RQMC is the same 
than in PIGS, that is, Eq.~(\ref{green1}). The main difference relies on the 
approximation that is used to the Green's function: RQMC adopts a short-time 
expression similar to the one used in DMC [see Eq.~\ref{dmc.eq7}] consisting 
of a drifted Gaussian that incorporates importance sampling. The ways in which
the paths are sampled are also different in the two methods. In the case of 
knowing a good trial wave function, RQMC may be advantageous as the resulting 
energy variance can be reduced significantly; otherwise, for the reasons 
explained in the paragraph above, PIGS may turn out to be a more reliable method
(Rossi \emph{et al.}, 2009; Rota \emph{et al.}, 2010).

%===============================================================
\subsection{Finite-temperature techniques}
\label{subsec:finiteT}
%===============================================================

\subsubsection{Path-integral Monte Carlo}
\label{subsub:pimc}
PIMC is based on the convolution property of the thermal density matrix 
shown in Eq.~(\ref{convolut}). This allows one to estimate the density matrix
at low temperature from its knowledge at higher temperatures, the latter 
being described by classical statistical mechanics. The partition 
function, $Z$, of a quantum system becomes then a multidimensional integral 
with a distribution law that resembles that of a closed classical polymer 
with an inter-bead harmonic coupling. If one assumes that all particles are 
bosons, the corresponding quantum statistical distribution is then positively 
defined and can be interpreted as a probability distribution function that  
can be sampled with standard Metropolis Monte Carlo techniques. The finite-$T$ 
mapping of a quantum system into a classical one composed of polymers was 
first proposed by Feynman (Feynman, 1972) and subsequently applied by Barker 
(1979), and Chandler and Wolynes (1981) to condensed-matter simulations.

The quantum partition function of a general Hamiltonian, $\hat{H}$, at temperature 
$T$ is: 
\begin{equation}
Z = \text{Tr} \, e^{-\beta \hat{H}}~.
\label{zeta}
\end{equation}
The non-commutativity of operators $\hat{K}$ and $\hat{V}$ makes it impractical a 
direct calculation of $Z$ in the quantum regime. Nevertheless, one can exploit 
the convolution property: 
 \begin{equation}
e^{-\beta (\hat{K} + \hat{V})} = \left( e^{-\varepsilon ( \hat{K} + \hat{V}
)} \right)^M ~,
\label{convolutionz}
\end{equation}
where $\varepsilon = \beta/M$, since now each of the terms in the right-hand
side of the equality effectively corresponds to a higher temperature, that is, 
$T^\prime = M \cdot T$. 
In the lowest order approximation, known as the primitive action (PA), the kinetic 
and potential contributions are factorised as:
\begin{equation}
e^{-\varepsilon ( \hat{K} + \hat{V})} \simeq e^{-\varepsilon  \hat{K}} \,
e^{-\varepsilon  \hat{V}}~,
\label{primitive}
\end{equation}
and the convergence to the exact result is guaranteed by Trotter's product
formula (Trotter, 1959):
\begin{equation}
e^{-\beta (\hat{K} + \hat{V})}  = \lim_{M \to \infty} \left(
e^{-\varepsilon  \hat{K}}  \, e^{-\varepsilon  \hat{V}}  \right)^M ~.
\label{trotter2}
\end{equation}

The PA approximation, however, is not accurate enough to reach proper convergence
at very low temperatures, when the number of terms involved, also called ``beads'', 
is large. In recent years, there has been relevant progress in achieving better 
convergence by using high-order splitting schemes of the exponential operator. 
Fourth-order algorithms can be developed by introducing double commutators (Chin 
and Chen, 2002; Sakkos, Casulleras, and Boronat, 2009) of the form:
\begin{equation}
[[\hat{V},\hat{K}],\hat{V}] = \frac{\hbar^2}{m} \, \sum_{i=1}^{N}
|\bm{F}_i|^2 ~,
\label{commutator}
\end{equation}
where $\bm{F}_i$ is the ``force'' acting on particle $i$, namely:
\begin{equation}
\bm{F}_i = \sum_{j \neq i}^{N} \bm{\nabla}_i V(r_{ij})~.
\label{force}
\end{equation}
One of the most efficient splitting schemes corresponds to:
\begin{eqnarray}
e^{-\varepsilon \hat{H}} &\simeq& e^{- v_1 \varepsilon \hat{W}_{a_1}}
e^{- t_1 \varepsilon \hat{K}} e^{- v_2 \varepsilon \hat{W}_{1-2 a_1}} \times \\ \label{chinaction}
& & e^{- t_1 \varepsilon \hat{K}} e^{- v_1 \varepsilon \hat{W}_{a_1}}
e^{- 2 t_0 \varepsilon \hat{K}}~, \nonumber
\end{eqnarray}
where $W(r)$ is a generalized potential that includes the double commutator in 
Eq.~(\ref{commutator}). We note that by optimising the value of the parameters 
in the expansion above, convergence with nearly sixth-order accuracy in $\varepsilon$
can be achieved. 

From the knowledge of the quantum partition function one can access the total 
and kinetic energies using the well-known thermodynamic relations: 
\begin{eqnarray}
E = \langle \hat{H} \rangle & = & - \frac{1}{Z} \, \frac{\partial Z}{\partial \beta} 
\label{ener} \\
E_{\rm k} = \langle \hat{K} \rangle & = & \frac{m}{\beta Z} \, \frac{\partial Z}{\partial m}~,
\label{kintherm}
\end{eqnarray}
where the potential energy comes from the difference $E_{\rm p} = E - E_{\rm k}$. The 
potential energy also can be computed using the expression:
\begin{equation}
\hat{O} (\bm{R}) = -  \frac{1}{\beta} \, \frac{1}{Z(\hat{V})} \, \left. \frac{d Z(\hat{V} + \lambda
\hat{O})}{d \lambda} \right|_{\lambda=0}~,
\label{generaloperator}
\end{equation}
that in general is suitable for estimating operators that depend only on particle 
coordinates. We note that the kinetic energy expression in Eq.~(\ref{kintherm}), 
which is known as the ``thermodynamic'' estimator, presents some technical drawbacks 
like for instance a diverging variance when the number of beads is large. Several 
solutions have been proposed to overcome this limitation, among which we highlight 
the ``virial'' estimator introduced by Cao and Berne (1989).

An alternative to the discussed decomposition scheme of the exponential
operator, is to use a pair product approximation (Ceperley, 1995). In 
this case, one approximates the density matrix by a factorization of 
correlations up to second order, resembling the Jastrow approximation 
used for the ground state, namely:
\begin{equation}
\rho({\bm R},{\bm R}^\prime;\varepsilon)= \prod_{i=1}^{N} \rho_1({\bm
r}_j,{\bm r}_j^\prime;\varepsilon)
\prod_{i<j}^{N} \hat{\rho}_2({\bm r}_{ij},{\bm r}_{ij}^\prime;\varepsilon)~.
\label{rho2}
\end{equation}  
In Eq.~(\ref{rho2}), $\rho_1$ represents the density matrix for a
non-interacting system and $\hat{\rho}_2$ the normalized pair density
matrix, that is:
\begin{equation}
\hat{\rho}_2({\bm r}_{ij},{\bm r}_{ij}^\prime;\varepsilon) = 
\frac{\rho_2({\bm r}_{ij},{\bm r}_{ij}^\prime;\varepsilon)}
{\rho_2^0({\bm r}_{ij},{\bm r}_{ij}^\prime;\varepsilon)}~,
\label{rho2tot}
\end{equation} 
in which $\rho_2$ and $\rho_2^0$ are the relative density matrices of the
interacting and non-interacting systems, respectively. The pair action is 
specially useful when the pair density matrix is known analytically or  
an accurate approximation of it is at hand. Application of this approach 
is particularly suitable for the study of central potentials, although  
it is not restricted to this type of interactions (Pierleoni and Ceperley, 
2006). 

The formalism already explained in this section applies only to distinguishable particles 
(i. e., ``boltzmanons''), since the symmetry requirement under exchange of particles has
been neglected systematically. In order to describe correctly quantum Bose crystals one 
needs to symmetrize the corresponding thermal density matrix, namely:
\begin{equation}
\rho_s({\bm R},{\bm R}^\prime;\varepsilon)= \frac{1}{N!} \sum_{\cal P} 
\rho ({\bm R},{\cal P} {\bm R}^\prime;\varepsilon)~,
\label{rsimetrica}
\end{equation}
where the summation runs over all possible $N!$ permutations involving the  
particles in the system. In contrast to the boltzmanon case, in which the number
of closed polymers equals the number of particles (${\bm R}_{M+1}={\bm R}_1$,
with $M$ the number of beads), the new boundary condition  ${\bm R}_{M+1}=
{\cal P}{\bm R}_1$ implies that each closed polymer can represent more 
than one particle. The acceptance rate for the proposed permutations then 
increases with the inverse of the temperature; when the thermal wavelength 
$\lambda_T$ is comparable to the mean interparticle distance the size of 
closed polymers becomes macroscopic, originating Bose-Einstein condensation 
and superfluidity (see Sec.~\ref{subsec:quantumvsclassical}).

The fraction of particles occupying the lowest momentum state in a bosonic 
system, i. e., the condensate fraction $n_{0} \equiv n(k=0)$, can be obtained 
from the long-range behavior of the one-body density matrix, defined as:
\begin{equation}
\varrho_{1}(r_{1 1^\prime})= \frac{V}{Z} \int d{\bm r}_2 \ldots d{\bm r}_N ~\rho_s({\bm R}, {\bm R}^\prime;\beta)~,
\label{onebodypimc}
\end{equation}
namely, $n_{0} = \lim_{r \to \infty} \varrho_{1}(r)$. In practice, $\varrho_{1}(r)$ 
is estimated by calculating frequency histograms over distances between ${\bm r}_1$ 
and ${\bm r}_1^\prime$.

Sampling the space of permutations is technically involved because one has to guarantee 
ergodicity. In recent years, the introduction of the worm algorithm has improved 
significantly the efficiency in this type of calculations (Boninsegni, Prokof'ev, 
and Svistunov, 2006a). The idea behind the worm algorithm is to work in an extended 
configuration space with two sectors. In the diagonal sector, termed $Z$, all  
paths are closed, which corresponds to conventional PIMC simulations. In the second 
sector, termed $Z_{G}$, all paths are closed except one, which is called the worm; 
this latter sector, therefore, is non-diagonal. The generalized partition function 
then can be written as:
\begin{equation}
Z_{W} = Z + C Z_{G} ~,
\label{worm1}
\end{equation}
where $C > 0$ is a dimensionless parameter that is fixed during the simulation. 
Parameter $C$ controls the relative statistics between sectors $Z$ and $Z_{G}$. 
In the non-diagonal sector one proposes swap movements that generate multi-particle 
permutations (i. e., by single pair permutations between the worm and closed 
paths), whereas in the diagonal sector particles evolve as boltzmanons.

\subsubsection{Path-integral molecular dynamics}
\label{subsub:rpmd}
In the PIMD formalism, the partition function of a quantum system is approximated with the 
Maxwell-Boltzmann expression:
\begin{eqnarray}
Z &&\approx \frac{1}{N!} \left(\frac{mL}{2 \pi \beta \hbar^{2}} \right)^{3NL/2} \nonumber \\ 
         && \times \int \prod_{j=1}^{N} \prod_{i=1}^{L} d{\bf r}_{ij}~ e^{-\beta \left( E_{\rm k} + E_{\rm p} \right)}~.  
\label{eq:zpimd}
\end{eqnarray}
The equation above completely disregards possible quantum atomic exchanges stemming from the indistinguishability 
of the atoms (in contrast to the PIGS and PIMC methods, see Secs.~\ref{subsub:pigs} and~\ref{subsub:pimc}); 
that is, particles are treated as boltzmanons. Nevertheless, in the case of quantum crystals it is well-known 
that the role of quantum statistics is secondary at moderate and high temperatures (e. g., $T > 100$~K in hydrogen at 
$P \sim 100$~GPa, see McMahon \emph{et al.}, 2012). In those situations, PIMD can be used to compute, for instance, 
quantum time-correlation functions and transition state rates in a very efficient manner (Gillan, 1990; Habershon 
\emph{et al.}, 2013; Herrero and Ram\'irez, 2014). 

The key idea behind PIMD is to formulate a Hamiltonian framework in which new space coordinates and momenta, 
$\left( {\bf u}_{ij} , {\bf p}_{ij} \right)$, are introduced for sampling the integral in Eq.(\ref{eq:zpimd}) 
with molecular dynamics techniques. In particular, the new space coordinates and momenta are referred to 
the staging modes, ${\bf u}_{ij}$, that diagonalize the harmonic energy term, namely:
\begin{eqnarray}
E_{\rm k}   &&=  \frac{m L}{2 \beta^{2} \hbar^{2}} \sum_{j=1}^{N} \sum_{i=1}^{L} \left({\bf r}_{ij}-{\bf r}_{(i+1)j}\right)^{2} \nonumber \\ 
            &&=  \sum_{j=1}^{N} \sum_{i=2}^{L} \frac{m_{i} L}{2 \beta^{2} \hbar^{2}} {\bf u}_{ij}^{2}~. 
\label{eq:hpimd}
\end{eqnarray}
For a given atom $j$, the staging mode coordinates are defined as ${\bf u}_{1j} = {\bf r}_{1j}$, and 
${\bf u}_{ij} = {\bf r}_{ij} - \frac{i-1}{i}  {\bf r}_{(i+1)j} - \frac{1}{i} {\bf r}_{1j} $ in the rest of  
cases; the corresponding staging mode masses are $m_{1} = 0$, and $m_{i} = \frac{i}{i-1}m$ in the rest of 
cases.

The momentum variables that are required for the molecular dynamics algorithm to work, are introduced 
through the substitution of the pre-factor in the partition function by a Gaussian integral of the form:    
\begin{equation}
\left( \frac{mL}{2 \pi \beta \hbar^{2}} \right)^{3NL/2} = 
C \int \prod_{j=1}^{N} \prod_{i=1}^{L} d{\bf p}_{ij}~ e^{\left( \frac{-\beta {\bf p}_{ij}^{2}}{2 \chi_{i}} \right)}~,
\label{eq:ppimd}
\end{equation}
where $C$ is a constant that depends on the staging momentum masses, but which has no influence on the calculation 
of the equilibrium properties; ${\bf p}_{ij}$ is the $i$ staging momentum of particle $j$. Masses, $\chi_{i}$, in 
Eq.~(\ref{eq:ppimd}) can be defined as $\chi_{1} = m$, and $\chi_{i} = m_{i}$ in the rest of cases; essentially, 
these must be chosen so that all $i > 1$ staging modes evolve in the same time scale. 
 
In either $(N,V,T)$ or $(N,P,T)$ PIMD simulations, control of the temperature is achieved through a massive thermostatting of 
the system that implies a chain of Nos\'e-Hoover thermostats coupled to each staging variable ${\bf u}_{ij}$ (Tuckerman and 
Hughes, 1998). The involved thermostats introduce friction terms in the corresponding dynamic equations and thus the dynamics 
of the quantum system is not longer Hamiltonian. Nevertheless, it is always possible to define a quantity with units of energy 
that is well conserved during the simulation and that can be used to check whether integration of the equations of motion is 
being done correctly (Martyna, Tuckerman, and Hughes, 1999). Finally, we note that equivalent estimators in the PIMD and 
PIMC frameworks may present some differences, but only in the terms involving momentum variables. Nevertheless, in those cases 
in which quantum atomic exchanges can be safely neglected, both PIMD and PIMC approaches should provide identical expectation 
values, as it follows from the equipartition theorem (Herrero and Ram\'irez, 2014).   

A detailed account of the path-integral molecular dynamics (PIMD) method certainly is out of the scope of the
present review. The details of this technique have been described thoroughly by Tuckerman and Hughes 
(1998) and Martyna, Tuckerman, and Hughes (1999), hence we refer the interested reader to those works.

\subsubsection{Quantum thermal baths}
\label{subsub:qtherbath}
The key idea behind quantum thermal baths (QTB) is to use a Langevin-type approach in which a dissipative force and a 
Gaussian random force are adjusted to have the power spectral density given by the quantum fluctuation-dissipation 
theorem (Dammak \emph{et al.}, 2009; Ceriotti, Bussi, and Parrinello, 2009; Barrat and Rodney, 2011). In doing this, 
the internal energy of the system can be mapped into that of an ensemble of harmonic oscillators whose vibrational modes 
follow a Bose-Einstein distribution. It is worth noticing that while such a quantum discretisation is applied to the 
energy, the atoms in the system are invariably treated as distinguisable particles. Consequently, QTB are not well
suited for describing physical phenomena in which quantum atomic exchanges are important, which typically occur in 
disordered and \emph{incommensurate} systems at low temperatures (in contrast to the PIGS and PIMC methods, see 
Secs.~\ref{subsub:pigs} and~\ref{subsub:pimc}).
  
In analogy to the classical Langevin thermostat method, each particle is coupled to a fictitious bath by introducing 
a random force and a dissipation term in the equations of motion of the form:
\begin{equation} 
m \frac{d^{2}{\bf r}}{dt^{2}} = {\bf F}({\bf r}) - \gamma m \frac{d{\bf r}}{dt} + \sqrt{2m\gamma}\Theta(t)~, 
\label{eq:qtb}
\end{equation}
where ${\bf r}$ and ${\bf F}$ represent the atomic positions and total forces exerted by the rest of particles, respectively. 
Function $\Theta(t)$ is a colored noise with a power spectral density that follows the Bose-Einstein distribution, namely:
\begin{eqnarray}
\tilde{\Theta}(\omega) &&= \int e^{-i \omega t} \langle \Theta(t) \Theta(t') \rangle dt \nonumber \\
                       &&= \hbar \omega \left( \frac{1}{2} + \frac{1}{e^{\hbar \omega/k_{\rm B}T}-1} \right)~,
\label{eq:theta}
\end{eqnarray}
which takes into account the zero-point energy of the system as given by the quasi-harmonic approximation 
(see Sec.~\ref{subsub:approxzeroT}).  

In practice, $\tilde{\Theta}(\omega)$ can be generated by using a signal-processing method based on filtering 
of white noise (Barrat and Rodney, 2011). The implementation of QTB in a discrete MD algorithm then is quite 
straightforward. QTB neither slow down the calculations appreciably nor are detrimental in terms of memory 
requirements. For these reasons, the use of QTB for simulation of QNE is becoming increasingly more popular 
in recent years (Hern\'andez-Rojas, Calvo, and Gonz\'alez-Noya, 2015). 

A word of caution, however, must be added here. QTB alone fail to reproduce the correct quantum behavior in highly 
anharmonic systems and processes (Ceriotti, Bussi, and Parrinello, 2009; Barrozo and de Koning, 2011; Bedoya-Mart\'inez, 
Barrat, and Rodney, 2014). Consequently, the conclusions attained with QTB-based methods should be always validated 
against results obtained with more accomplished quantum approaches (e. g., PIMC and PIMD). Meanwhile, it has been 
recently demonstrated that QTB can be used to accelerate noticeably the convergence in PIMD calculations (Ceriotti, 
Manolopoulos, and Parrinello, 2011; Ceriotti and Manolopoulos, 2012; Brieuc, Dammak, and Hayoun, 2016). In particular, 
generalized Langevin thermostats allow to sample the canonical distribution more efficiently by reducing the usual 
ergodic problems encountered in path-integral simulations performed with a large number of beads. It is probably in 
this latter context that QTB techniques can be particularly useful.  

\begin{table*}
\centering
\begin{tabular}{c c c c c}
\hline
\hline
$ $ & $ $ & $ $ & $ $ & $ $  \\
${\rm Package}$ &\quad ${\rm Capabilities}$ \quad& \quad ${\rm Parallelisation}$ \quad& \quad${\rm License}$ \quad & ${\rm Reference}$ \\
$ $ & $ $ & $ $ & $ $ & $ $  \\
\hline
$ $ & $ $ & $ $ & $ $ & $ $  \\
${\rm ABINIT}$  & ${\rm PIMD}$        & ${\rm CPU} $     & ${\rm Free}$ & ${\rm (Gonze ~\emph{et al.}, ~2016)} $ \\
$ $             & $ $ & $ $ & $ $ & $ $ \\
${\rm CASINO}$  & ${\rm VMC/DMC}$     & ${\rm CPU} $     & ${\rm Free}$ & ${\rm (Needs ~\emph{et al.}, ~2010)} $ \\
$ $             & $ $ & $ $ & $ $ & $ $ \\
${\rm CHAMP}$  & ${\rm VMC/DMC}$     & ${\rm CPU} $     & ${\rm Free}$ & ${\rm (Umrigar ~\emph{et al.}, ~2007)} $ \\
$ $             & $ $ & $ $ & $ $ & $ $ \\
${\rm CP2K}$    & ${\rm PIMD}$        & ${\rm CPU} $     & ${\rm Free}$ & ${\rm (Hutter ~\emph{et al.}, ~2014)} $ \\
$ $             & $ $ & $ $ & $ $ & $ $ \\
${\rm CPMD}$    & ${\rm PIMD}$        & ${\rm CPU} $     & ${\rm Free}$ & ${\rm (Marx, ~Tuckerman, ~and~Martyna, ~1999)} $ \\
$ $             & $ $ & $ $ & $ $ & $ $ \\
${\rm i-PI}$    & ${\rm PIMD}$        & ${\rm CPU} $     & ${\rm Free}$ & ${\rm (Ceriotti, ~More, ~and~Manolopoulos, ~2014)} $ \\
$ $             & $ $ & $ $ & $ $ & $ $ \\
${\rm openMM}$  & ${\rm PIMD}$        & ${\rm CPU/GPU}$  & ${\rm Free}$ & ${\rm (Ceriotti~\emph{et al.}, ~2010a)} $ \\
$ $             & $ $ & $ $ & $ $ & $ $ \\
${\rm PIMC++}$  & ${\rm PIMC}$        & ${\rm CPU} $     & ${\rm Free}$ & ${\rm (Clark~and~Ceperley, ~2008)} $ \\
$ $             & $ $ & $ $ & $ $ & $ $ \\
${\rm pi-QMC}$  & ${\rm PIMD}$        & ${\rm CPU} $     & ${\rm Free}$ & ${\rm (Shumway, ~2005)} $ \\
$ $             & $ $ & $ $ & $ $ & $ $ \\
${\rm QL}$      & ${\rm VMC}$         & ${\rm GPU} $     & ${\rm Free}$ & ${\rm (Lutsyshyn, ~2015)} $ \\
$ $             & $ $ & $ $ & $ $ & $ $ \\
${\rm QMCPACK}$ & ${\rm VMC/DMC}$ & ${\rm CPU/GPU}$      & ${\rm Free}$ & ${\rm (Kim~\emph{et~ al.}, ~2012;~Esler~\emph{et~ al.}, ~2012)} $ \\
$ $             & $ $ & $ $ & $ $ & $ $ \\
${\rm QSATS}$ & ${\rm PIGS}$      & ${\rm CPU}$          & ${\rm Free}$ & ${\rm (Hinde, ~2011)} $ \\
$ $             & $ $ & $ $ & $ $ & $ $ \\
${\rm QWALK}$ & ${\rm VMC/DMC}$     & ${\rm CPU}$       & ${\rm Free}$ & ${\rm (Wagner, ~Bajdich, ~and~Mitas, ~2009)} $ \\
$ $             & $ $ & $ $ & $ $ & $ $ \\
\hline
\hline
\end{tabular}
\label{tab:codes}
\caption{List of computer simulation packages that allow to simulate quantum nuclear effects in periodic systems.
         PIMD, PIMC, VMC, DMC, and PIGS in the ``Capability'' row stand for, path-integral molecular dynamics, 
         path-integral Monte Carlo, variational Monte Carlo, diffusion Monte Carlo, and ground-state path-integral
         Monte Carlo, respectively. CPU and GPU in the ``Parallelisation'' row stand for central and graphical 
         processing units.}
\end{table*}

%===============================================================
\subsection{Computer packages}
\label{subsec:codes}
%===============================================================
While the number of classical simulation packages, either commercially or freely available, is 
practically countless, the number of computer packages that allow to simulate QNE is very 
limited. In Table~I, we list those computer packages that, to the best of our knowledge, 
are publicly available and can be used to simulate QNE in periodic systems, along with a brief 
description of their basic capabilities. In total, they amount to a bit more than ten.   

We note that PIMD (see Sec.~\ref{subsub:rpmd}) is the method that is implemented most frequently. 
On the other hand, quantum Monte Carlo techniques (i. e., VMC and DMC) are available in fewer 
codes. Although it is not indicated in Table~I, most of the listed simulation packages also 
allow to describe the interactions between atoms through \emph{ab initio} methods (see next 
Sec.~\ref{subsec:firstprinciples}). In addition to this, they are all designed to run in 
high-performance computing architectures and can be downloaded free of charge from internet
or made available on request.  

A likely reason behind the scarcity of studies considering QNE may be, apart from the increased 
computational and technical burdens, the limited number of available quantum simulation packages. 
We note that most of the codes in Table~I are relatively new, hence until recently any researcher 
interested in simulating QNE had to craft his/her own quantum implementation. Nevertheless, we 
expect that due to the steady growth in computing power and the increasing awareness of the 
importance of QNE in condensed matter systems and materials, the availability and user-friendliness
of quantum simulation packages will increase in the next years.

%===============================================================
%===============================================================
\section{Modeling of atomic interactions}
\label{sec:modeling}
%===============================================================
%===============================================================
The simulation techniques that are used to describe the atomic interactions in quantum crystals,
and materials in general, can be classified in two major categories: ``semi-empirical'' and 
``first-principles''. In semi-empirical approaches, the interparticle forces are typically modeled 
with analytical functions, known as force fields or classical potentials, that are devised to 
reproduce a particular set of experimental data or the results of highly accurate calculations. The 
inherent simplicity of classical potentials makes it possible to address the study of quantum 
solids within ample thermodynamic intervals and large length/time scales, with well-established quantum 
simulation techniques like the ones discussed in Sec.~\ref{sec:simulation}. By using semi-empirical 
potentials and exploiting the current computational power and algorithm development, quantum simulations 
of condensed matter systems can be routinely performed nowadays in multi-core processors. Nevertheless, 
in spite of their great versatility, classical potentials may sometimes present some impeding transferability 
issues. Transferability issues are related to the impossibility of mimicking the targeted systems at 
conditions different from those in which the setup of the corresponding force field was performed. 
An illustrative example of such a failure is given by the unreliable description of highly compressed
rare-gas crystals with pairwise potentials (Cazorla and Boronat, 2008a; Cazorla and Boronat, 2015a; 
Cazorla and Boronat, 2015b). In addition to this, there are many physical phenomena that simply cannot 
be reproduced accurately with straightforward force fields (e. g., magnetic spin interactions, 
electronic screening effects, and oxidation-state changes, to cite just a few examples).  

In this context, the output of first-principles calculations, also known as \emph{ab initio}, turns 
out to be crucial. In first-principles approaches, as the name indicates, no empirical information is 
assumed on the derivation of the atomic interactions: these are directly obtained from applying the 
principles of quantum mechanics to the electrons and nuclei. Transferability issues, therefore, are 
absent. First-principles approaches are in general very accurate, but they can be also very demanding 
in terms of computational expense. This circumstance makes the full \emph{ab initio} study of quantum 
crystals, that is, in which both the electronic and nuclear degrees of freedom are treated quantum 
mechanically, intricate and computationally very demanding (see, for instance, Pierleoni, Ceperley, 
and Holzmann, 2004; Pierleoni and Ceperley, 2005; Pierleoni and Ceperley, 2006; McMahon \emph{et al.}, 
2012). Common acceleration schemes within first-principles schemes involve the use of pseudopotentials 
(see, for instance, Vanderbilt, 1990; Troullier and Martins, 1991), which avoids to treat explicitly 
the core electrons. This approximation is based on the fact that many materials properties can be 
predicted by focusing exclusively on the behavior of valence electrons. Nonetheless, pseudopotentials 
can actually be the source of potential errors. Fortunately, some strategies can be used to minimize the 
impact of the approximations introduced by pseudopotentials like, for instance, the projector augmented 
wave method (Bl\"{o}chl, 1994) and linearized augmented plane waves (Andersen, 1975). Next, we concisely 
explain some basic aspects of first-principles and semi-empirical methods as related to the study 
of quantum solids.   

%===============================================================
\subsection{First-principles methods}
\label{subsec:firstprinciples}
%===============================================================
In solids, the dynamics of electrons and nuclei can be decoupled to a good approximation because 
their respective masses differ by several orders of magnitude. The wave function of the corresponding 
many-electron system, $\Psi ({\bf r}_{1}, {\bf r}_{2},...,{\bf r}_{N})$, therefore can be determined 
by solving the Schr\"{o}dinger equation involving the non-relativistic Born-Oppenheimer Hamiltonian:
\begin{eqnarray}
H = -\frac{1}{2} \sum_{i} {\bf \nabla}^{2}_{i} - \sum_{I} \sum_{i} \frac{Z_{I}}{|{\bf R}_{I} - {\bf r}_{i}|} \nonumber \\
        + \frac{1}{2} \sum_{i} \sum_{j \neq i} \frac{1}{|{\bf r}_{i} - {\bf r}_{j}|}~,
\label{eq:BO-hamilton}
\end{eqnarray}
where $Z_{I}$ are the nuclear charges, ${\bf r}_{i}$ the positions of the electrons, and ${\bf R}_{I}$ 
the positions of the nuclei, which are considered fixed. (We note that non-adiabatic effects beyond 
the Born-Oppenheimer approximation in principle can be also treated within first-principles methods 
by using wave functions that explicitly depend on the electronic and nuclear degrees of freedom; see, 
for instance, Ceperley and Alder, 1987; Tubman \emph{et al.}, 2014; Yang \emph{et al.}, 2015.) In real 
materials $\Psi$ is a complex mathematical function that in most cases is unknown. Electrons are fermion 
particles, hence their wave function must change sign when two of them exchange orbital states. This quantum 
antisymmetry leads to an effective repulsion between electrons, called the Pauli repulsion, that helps in 
lowering their total Coulomb energy. At the heart of any first-principles method is to find a good approximation 
to $\Psi$, or an equivalent solution, that is manageable enough to perform calculations and simultaneously 
describes the system of interest correctly. Examples of \emph{ab initio} methods include density functional 
theory (DFT), M\o ller-Plesset perturbation theory (MP2), the coupled-cluster method with single, double and 
perturbative triple excitations [CCSD(T)], and electronic quantum Monte Carlo (eQMC), to cite just a few. 
From these, DFT and eQMC have been most intensively applied to the study of quantum solids and for this 
reason we summarise their foundations in what follows.   

\subsubsection{Density functional theory}
\label{subsubsec:dft}
In 1965, Kohn and Sham developed a pioneering theory to effectively calculate the energy and properties 
of many-electron systems without the need of explicitly knowing $\Psi$ (Kohn and Sham, 1965; Sham and Kohn, 
1966). The main idea underlying this theory, called density functional theory (DFT), is that the exact 
ground-state energy, $E$, and electron density, $n({\bf r})$, can be determined by solving an effective 
one-electron Schr\"odinger equation of the form:
\begin{equation}
H_{eff} \psi_{i \sigma} = \epsilon_{i \sigma} \psi_{i \sigma}~, 
\label{eq:onelectron}
\end{equation}
where index $i$ labels different one-electron orbitals and $\sigma$ the corresponding spin state.
In particular,
\begin{equation}
H_{eff} = -\frac{1}{2}\nabla^{2} + V_{ext}({\bf r}) + \int \frac{n({\bf r'})}{|{\bf r} - {\bf r'}|} d{\bf r'} + V_{xc}({\bf r})~,
\label{eq:heff}
\end{equation}
and
\begin{equation}
n({\bf r}) = \sum_{i \sigma} |\psi_{i \sigma} ({\bf r})|^{2}~,
\label{eq:density}
\end{equation}
where $V_{ext}$ represents an external field and
$V_{xc} ({\bf r}) = \delta E_{xc} / \delta n ({\bf r})$
is the exchange-correlation potential.

The exchange-correlation energy has a purely quantum mechanical origin and can be defined as the
interaction energy difference between a quantum many-electron system and its classical counterpart.
Despite $E_{xc}$ represents a relatively small fraction of the total energy, this contribution is
extremely crucial for all materials and molecules because it acts directly on the bonding between
atoms. In general, $E_{xc} [n]$ is unknown and needs to be approximated. This is the only source
of fundamental error in DFT methods. The exact form of the exchange-correlation energy can be readily 
expressed through the adiabatic connection fluctuation-dissipation theorem as (Langreth and Perdew, 
1975; Nguyen and de Gironcoli, 2009):
\begin{equation}
E_{xc} [n] = \int n({\bf r})~d{\bf r}  \int \frac{n_{xc}({\bf r} , {\bf r'})}{|{\bf r} - {\bf r'}|}~d{\bf r'}~,
\label{eq:exc}
\end{equation}
where $n_{xc}({\bf r} , {\bf r'}) = n_{x} ({\bf r} , {\bf r'}) + n_{c} ({\bf r} , {\bf r'})$ is the 
exchange-correlation hole density at position ${\bf r'}$ surrounding an electron at position {\bf r}. 
Some important constraints on $n_{xc}({\bf r} , {\bf r'})$ are already known. For instance, $n_{x} ({\bf r} 
, {\bf r'})$ must be non-positive everywhere and its space integral is equal to $-1$. Also, the space 
integral of the correlation hole density is zero. These constraints can be employed in the construction 
of approximate $E_{xc} [n]$ functionals.

In standard DFT approaches $E_{xc} [n]$ is approximated with the expression:
\begin{equation}
E_{xc}^{approx}[n] = \int \epsilon_{xc}^{approx}({\bf r}) n({\bf r}) d{\bf r}~,
\label{eq:excapprox}
\end{equation}
where $\epsilon_{xc}^{approx}$ is made to depend on $n({\bf r})$, $\nabla n({\bf r})$,
and/or the electronic kinetic energy
$\tau ({\bf r}) = \frac{1}{2} \sum_{i \sigma} |\nabla \psi_{i \sigma} ({\bf r})|^{2}$~.

Next, we summarise the basic aspects of the most popular $E_{xc} [n]$ functionals found in computational 
studies of condensed matter systems and materials. Additional details on these topics can 
be found in recent and more specialized reviews (see, for instance, Perdew, 2013; Klime\v{s} and Michaelides 
2012; Dobson and Gould, 2012; Cazorla, 2015). We note that the current number of commercially available and 
open-source DFT computer packages is huge (at least in comparison to that of eQMC codes); a reference to some 
of them can be found, for instance, in Cazorla (2015).

\paragraph{Local and Semi-Local Functionals.}
In local approaches (e. g., local density approximation -LDA-), $E_{xc}$ is approximated
with Eq.~(\ref{eq:excapprox}) and the exchange-correlation energy is taken to be equal to
that in an uniform electron gas of density $n ({\bf r})$, namely $\epsilon_{xc}^{unif}$.
The exact $\epsilon_{xc}^{unif} [n]$ functional is known numerically from quantum
Monte Carlo calculations (Perdew and Zunger, 1981; Ceperley and Alder, 1980). In order to deal 
with the non-uniformity in real electronic systems, the space is partitioned into infinitesimal 
volume elements that are considered to be locally uniform. In semi-local approaches (e. g., 
generalized gradient approximation -GGA-), $E_{xc}$ is approximated also with Eq.~(\ref{eq:excapprox}) 
but $\epsilon_{xc}^{approx}$ is made to depend on $n({\bf r})$ and its gradient 
$\nabla n({\bf r})$ (Perdew \emph{et al.}, 1992; Perdew \emph{et al.}, 1996).
Both local and semi-local approximations satisfy certain exact $E_{xc}$ constraints
(e. g., some exact scaling relations and the exchange-correlation hole sum rules)
and can work notably well for systems in which the electronic density varies slowly
over the space (e. g., bulk crystals at equilibrium conditions). However, by construction
local and semi-local functionals cannot account for long-range electronic correlations,
otherwise known as dispersion interactions, which certainly are ubiquitous in quantum
crystals.

\paragraph{Hybrid Exchange Functionals.}
Hybrid functionals comprise a combination of non-local exact Hartree-Fock and local
exchange energies, together with semi-local correlation energies. The proportion in which both
non-local and local exchange densities are mixed generally relies on empirical rules.
The popular B3LYP approximation (Becke, 1993), for instance, takes a $20$~\% of the exact HF exchange 
energy and the rest from the GGA and LDA functionals. Other well-known hybrid functionals
are the HSE proposed by Heyd-Scuseria-Ernzerhof (Heyd \emph{et al.}, 2003), PBE0 (Adamo and Barone, 
1999), and the family of Minnesota meta hybrid GGA (Zhao \emph{et al.}, 2005).
In contrast to local and semi-local functionals, hybrids can describe to some extent the delocalisation
of the exchange-correlation hole around an electron. This characteristic is specially useful when
dealing with strongly correlated systems containing $d$ and $f$ electronic orbitals (e. g., perovskite 
oxides). Hybrid functionals, however, do not account for the long range part of the correlation hole 
energy and thus cannot reproduce dispersion forces. Effective ways to correct for these drawbacks have 
been proposed by several authors (Chai and Head-Gordon, 2008; Lin \emph{et al.}, 2013; Mardirossian 
and Head-Gordon, 2014).

\paragraph{Dispersion-Corrected Functionals.}
DFT-based dispersion schemes reproduce the asymptotic $1/r^{6}$ interaction between two particles 
separated by a distance $r$ in a gas. The most straightforward way of achieving this consists 
in adding an attractive energy term to the exchange-correlation energy of the form $E_{\rm disp} = 
-\sum_{i,j} C_{ij} / r^{6}_{ij}$ (indexes $i$ and $j$ label different particles). This approximation 
represents the core of a suite of methods named DFT-D that, due to their simplicity and low computational 
cost, are being employed widely (Grimme, 2004). Nevertheless, DFT-D methods present some inherent 
limitations. For instance, many-body dispersion effects and faster decaying terms such as the $B_{ij} 
/ r^{8}_{ij}$ and $C_{ij} / r^{10}_{ij}$ interactions are completely disregarded. Also, it is not totally 
clear from where one should obtain the optimal $C_{ij}$ coefficients. Several improvements on DFT-D 
methods have been proposed, in which the value of the dispersion coefficients are made to depend somehow 
on the specific atomic environment. Examples of those include the DFT-D3 method by Grimme (Grimme 
\emph{et al.}, 2010), the vdW(TS) approach by Tkatchenko and Scheffler (2009), and the BJ model by Becke 
and Johnson (2007). A further degree of elaboration exists in which no external input parameters are needed 
and the dispersion interactions are directly computed from the electron density. In this context, the 
exchange-correlation energy is expressed as $E_{xc} = E_{x}^{\rm GGA} +  E_{c}^{\rm LDA} + E_{c}^{\rm nl}$, 
where $E_{c}^{\rm nl}$ is the non-local correlation energy. $E_{c}^{\rm nl}$ can be calculated as a double 
space integral involving the electron density and a two-position integration kernel. This approach, 
introduced by Dion \emph{et al.} (2004), represents a key development in DFT methods as it combines all 
types of interaction ranges within a same formula. Refinements of this scheme have been proposed recently
in which the original two-position integration kernel is modified (Vydrov and Voorhis, 2012), or the exchange 
term in $E_{xc}$ is replaced with other more accurate functionals (Lee \emph{et al.}, 2010; Carrasco 
\emph{et al.}, 2011).

\subsubsection{Electronic quantum Monte Carlo}
\label{subsubsec:qmc}
Here we explain the basics of the diffusion Monte Carlo (DMC) method (see Sec.~\ref{subsub:dmc}) 
as applied to the study of many-electron systems [for a more technical and complete discussion on
this topic see, for instance, Foulkes \emph{et al.} (2001) and Towler (2006)]. In electronic quantum
Monte Carlo (eQMC) methods one deals explicitly with the solution to the imaginary-time dependent 
Schr\"{o}dinger equation (in contrast to DFT methods). The quantum antisymmetry of the electrons leads 
to the so-called ``sign problem'', that is related to the fact that the probability distribution function 
$f = \Psi_{T} \Psi_{0}$ is not positive definite everywhere (see Sec.~\ref{subsub:dmc}). If the 
nodes of the guiding and true ground-state wave functions [that is, the $3N-1$-dimensional surfaces 
at which $\Psi ({\bf r}_{1}, {\bf r}_{2},...,{\bf r}_{N}) = 0$] were coincident, the sign problem  
would disappear. However, in most many-electron problems this condition is never satisfied. Several 
approaches have been proposed in the literature to tackle the sign problem, among which we highlight 
the ``fixed-node'' and ``released-node'' methods. 

\paragraph{Fixed-node method.}
In this method the nodes of the ground-state function $\Psi_{0}$ are forced to be equal to those of 
the guiding wave function $\Psi_{T}$ [see, Anderson (1975) and (1976)]. As a result, the probability 
distribution function that is asymptotically sampled is always positive because a change of sign in 
$\Psi_{T}$ is replicated by a change of sign in $\Psi_{0}$. By using this approximation, however, one
always obtains results that are upper bounds to the exact ground-state energy (Reynolds \emph{et al.}, 
1982). When dealing with fermionic systems, therefore, it is critically important to choose guiding 
wave functions with high quality nodal surfaces. This requirement is also necessary for guaranteeing 
numerical stability in the simulations, since the divergence of the drift force, 
${\bf F} = 2 {\bf \nabla} \Psi_{T}/ \Psi_{T}$, close to a node cannot always be counteracted by the 
replication of energetically favorable configurations.

The fixed-node electronic DMC (FN-DMC) method was first applied to the electron gas by Ceperley and Alder 
(1980). Subsequently, it was employed to study solid hydrogen (Ceperley and Alder, 1987) and other crystals 
containing heavier atoms (Fahy \emph{et al.}, 1988, 1990; Li \emph{et al.}, 1991). Wigner crystals in two 
and three dimensions have been also investigated thoroughly with eQMC methods (Tanatar and Ceperley, 1989;
Drummond \emph{et al.}, 2004; Drummond and Needs, 2009). Important FN-DMC developments include the introduction 
of variance minimization techniques to optimize wave functions (Umrigar \emph{et al.}, 1988) and the use of 
non-local pseudopotentials (Hammond \emph{et al.}, 1987; Hurley and Christiansen, 1987; Fahy \emph{et al.}, 
1988; Mitas \emph{et al.}, 1991; Trail and Needs, 2013; Lloyd-Williams, Needs, and Conduit, 2015; Trail
and Needs, 2015). We also highlight the generalisation of eQMC methods to systems with broken time-reversal 
symmetry (e. g., interacting electrons in an applied magnetic field or states with non-zero angular momentum), 
which is known as the ``fixed-phase'' approximation (Ortiz, Ceperley, and Martin, 1993). These improvements, 
together with a certain availability of commercial and open source simulation packages (see Sec.~\ref{subsec:codes}), 
have stimulated the study of a wide range of electronic systems with DMC like, for instance, strongly correlated 
oxide materials (Huihuo and Wagner, 2015; Wagner, 2015), hydrates (Alf\`e \emph{et al.}, 2013; Cox \emph{et al.}, 
2014), and organic molecules (Purwanto \emph{et al.}, 2011; Jiang \emph{et al.}, 2012). 

\paragraph{Released-node method.}
In the released node (RN) method the nodal constraints imposed by the guiding function are relaxed in order 
to adapt to those of the exact wave function (Ceperley and Alder, 1980; Ceperley and Alder, 1984; Hammond 
\emph{et al.}, 1994; Tubman \emph{et al.}, 2011). As we explain next, this technique provides a 
solution that is not stable in imaginary time and thus its use is restricted to systems in which (i)~the nodes 
of the guiding function are relatively accurate, or (ii)~the Pauli principle is relatively unimportant (that 
is, the energy difference between the many-fermion system and its many-boson counterpart is small).

An arbitrary antisymmetric wave function, $\Psi_{A}$, can always be expressed as a linear combination of two 
positive functions like: 
\begin{equation}
\Psi_{A} \left( {\bf r}, \tau  \right) = \phi_{+} \left( {\bf r}, \tau  \right) - \phi_{-} \left( {\bf r}, \tau  \right)~.
\label{eq:phiA}
\end{equation}
As both $\phi_{+}$ and $\phi_{-}$ are positively defined, each one can be interpreted as a probability density 
and be propagated individually. A convenient definition of $\phi_{\pm}$ at $\tau = 0$ is:  
\begin{equation}
\phi_{\pm} \left( \tau = 0 \right) = \frac{1}{2} \left(  |\Psi_{A}| \pm \Psi_{A} \right)~,
\label{eq:phit0}
\end{equation}
since at large imaginary time the corresponding projected states are:  
\begin{equation}
\phi_{\pm} \left( \tau \to \infty \right) = \pm C_{F} \Psi_{0}^{F} + C_{B} \Psi_{0}^{B} e^{\left( E_{0}^{F} - E_{0}^{B} \right) \tau}~,
\label{eq:phitinf}
\end{equation}
where $\Psi_{0}^{F}$ and $\Psi_{0}^{B}$ are the ground-state fermion and boson wave functions of the Hamiltonian, 
respectively. $\Psi_{A}$ consistently renders the ground-state energy of the fermionic system, namely: 
\begin{equation}
E_{0}^{RN} = \frac{\int \Psi_{A} \left(\tau \to \infty \right) {\rm H} \Psi_{0} d{\bf r}}{\int \Psi_{A} \left(\tau \to \infty \right) \Psi_{0} d{\bf r}} =E_{0}^{F}~. 
\label{eq:energyrn}
\end{equation}
However, since the $E_{0}^{F} - E_{0}^{B}$ energy difference is always positive, the bosonic parts in $\phi_{\pm}$ 
grow exponentially with imaginary time (see Eq.~\ref{eq:phitinf}), leading to an energy variance of: 
\begin{equation}
\sigma \left( E_{0}^{RN} \right) \propto e^{\left( E_{0}^{F} - E_{0}^{B} \right) \tau}~, 
\label{eq:varn}
\end{equation}
that is divergent. For this reason the RN method is classified as a ``transient estimator'' approach. 
Cases in which application of the RN method is judicious, numerical weighting techniques can be used
to reduce considerably the variance of the energy and other quantities (Hammond \emph{et al.}, 1994; 
Tubman \emph{et al.}, 2011). The RN approach can be employed also as a measure of the quality of the 
upper bounds provided by the fixed-node method (Casulleras and Boronat, 2000; Sola, Casulleras, and 
Boronat, 2006).   

\paragraph{Electronic guiding wave functions.}
In eQMC methods, the choice of the guiding function is particularly important because it determines the 
degree of accuracy in the calculations. The most widely used $\Psi_{T}$ model is the Slater-Jastrow wave 
function, that is expressed as:
\begin{equation}
\Psi_{T}({\bf X}) = e^{J({\bf X})} \sum_{j} c_{j}D_{j}({\bf X})~, 
\label{eq:sjwf}
\end{equation}
where ${\bf X} = \left( {\bf x}_{1}, {\bf x}_{2}, \cdots, {\bf x}_{N} \right)$ and ${\bf x}_{i} = \lbrace {\bf r}_{i}, \sigma_{i} \rbrace$
represent the space and spin coordinates of electron $i$, $e^{J}$ is the Jastrow factor, $c_{j}$ are coefficients, and $D_{i}$ 
Slater determinants of single-particle orbitals of the form:
\begin{equation}
 D_{j}({\bf X}) = \left| \begin{array}{cccc}
\phi_{1}^{j}({\bf x}_{1}) & \phi_{1}^{j}({\bf x}_{2}) & \ldots & \phi_{1}^{j}({\bf x}_{N}) \\
\phi_{2}^{j}({\bf x}_{1}) & \phi_{2}^{j}({\bf x}_{2}) & \ldots & \phi_{2}^{j}({\bf x}_{N}) \\
 \vdots & \vdots &  \vdots & \\
\phi_{N}^{j}({\bf x}_{1}) & \phi_{N}^{j}({\bf x}_{2}) & \ldots & \phi_{N}^{j}({\bf x}_{N}) \end{array} \right|~. 
\label{eq:Slaterd}
\end{equation}
The orbitals $\lbrace \phi_{i}^{j} \rbrace$ often are obtained from DFT or Hartree-Fock calculations, 
and are assumed to be products of factors that depend either on the space or spin coordinates. It is
common practice in eQMC calculations to replace $D_{j}$ with products of separate up- and down-spin 
determinants, since this improves computational efficiency (Foulkes \emph{et al.}, 2001). 
 
The Jastrow factor in Eq.~(\ref{eq:sjwf}) normally contains one and two-body terms, namely:
\begin{equation}
J({\bf X}) = \sum_{i}^{N} \chi({\bf x}_{i}) - \frac{1}{2}\sum_{i}^{N}\sum_{j \neq i}^{N} u({\bf x}_{i},{\bf x}_{j})~,
\label{eq:jastrowfacele}
\end{equation}
where functions $u$ describe the electron-electron correlations and $\chi$ the electron-nuclear correlations.
The two-electron terms in Eq.~(\ref{eq:jastrowfacele}) reduce the value of the wave function whenever two electrons 
approach each to the other, hence reducing the repulsive electron-electron interaction energy. However, the 
introduction of $u$ terms also has the unwanted effect of pushing electrons away from regions of high-charge density 
into regions of low-charge density, thus depleting the electronic density in the atomic bonds. By introducing
the one-body functions $\chi$ in the Jastrow factor this problem is overcome. 

Another approach that can be employed to improve the description of electron-electron correlations is to consider 
backflow correlations within the Slater determinants. Backflow correlations were originally derived from a current 
conservation argument due to Feynman and Cohen (1956) to provide a picture of excitations in liquid $^{4}$He; they 
represent the characteristic flow pattern in a quantum fluid where particles in front of a moving one go on filling 
the space left behind it. The introduction of backflow correlations may relax in a practical way the constraints 
associated to the fixed-node approximation. For instance, it has been demonstrated that the use of backflow wave 
functions in homogeneous electron systems reduces significantly the corresponding VMC and DMC energies (Kwon, 
Ceperley, and Martin, 1993; Kwon, Ceperley, and Martin, 1998; L\'opez-R\'ios \emph{et al.}, 2006).

\subsubsection{DFT vs. eQMC}
\label{subsubsec:dftvsqmc}
In the last decade, important methodological progress has been made in the context of DFT calculations that 
allow now to describe the electronic features of many materials adequately. Examples of these advancements 
are explained in Sec.~\ref{subsubsec:dft} and essentially are related to the construction of accurate and 
computationally efficient hybrid exchange and dispersion-corrected functionals. A pending challenge in DFT 
methods, however, is posed by the difficulties encountered in the reproduction of many-body and Coulomb 
screening effects. This type of shortcomings stems from the pairwise additivity that is assumed in the construction 
of most DFT functionals. Essentially, the interaction energy between two atoms completely neglects the effects 
introduced by the medium that separates them (Misquitta \emph{et al.}, 2010; Tkatchenko, Alf\`e, and Kim, 2012; 
Gobre and Tkatchenko, 2013). In this context, the adiabatic connection fluctuation-dissipation theorem has been 
exploited to calculate correlation DFT energies that incorporate many-body terms beyond pairwise. This is the case 
of the random phase approximation to DFT (Dobson, White, and Rubio, 2006) and DFT+MBD methods (Ruiz \emph{et al.}, 
2012; Tkatchenko \emph{et al.}, 2012; Ambrosetti \emph{et al.}, 2014), which at the moment are receiving the highest 
attention. In the latest DFT+MBD versions, for instance, the Schr\"{o}dinger equation of a set of fluctuating and 
interacting quantum harmonic oscillators is solved directly within the dipole approximation, and the resulting 
many-body energy is coupled to an approximate semilocal DFT functional (Tkatchenko \emph{et al.}, 2012; Ambrosetti 
\emph{et al.}, 2014). Many-body DFT-based methods, however, are still in their infancy and the associated computational 
expenses are elevated, hence their applicability yet is limited.

Electronic QMC methods, on the other hand, are inherently exact as they account for any type of electronic 
correlation, exchange, or many-body screening effect (although they are affected by the ``sign problem'' 
explained in previous sections). A further advantage of using eQMC methods is that it is possible to treat 
the zero-point motion of the nuclei beyond the Born-Oppenheimer approximation, that is, considering 
non-adiabatic effects (Ceperley and Alder, 1987; Tubman \emph{et al.}, 2014; Yang \emph{et al.}, 2015). 
This can be done by using wave functions that explicitly depend on both the electronic and nuclear degrees 
of freedom in projector MC schemes (e. g., DMC and GFMC). Another interesting feature of eQMC, in contraposition 
to DFT methods, is that in the case of light atoms the use of pseudopotentials can be avoided; this aspect is 
specially desirable for the study of quatum solids like H$_{2}$ and $^{4}$He, since core electrons then can 
be simulated without any constraints (Morales, Pierleoni, and Ceperely, 2009; Morales \emph{et al.}, 2013). 

On the down side, eQMC methods present some technical difficulties that are absent in DFT calculations. For instance, 
the periodic Ewald sum that is used to estimate the electron-electron interactions introduces a finite-size error in 
the exchange-correlation energy, since it depends on the size and shape of the simulation cell (Foulkes \emph{et al.}, 
2001). Consequently, the use of either increasingly large simulation cells or effective correction schemes (Fraser 
\emph{et al.}, 1996; Hood \emph{et al.}, 1997; Chiesa \emph{et al.}, 2006) is necessary to guarantee proper 
convergence. For a detailed description of finite-size errors treatment in eQMC methods we refer 
the interested reader to the recent and specialised articles by Drummond\emph{et al.} (2008), Ma, Zhang, and Krakauer
(2011), and Holzmann \emph{et al.} (2016). Another intricacy is found in the calculation of the atomic forces. Calculating 
forces using a stochastic algorithm turns out to be very difficult because straightforward derivation of the total energy 
with respect to the atomic positions, as it follows from the Hellmann-Feynman principle, leads to estimators with very 
large variances. Correlated sampling techniques have been proposed to make the statistical errors in the relative energy 
of different geometries much smaller than the errors in the separate energies (Filippi and Umrigar, 2000). Finite difference 
methods, however, become already impractical when considering systems containing a few tens of atoms. Alternative approaches 
based on ``zero-variance'' Hellmann-Feynman estimators (Assaraf and Caffarel, 2000; Chiesa, Ceperley, and Zhang, 2005; Per, 
Russo, and Snook, 2008; Clay III \emph{et al.}, 2016) and sampling of ``pure'' probability distributions (Badinski 
\emph{et al.}, 2010) have been introduced more recently; nevertheless, the central problem of calculating accurate forces 
in extended systems efficiently yet remains. 

The great accuracy of eQMC methods neither comes free of cost. Although the scaling with respect to the number of electrons
is the same than in DFT methods, namely $N^{3}$ in standard cases, the pre-factors in eQMC are considerably larger (e. g., 
roughly $10$ and $100$ times larger in VMC and DMC, respectively; Foulkes \emph{et al.}, 2001; Towler, 2006). Also, the 
convergence of the total energy is achieved more slowly than in DFT due to the usual MC propagation and sampling 
procedures. In spite of this, thanks to the escalating increase in computing efficiency and recent algorithmic advances, 
the use of eQMC methods is transitioning from that of benchmark calculations in few-atoms systems to that of production 
runs in hundreds-of-atoms systems (Kim \emph{et al.}, 2012; Esler \emph{et al.}, 2012; Wagner, 2014). Actually, efficient 
QMC-based methods have been already developed that allow to simulate both the electrons and nuclei in crystals 
quantum mechanically (Grossman and Mitas, 2005; Wagner and Grossman, 2010). Among those, we highlight 
the coupled electron-ion Monte Carlo method due to Pierleoni, Ceperley, and collaborators (Pierleoni, Ceperley, and 
Holzmann, 2004; Pierleoni and Ceperley, 2005; Pierleoni and Ceperley, 2006), for its special relevance to the field of 
quantum solids (see, for instance, Sec.~\ref{subsec:h2extreme}). In view of this progress, we foresee that in the next 
years the use of eQMC techniques will become more popular within the community of computational condensed matter scientists.

%===============================================================
\subsection{Effective interaction models}
\label{subsec:ff}
%===============================================================
Using first-principles methods to describe the interactions between atoms in quantum crystals normally requires 
intensive computational resources. Fortunately, the interactions between particles sometimes are so \emph{simple} 
that they can be approximated with analytical functions known as classical interatomic potentials or force fields. 
In those particular cases one can concentrate in solving the quantum mechanical equations for the nuclear degrees 
of freedom only, hence accelerating the calculations dramatically. Classical interaction models are constructed 
by following physical knowledge and intuition; they normally contain a set of parameters that are adjusted to 
reproduce experimental or \emph{ab initio} data. The force matching method due to Ercolessi and Adams (1994), 
for instance, is a well-established force field fitting technique that is widely employed in computational 
physics and materials science (Masia, Gu\`ardia, and Nicolini, 2014). Nevertheless, the ways in which classical 
interatomic potentials are constructed are neither straightforward nor uniquely defined, and the thermodynamic 
intervals over which they remain reliable are not known \emph{a priori}.

In situations where the use of first-principles methods is prohibitive and the available classical potentials 
are not versatile enough to reproduce the physical phenomena of interest, machine learning techniques 
can be very useful. Machine learning (ML) is a subfield of artificial intelligence that exploits the systematic 
identification of correlation in data sets, to make predictions and analysis (Behler, 2010; Rupp, 2015).  
Effective potentials resulting from ML are not built on physically motivated functional forms but obtained 
from purely mathematical fitting techniques that reproduce a set of reference data as closely as possible. Some of
these fitting procedures strongly rely on the concept of artificial neural networks, which can ``learn'' the topology 
of a potential-energy surface from a set of reference points. ML techniques are common tools in mathematics and 
computer science, and are starting to be applied with confidence in Chemistry (Raghunathan \emph{et al.}, 2015) 
and Physics (Manzhos, Yamashita, and Carrington, 2009; Li, Kermode, and De Vita, 2015).

\subsubsection{Classical potentials}
\label{subsubsec:classicalpot}
The interactions between atoms in quantum solids have been traditionally modeled with two-body potentials. 
The most popular of all them is the Lennard-Jones (LJ) potential, which is expressed as:  
\begin{equation}
V_{2}^{\rm LJ} (r) = 4 \epsilon \left[ \left( \frac{\sigma}{r}\right)^{12} - \left(\frac{\sigma}{r}\right)^{6} \right]~,
\label{eq:lj}
\end{equation}
where $\epsilon$ and $\sigma$ are free parameters, and $r$ is the distance between two particles. The first 
term in Eq.~(\ref{eq:lj}) represents  repulsive short-ranged electrostatic and Pauli-like interactions 
acting between electrons; the second term represents the attractive long-ranged van der Waals interactions 
resulting from instantaneous electronic dipoles. In spite of its simplicity, the LJ potential has been 
used in the study of condensed matter systems with great success; it was the first interaction model
to be systematically employed in variational Monte Carlo simulations of quantum solids (Hansen and Levesque, 1968; 
Hansen, 1968; Bruce, 1972). The LJ potential is convenient also for simulating atomic systems composed of several 
chemical species for which the corresponding $\sigma$'s and $\epsilon$'s are already known; the resulting crossed 
interactions then can be approximated to a good extent with the LJ parameters given by the Lorentz-Berthelot rules: 
$\sigma_{ij} = \frac{\sigma_{ii} + \sigma_{jj}}{2}$ and $\epsilon_{ij} = \sqrt{\epsilon_{ii}\epsilon_{jj}}$. 
 
When two atoms are brought together, however, the value of the repulsive LJ term in general increases too rapidly. 
In quantum solids particles can be close to each other due to their zero point motion, hence an accurate description 
of the atomic interactions at short distances is necessary even at low densities (Ceperley and Partridge, 1986; Boronat 
and Casulleras, 1994). In this context, the pairwise interaction model originally proposed by Ahlrichs, Penco and 
Scoles (1977) is more appropriate since it reproduces \emph{ab initio} results for the repulsive interactions between 
closed shell atoms (Hepburn, Scoles, and Penco, 1975). The form of this potential is: 
\begin{eqnarray}
V_{2}^{\rm Aziz} (r) = A e^{-a r + b r^{2}} -f(r) \sum_{i=6,8,10} \frac{C_{i}}{r^{i}}~,
\label{eq:aziz}
\end{eqnarray}
where $A$, $a$, $b$, and $C_{i}$ are free parameters, and $f(r)$ is an exponential damping function that is 
introduced to avoid the divergence of the $1/r^{n}$ terms at small distances. Aziz and collaborators worked extensively  
in this model to deliver an accurate description of the atomic interactions in many rare-gas systems (Aziz \emph{et al.}, 
1979; Aziz, Meath, and Allnatt, 1983; Aziz, McCourt, and Wong, 1989), hence the notation employed. Equation~(\ref{eq:aziz}) 
also yields an improved description of the long-range dispersion forces as compared to the LJ model, since it contains several 
types of multi-pole interactions.  

In some situations, a well balanced description of solids cannot be attained with pairwise potentials only. This is the case, 
for instance, of crystals at extreme thermodynamic conditions (Loubeyre, 1987; Cazorla and Boronat, 2008a; Cazorla and Errandonea, 
2014; Cazorla and Boronat, 2015b). A possible solution to overcome this modeling difficulty is to go beyond pairwise additivity, 
that is, to consider higher order terms in the approximation to the atomic interactions. Several three-body interatomic potentials 
have been proposed in the literature (Axilrod and Teller, 1943; Cencek, Patkowski, and Szalewicz, 2009), and the most popular in 
the context of quantum solids is (Bruch and McGee, 1973):  
\begin{eqnarray} 
V_{3} \left({\bf x}, {\bf y}, {\bf z} \right) &&= \frac{\nu}{x^{3} y^{3} z^{3}} - B e^{-c ( x + y + z )} \nonumber \\
&& \times \left( 1 + 3 \cos{\alpha} \cos{\beta} \cos{\gamma} \right)~,  
\label{eq:bmpot}
\end{eqnarray}
where $\nu$, $B$, $c$ are free parameters, $\lbrace x,y,z \rbrace$ the distance between particles in a trimer, and $\lbrace 
\alpha,\beta,\gamma \rbrace$ the corresponding interior angles. $V_{3}$ is an interatomic potential that renders triple dipole 
and exchange interactions; inclusion of this type of forces appears to be necessary for obtaining a realistic description of 
the energy and elastic properties of very dense quantum solids (Grimsditch, Loubeyre, and Polian, 1986; Pechenik, Kelson, and 
Makov, 2008; Cazorla and Boronat, 2015b).  

\begin{figure}
\centerline
        {\includegraphics[width=1.0\linewidth]{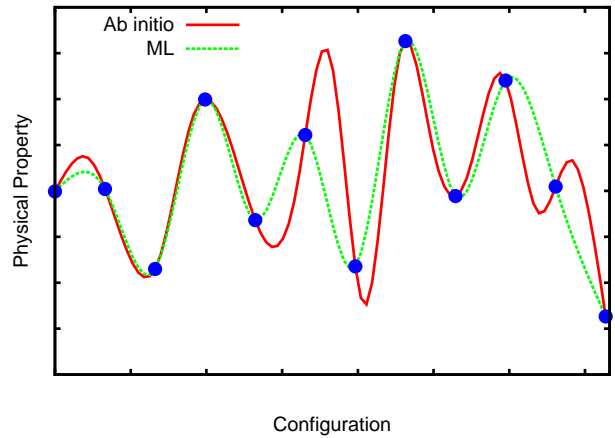}}
\caption{(Color online) Sketch of the key idea behind machine learning for the modeling of 
         atomic interactions. \emph{Ab initio} results, which are obtained
         at high computational cost, are approximated by interpolating with
         ML potentials between selected reference configurations (blue dots).}
\label{fig:ml}
\end{figure}

\subsubsection{Machine learning}
\label{subsubsec:machinelearn}
When calculations are performed in a series of similar systems or a 
number of configurations involving a same system, the results contain redundant 
information. One example is to run a molecular dynamics simulation in which the 
total internal energy and atomic forces are calculated at each time step; after a 
sufficiently long time, points which are close in configurational space and have 
similar energies are visited during the sampling of the potential-energy surface. 
Such a redundancy can be exploited to perform computationally intensive calculations 
(that is, of first-principles type) only in few selected configurations and to use 
machine learning (ML) to interpolate between those, obtaining so approximate solutions 
for the remaining of configurations (see Fig.~\ref{fig:ml}). The success of this 
approach depends on a balance between incurred errors due to interpolation and 
invested computational effort.     

ML modeling tools can provide both the energy and atomic forces directly from the 
atomic positions, hence they can be regarded as a particular class of atomistic potentials. 
ML potentials, however, rely on very flexible analytic functions rather than on physically
motivated functionals. Promising analytic approaches that have been recently proposed to 
construct ML potentials include permutation invariant polynomials (Brown \emph{et al.}, 2003), 
the modified Shepard method using Taylor expansions (Bowman, Czak\'o, and Fu, 2011), Gaussian 
processes (Bart\'ok \emph{et al.}, 2010; Bart\'ok \emph{et al.}, 2013), interpolating moving 
least squares (Dawes \emph{et al.}, 2007), and artificial neural networks (Lorenz, Gro{\ss}, 
and Scheffler, 2004). Artificial neural networks, for instance, have been demonstrated to be
``universal approximators'' (Behler, 2015) since they allow to approximate unknown multidimensional 
functions to within arbitrary accuracy given a set of known function values. 

To the best of our knowledge, ML potentials have not been applied yet to the study of quantum 
solids. However, the great versatility of ML approaches (Behler, 2010; Rupp, 2015; Behler, 
2015) could be exploited to describe such systems in specially challenging situations 
like, for instance, molecular solids (e. g., H$_{2}$, N$_{2}$, and CH$_{4}$) under extreme 
thermodynamic conditions. Classical interaction models normally disregard the orientational 
degrees of freedom in molecules and require the specification of bond connectivity between 
atoms. Therefore, they are not able to describe the orientational phase transitions and 
breaking/formation of atomic bonds occurring at high-$P$ and high-$T$ conditions (see 
Sec.~\ref{sec:molsol}). ML potentials could represent an intermediate solution between classical 
potentials and first-principles methods, both in terms of numerical accuracy and computational 
burden.

%===============================================================
\section{Archetypal quantum crystals}
\label{sec:archetypal}
%===============================================================
Helium and hydrogen are the lightest elements in Nature and the
paradigm of quantum solids. The classical picture of a crystal at low
temperature, with all the atoms strongly localised around their equilibrium
lattice positions, breaks completely in solid helium and hydrogen. In archetypal 
quantum crystals atoms move noticeably around the equilibrium lattice positions 
even in the limit of zero temperature, and exchanges between few particles occur 
with frequency. Consequently, the degree of anharmonicity in these systems 
is very high. Quantum simulation methods beyond the harmonic approximation 
(see Sec.~\ref{sec:simulation}) in fact are necessary for describing archetypal 
quantum solids correctly. 

\subsection{Helium}
\label{subsec:helium}
\begin{figure}
\centerline
        {\includegraphics[width=1.00\linewidth,angle=0]{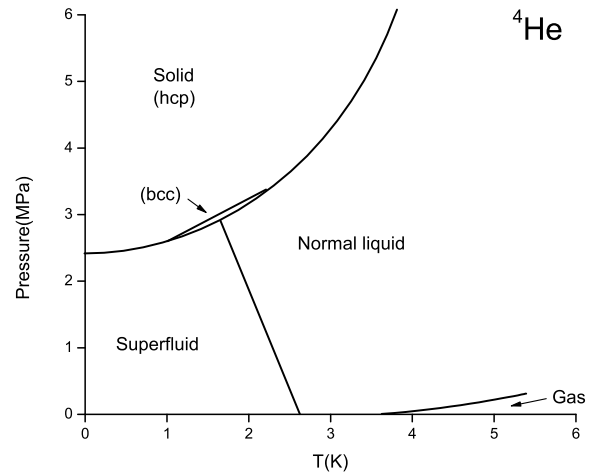}}
        \caption{Phase diagram of $^4$He at low pressures and temperatures.}
\label{phase4he}	
\end{figure}

Helium has two stable isotopes, $^4$He and $^3$He, which are a bosonic and a fermionic
particles, respectively. Both isotopes solidify under moderate pressures in the $T \to 0$  
limit, namely at $P \simeq 25$~bar in $^4$He and $30$~bar in $^3$He. $^4$He 
solidifies in the hexagonal hcp phase except for a small region at low pressures in 
which the stable phase is cubic bcc (see Fig.~\ref{phase4he}). Meanwhile, $^3$He 
solidifies in the cubic bcc phase with a relatively large molar volume of 
$V \simeq 24.5$~cm$^3$mol$^{-1}$. Under specific $P-T$ conditions, both isotopes 
transform into the cubic fcc phase (Glyde, 1994).

\begin{figure}
\centerline
       {\includegraphics[width=1.00\linewidth,angle=0]{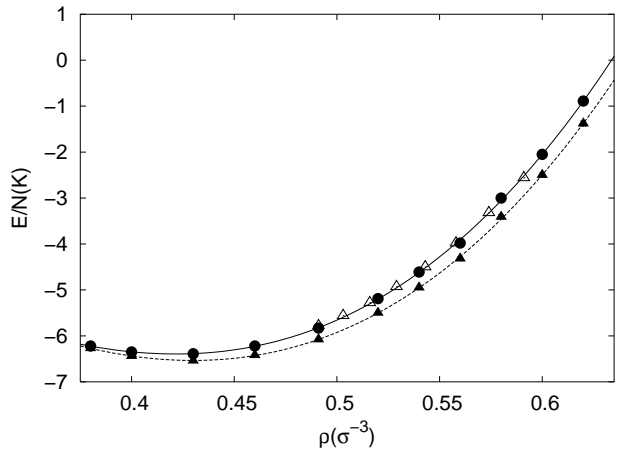}}
        \caption{Energy per particle in solid $^4$He expressed as a function of 
	density. Open triangles represent experimental data from Edwards and Pandorf 
        (1965); other symbols and lines correspond to the DMC results. Solid circles 
        and line represent results in which the bias introduced by finite size effects 
        has been reduced significantly. Solid triangles and the dashed line represent 
        results which have been corrected only partially for the same type of bias. 
        Adapted from Vranje\v{s} \emph{et al.} (2005).}
\label{eos4he}	
\end{figure}

At low pressure ($P < 1$~GPa), the thermodynamic properties of solid $^{4}$He are well known 
from experiments and accurately reproduced by QMC methods. In Fig.~\ref{eos4he}, we compare 
experimental and computational results for the dependence of the energy per particle 
on density. The theoretical results correspond to DMC simulations performed by Vranje\v{s} 
\emph{et al.} (2005) using a semi-empirical pair potential (Aziz, McCourt, and Wong, 1987); 
the agreement between observations and theory is excellent. Likewise, accurate results have been 
obtained also with the PIGS method (Rossi \emph{et al.}, 2012). From function $E/N(\rho)$ one 
can easily work out the pressure, $P(\rho)=\rho^2d(E/N)/d \rho$, obtaining so the corresponding 
equation of state (e.o.s.). Excellent agreement between theory and experiment has been demonstrated 
also for this quantity [see, for instance, Cazorla and Boronat (2008a)]. The quantum nature of 
solid $^4$He is thermodynamically reflected on its high compressibility; for instance, the corresponding 
experimental molar volume is reduced from $21$~cm$^3$mol$^{-1}$ at $25$~bar to $9$~cm$^3$mol$^{-1}$ 
at $5$~kbar. The location of the first-order liquid-solid phase transition also is accurately 
reproduced by QMC calculations. Recent DMC estimations provide a transition pressure of $27.3$~atm, with
freezing and melting densities equal to $\rho_f = 0.437~\sigma^{-3}$ and $\rho_m = 0.481~\sigma^{-3}$ 
($\sigma=2.556$~\AA), respectively (Vranje\v{s} \emph{et al.}, 2005). The corresponding  
experimental values are, $P_{t}^{\rm expt} = 25$~atm, $\rho_f^{\rm expt} = 0.434~\sigma^{-3}$, 
and $\rho_m^{\rm expt} = 0.479~\sigma^{-3}$ (Glyde, 1994).    

Valuable information on the quantum nature of a solid is obtained from its 
Lindemann ratio:
\begin{equation}
\gamma= \frac{1}{a} \left[ \frac{1}{N} \sum_{i=1}^N ({\bm r}_i-{\bm R}_i)^2
\right]=\frac{\langle {\bm u}^2 \rangle^{1/2}}{a}~,
\label{lindemann}
\end{equation}
where ${\bm R}_i $ represent the coordinates of the perfect lattice sites, and $a$ 
the corresponding lattice constant. Parameter $\gamma$ quantifies the displacement of 
particles around their equilibrium positions. The quantum character of a solid can be 
said to be proportional to the value of its Lindemann ratio. In solid $^{4}$He and $^{3}$He 
at ultra-low temperatures, for instance, $\gamma$ amounts to $\sim 0.3$ (Glyde, 1994), which 
are the largest values known in any material at those thermodynamic conditions. The large 
excursions of helium atoms around their lattice positions allow them to explore the 
non-harmonic part of the potential energy surface, leading to high anharmonicity. 
Another singular aspect in solid helium is the large kinetic energy per particle. 
At $P = 50$~atm, for instance, $E_{\rm k}$ amounts to $\sim 24$~K (Diallo \emph{et al.}, 
2007), which is of the same order of magnitude than the corresponding potential 
energy, namely $\sim -31$~K (resulting from a cancellation between large repulsive
and attractive terms). 

\begin{figure}
\centerline
        {\includegraphics[width=1.00\linewidth,angle=0]{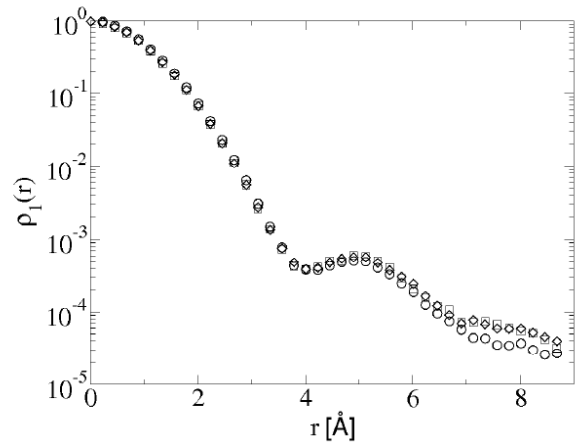}}
        \caption{One-body density matrix calculated in hcp $^4$He 
        at density $\rho = 0.0294$~\AA$^{-3}$. Circles, squares, and
	diamonds stand for results obtained at $T = 0$, $1$, and $2$~K, 
        respectively. Adapted from Rota and Boronat (2012).}
\label{rhosol}	
\end{figure}

The influence of Bose-Einstein statistics on the energy and structural properties
of solid $^{4}$He is negligible [$\sim 1~\mu$K/atom, Clark and Ceperley (2006)]. 
In fact, many of the results just presented have been obtained with non-symmetric wave 
functions and as it has been explained the agreement with the experiments is excellent. 
However, quantum atomic exchanges play a pivotal role in other intriguing properties like, 
for instance, Bose-Einstein condensation and superfluidity (Ceperley, 1995). These phenomena 
occur in liquid $^{4}$He at ultra-low temperatures and, due to the extreme quantum nature of 
helium, it was wondered long time ago whether the same effects could be observed also in the 
crystal phase (see Sec.~\ref{sec:defects} for a historical overview of this topic). In recent 
years, there have been several theoretical works aimed at clarifying 
these questions. In particular, the one-body density matrix, $\varrho_{1}(r)$ 
[Eq.~(\ref{onebodypimc})], of solid $^{4}$He has been calculated with different methods. 
Initial zero-temperature estimations based on symmetrized wave functions (Cazorla \emph{et al.}, 
2009; Galli and Reatto, 2006) provided a non-zero but small plateau at long distances. However, unbiased 
$\varrho_{1}(r)$ results obtained with the PIGS and PIMC methods have unequivocally demonstrated 
that the condensate fraction in perfect solid $^{4}$He is actually zero (Ceperley and Bernu, 2004; 
Bernu and Ceperley, 2005; Clark and Ceperley, 2006; Boninsegni, Prokof'ev, and Svistunov, 2006b). 
In particular, the tail of $\varrho_{1}(r)$ decays exponentially at long distances, as it is 
illustrated in Fig.~\ref{rhosol} (Rota and Boronat, 2012). Actually, the exchange frequency between 
particles at different lattice sites is very small as compared to that in the liquid phase [for 
instance, the exchange frequency for $2$, $3$, and $4$ atom exchanges is of $\sim 3~\mu$K/atom, 
Ceperley and Bernu (2004); Clark and Ceperley (2006)], and long permutation cycles able to trigger 
superfluidity are highly improbable. Nevertheless, we note that when a finite and stable concentration 
of defects is assumed to exist in the crystal these conclusions change drastically (see 
Sec.~\ref{sec:defects}). 

\begin{figure}
\centerline
        {\includegraphics[width=1.00\linewidth,angle=0]{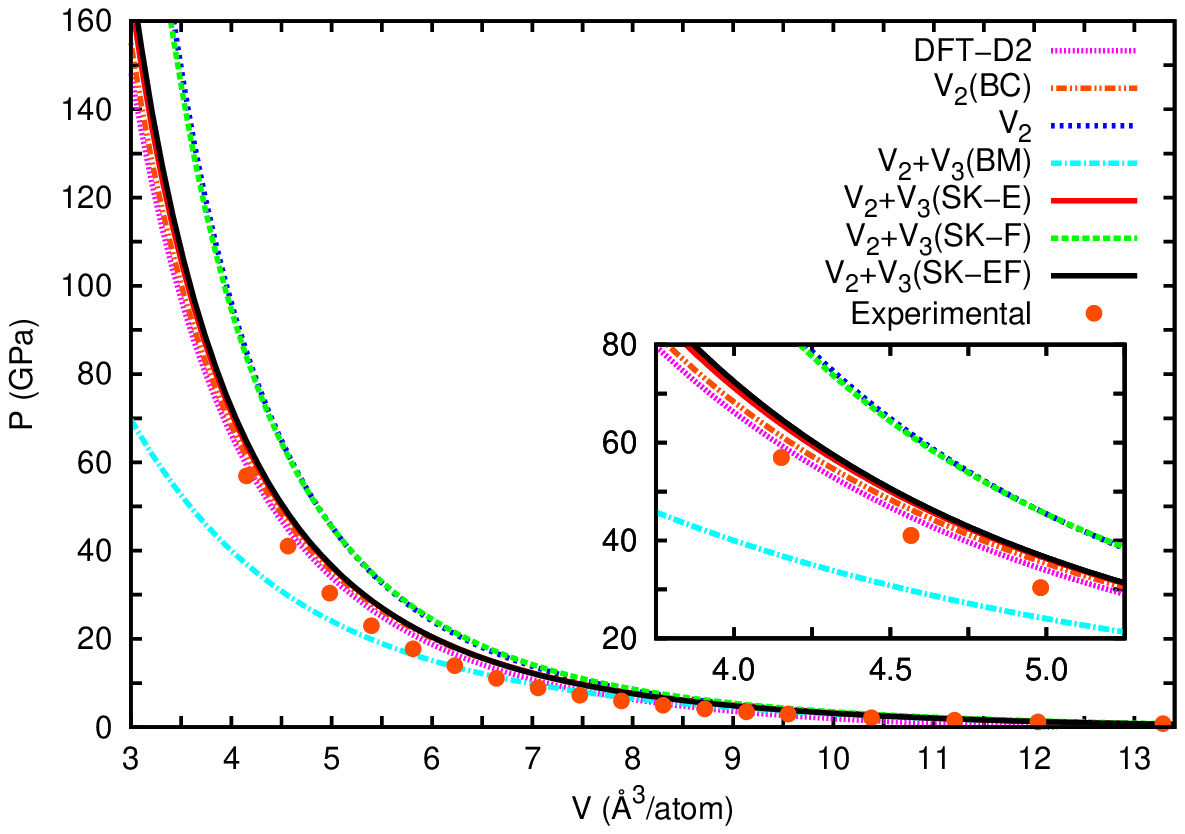}}
        \caption{Zero-temperature equation of state of hcp $^4$He at high
	pressures calculated with several effective three-body interaction 
        models fitted to reproduce \emph{ab initio} data, and the DFT-D
        method (see text). Experimental data from Loubeyre \emph{et al.} (1993) 
        are shown for comparison. \emph{Inset}: $P(V)$ curves in the high-$P$ 
        region are magnified in order to appreciate better the differences. 
        From Cazorla and Boronat (2015b).}
\label{he4hpress}	
\end{figure}

At high pressures, the atoms in a crystal experience strong short-range repulsions 
due to electrostatic forces and the Pauli exclusion principle. Customary semi-empirical 
potentials that at low densities provide a good description of the crystal, start then 
to be unreliable due to severe transferability issues (see Sec.~\ref{subsec:ff}). This is 
the case of the Aziz potential for $^4$He (Aziz, McCourt, and Wong, 1987), which possesses 
a too steep repulsive core and leads to inaccurate results at pressures $P \ge 1$~GPa (Cazorla 
and Boronat, 2008a). Recently, dispersion-corrected density functional theory (DFT) has been 
used in combination with the DMC method to study the quantum behavior of solid helium at 
pressures up to $\sim 150$~GPa (Cazorla and Boronat, 2015a; Cazorla and Boronat, 2015b). 
Essentially, analytical potentials have been constructed to reproduce sets of atomic energies 
and forces calculated with first-principles methods. In a first approximation (Cazorla and Boronat, 
2015a), the effective pair interaction has been obtained by fitting the static compression curve 
calculated with DFT-D to an analytical function based on the Aziz potential [see Eq.~(\ref{eq:aziz})] 
and an attenuation repulsion factor proposed by Moraldi (Moraldi, 2012). This has allowed for a 
sizable improvement in the description of the high-$P$ e.o.s. as compared to the available experimental 
data. However, it has been shown that such a simple approach provides unphysical results for the 
elastic constants and pressure dependence of the kinetic energy. In a posterior work, Cazorla 
and Boronat (2015b) have introduced a family of three-body interaction potentials based on 
Eq.~(\ref{eq:bmpot}) that allow to overcome (in part) these modeling shortcomings while still 
providing an accurate e.o.s. up to $\sim 60$~GPa (see Fig.~\ref{he4hpress}). 

With regard to solid $^3$He, the number of related studies is very limited. Besides some old 
variational calculations, the most recent and accurate investigation of its thermodynamic 
properties has been performed by Moroni \emph{et al.} (2000) with the DMC method. In Moroni 
\emph{et al.}'s (2000) work the quantum antisymmetry of the system is neglected, that is, 
particles are treated as bosons rather than as fermions. Nevertheless, since the exchange 
energy in the crystal is very small (of the order of mK, see Ceperley and Jacucci, 1987; 
C${\rm \hat{a}}$ndido, Hai, and Ceperley, 2011) it can be expected that quantum symmetry effects 
will play an insignificant role on the energy. The results obtained for the dependence of 
the energy on density show a discrepancy with the experimental data, which Moroni \emph{et al.} 
(2000) have attributed to a wrong reference in the integration of the experimental equation 
of state. After correction of such an error, the agreement between theory and experiments 
becomes excellent, namely, of the same quality than achieved in solid $^4$He.

\subsection{Hydrogen}
\label{subsec:hydrogen}
Bulk molecular hydrogen (deuterium) at zero pressure, in contrast to $^{4}$He, solidifies at a 
temperature of $\sim 14$~K ($\sim 19$~K) due to the stronger attractive interactions between 
particles. ${\rm H_{2}}$ (${\rm D_{2}}$) molecules are composed of two hydrogen (deuterium) atoms 
joined by a covalent bond, which in the para-hydrogen (ortho-deuterium) state have zero angular 
momentum and spherically symmetric wave functions. Both types of particles, therefore, are bosons and 
the interactions between molecules of the same species can be modelled with radial pairwise potentials 
(at high pressures, however, the molecular angular momentum is not longer zero and thereby pairwise 
approximations to the intermolecular interactions become invalid, see Sec.~\ref{subsec:h2extreme}). 
Actually, in most quantum simulation studies of ${\rm H_{2}}$ and ${\rm D_{2}}$ crystals at low 
pressure (i. e., $P \le 0.1$~GPa) the intermolecular forces have been modelled with the semi-empirical 
Silvera-Goldman (Silvera and Goldman, 1978) and Buck (Buck \emph{et al.}, 1983; Norman \emph{et al.}, 
1984) pair potentials. The role of three-body forces on the corresponding low-$P$ equation of state 
has been explored, but their net effects have been found to be negligible (Operetto and Pederiva, 
2006). Meanwhile, anisotropic corrections to the pair potential have been tested against experiments 
and found to be significant only at pressures higher than $\sim 10$~GPa (Cui \textit{et al.}, 1997).

\begin{figure}
\centerline
        {\includegraphics[width=1.00\linewidth,angle=0]{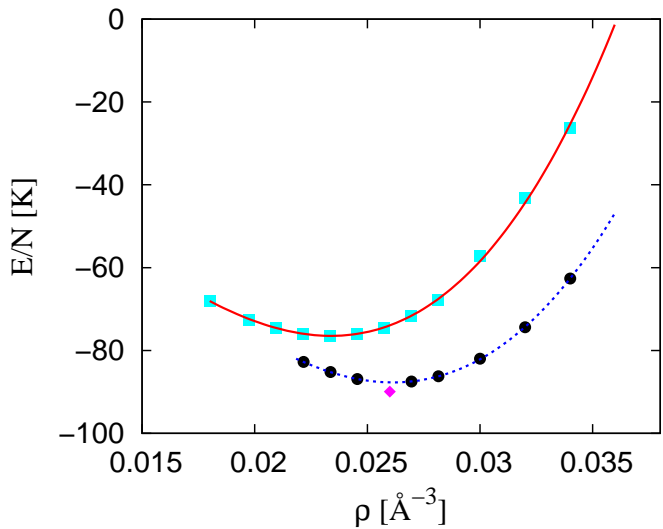}}
        \caption{DMC energy per H$_2$ molecule as a function of density. Squares
        and circles correspond to the liquid and solid phases, respectively. Solid
        and dashed lines are polynomial fits to the DMC energies for the liquid
        and solid phases, respectively. The diamond represents the experimental energy of
        hcp molecular hydrogen from Schnepp (1970). From Osychenko, Rota, and Boronat
        (2012).}
\label{h2eos}
\end{figure}

\begin{figure}
\centerline
       {\includegraphics[width=1.00\linewidth,angle=0]{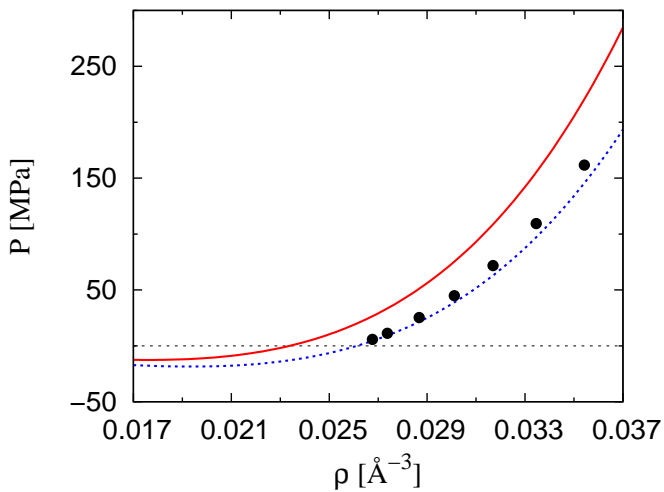}}
        \caption{Equation of state of liquid (solid line) and solid (dashed line) H$_2$.
        Experimental points for the solid phase are represented with solid circles 
        (Driessen, de Waal, and Silvera, 1979). From Osychenko, Rota, and Boronat (2012).}
\label{h2pres}
\end{figure}

In Fig.~\ref{h2eos}, we show the energy per molecule in hexagonal hcp H$_2$ calculated in 
the limit of zero temperature with the DMC method and the Silvera-Goldman potential (Osychenko, 
Rota, and Boronat, 2012). We note that the experimental energy per particle is $E/N=-89.9$~K, 
which is underestimated (overestimated) by the Silvera-Goldman (Buck) potential model. Close 
to the equilibrium density the H$_2$ kinetic energy is $89.5$~K, which roughly amounts to
half of the potential energy. In comparison to solid $^4$He, in which both types of energies 
are nearly equal, quantum nuclear effects in solid hydrogen turn out to be smaller (see also 
Fig.~\ref{fig:deboer}). The energy curve of the metastable liquid is shown in Fig.~\ref{h2eos} 
for comparison. Results for the equation of state of solid and liquid H$_2$ are enclosed in 
Fig.~\ref{h2pres}. The agreement between theory and experiments (Driessen, de Waal, and Silvera, 
1979) is quite satisfactory in the solid phase, although at pressures beyond $\sim 100$~MPa this 
starts to worsen due to the limitations of the employed intermolecular potential (Moraldi, 2012; 
Omiyinka and Boninsegni, 2013). 

As a by-product of the recent experimental activity on the search for a supersolid state of 
matter (see Sec.~\ref{sec:defects}), an interest has developed in studying highly disordered 
solids like, for instance, amorphous or glassy systems. A glassy state in solid $^4$He, termed 
as ``superglass'', has been predicted to exhibit superfluid behavior by Boninsegni, Prokof'ev, 
and Svistunov (2006b). An analogous study has been carried out more recently in solid H$_2$ by 
Osychenko, Rota, and Boronat (2012). In this case, PIMC simulations have shown that glassy 
molecular hydrogen eventually becomes superfluid at temperatures below $\sim 1$~K. The critical
temperature for this transition, however, is so small that it is unlikely to be observed 
in experiments (K\"uhnel \emph{et al.}, 2011).   

Interestingly, the free surface of bulk H$_{2}$ has been also investigated with experiments (Brewer 
\emph{et al.}, 1990; Vilches, 1992; Kinder, Bouwen, and Schoemaker, 1995) and PIMC simulations
(Wagner and Ceperley, 1994; Wagner and Ceperley, 1996). It has been found that the melting temperature 
of the bare hydrogen surface is reduced down to $\sim 6$~K and that zero-point molecular fluctuations are 
considerably enhanced with respect to bulk. For instance, at low temperatures the corresponding Lindemann 
ratio increases from $\sim 0.1$ in the inner layers up to $\sim 0.2$ in the outer surface (Wagner and 
Ceperley, 1996). Yet, the corresponding melting temperature is still too high to expect that liquid 
H$_{2}$ will become superfluid (that is, well above the predicted critical temperature $T_{c} \sim 
1-2$~K; Apenko, 1999).

\subsection{Neon}
\label{subsec:neon}
Solid neon behaves more ``classically'' than solid helium but more ``quantumly'' than the rest of rare-gas species 
(see Fig.~\ref{fig:deboer}). The study of this crystal is useful to understand the transition from the quantum 
regime to the classical one in solid-state systems. The interest in solid neon as a case study of a moderate quantum 
system dates back to the $1960$'s. Bernades (1958) and Nosanow and Shaw (1962) were the first to attempt an estimation 
of the kinetic energy in solid neon using theoretical methods. By relying on variational and self-consistent Hartree 
calculations performed with uncorrelated single-particle wave functions, they reported ground-state $E_{\rm k}$ 
values of $\sim 41$~K. However, the binding energies reported in those early works were in strong disagreement with 
contemporary experiments, evidencing the need to go beyond uncorrelated microscopic approaches. Few years later, 
Koehler (1966) applied the self-consistent phonon approach to the same system and obtained results for the cohesive 
energy that were in better agreement with the experiments; Koehler's estimation of the kinetic energy was $42.6$~K. 

\begin{figure}
\centerline
        {\includegraphics[width=1.0\linewidth]{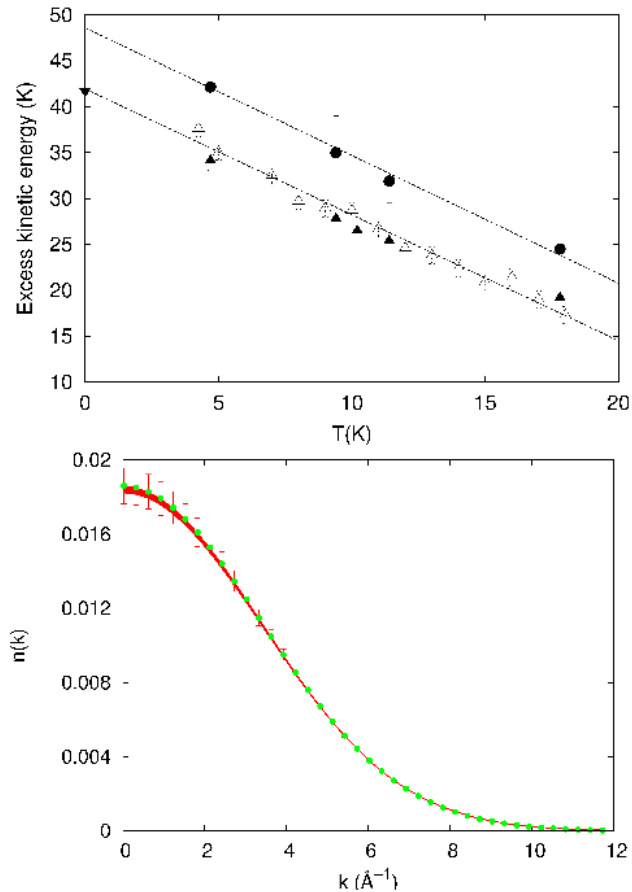}}
\caption{(Color online) Quantum nuclear effects in solid neon at low temperatures.
         \emph{Top panel} Excess kinetic energy expressed as a function of temperature. Experimental data from Timms, 
         Simmons, and Mayer (2003) are represented with $\vartriangle$; measurements from Peek \emph{et al.} (1992) 
         with $\bullet$; PIMC calculations from Timms \emph{et al.} (1996) with $\blacktriangle$; DMC ground-state 
         calculations from Cazorla and Boronat (2008b) with $\blacktriangledown$. The lines in the plot are linear 
         fits to the experimental data. \emph{Bottom panel} Ground-state momentum distribution in solid Ne calculated 
         with the DMC method (green dots). The solid line (red) is a Gaussian fit to the results, the width of which 
         represents its uncertainty. Adapted from Cazorla and Boronat (2008b).}
\label{fig:neon}
\end{figure}

It was not until the $1990$'s that, with the development of the deep inelastic neutron scattering technique, the kinetic 
energy in quantum crystals could be measured precisely. Peek \emph{et al.} (1992) were the first to perform those 
measurements in solid neon, reporting a ground-state kinetic energy of $49.1~(2.8)$~K. In view of the large discrepancies 
found with respect to previous estimations based on harmonic models, the authors of that study suggested that solid neon 
was highly anharmonic. Later on, Timms \emph{et al.} (1996) carried out a new series of neutron scattering experiments in 
which higher momentum and energy transfers were considered. They found that in the temperature interval $4-20$~K their excess 
kinetic energy measurements, defined as $E_{\rm exc} \equiv E_{\rm k} - \frac{3}{2}k_{\rm B}T$, were systematically lower 
than Peek's results by few Kelvin (see Fig.~\ref{fig:neon}). The validity of Timms \emph{et al.}'s results (1996) is supported 
by the outcomes of several PIMC studies based on classical interatomic potentials (Timms \emph{et al.}, 1996; Cuccoli 
\emph{et al.}, 2001; Neumann and Zoppi, 2002). More recently, Timms, Simmons, and Mayer (2003) have performed additional 
neutron scattering measurements and reported that the ground-state kinetic energy in solid neon is $41(2)$~K. The accuracy 
of this result has been confirmed by recent DMC calculations performed by Cazorla and Boronat (2008b), which provide 
$E_{\rm k} = 41.51(6)$~K in the $T \to 0$ limit (see Fig.~\ref{fig:neon}). We note that the Lindemann ratio (see 
Sec.~\ref{subsec:quantumvsclassical}) in solid neon is approximately three times smaller than in helium, namely 
$\gamma_{\rm Ne} \sim 0.08$ (Cazorla and Boronat, 2008b), pointing to a moderate degree of quantumness. 

An interesting topic in the study of solid neon is related to the shape of its momentum distribution, $n({\bf k})$
(see Fig.~\ref{fig:neon}). Withers and Glyde (2007) have shown that when a crystal has a momentum distribution 
that is not well described by a Gaussian, it may be that the system is highly anharmonic or that quantum atomic exchanges 
occur frequently in it. While quantum atomic exchanges are very likely to be negligible, anharmomic effects seem to 
be fairly important in solid neon. This has been demonstrated by Cazorla and Boronat (2008b), who have performed different 
types of harmonic-based calculations and compared their results to those obtained with full anharmonic methods. For instance, 
the ground-state kinetic energy that is predicted with the self-consistent average phonon approach (Shukla \emph{et al.}, 1981) 
amounts to $\sim 47$~K and the corresponding Lindemann ratio to $0.06$, which are appreciably different from the experimental 
and DMC results. 

Nevertheless, there is widespread agreement among experimentalists and theorists in that the momentum distribution in solid Ne 
is well approximated by a Gaussian function (Timms \emph{et al.}, 1996; Neumann and Zoppi, 2002; Cazorla and Boronat, 2008b). 
The question then arises about how anharmonic (or quantum) a crystal must be for its $n({\bf k})$ to differ appreciably from 
a Gaussian. The momentum distribution in solid $^{4}$He is non-Gaussian as it has a large occupation of low momentum 
states as compared to a Maxwell-Boltzmann distribution [Diallo \emph{et al.}, 2004; Rota and Boronat, 2011]. 
However, little is known about the relation between anharmonicity and $n({\bf k})$ in other quantum crystals lying in 
between helium and neon, in terms of quantumness, which are mostly molecular systems (see Fig.~\ref{fig:deboer}). In the case of 
solid para-H$_{2}$, for instance, inelastic neutron scattering experiments have provided a translational momentum distribution that 
is Gaussian (Langel \emph{et al.}, 1988); however, it is not clear whether this result can be attributed to the intensive 
use of Gaussian approximations during the refinement of experimental data (Colognesi \emph{et al.}, 2015). Is it then
solid helium the only quantum crystal with a non-Gaussian $n({\bf k})$? The answer, so far, seems to be yes.  

Free-energy calculations based on path-integral simulations and semi-empirical pairwise potentials have been 
employed also to study the quantum phase diagram of neon up to pressures of $2-3$~kbar and temperatures of 
$50$~K (Ram\'irez and Herrero, 2008; Ram\'irez \emph{et al.}, 2008; Brito and Antonelli, 2012). Significant QNE 
have been found in the solid-gas and liquid-gas $P-T$ coexistence lines, which consist in a shift of about $1.5$~K 
towards lower temperatures as compared to the classical phase diagram. Moderate quantum isotopic effects have been 
observed also in the triple solid-liquid-gas coexistence point both in experiments (Furukawa, 1972) and 
path-integral calculations (Ram\'irez and Herrero, 2008).

%===============================================================
\section{Elasticity and mechanical properties}
\label{sec:elasticity}
%===============================================================
The free energy of a crystal subjected to a homogeneous elastic deformation is:
\begin{equation}
F(V, T, \epsilon) = F_{0}(V,T) + \frac{1}{2} V \sum_{ij} C_{ij} \epsilon_{i} \epsilon_{j}~,
\label{eq:free-elastic}
\end{equation}
where $F_{0}$ is the free energy of the undeformed solid, $C_{ij}$ the corresponding elastic 
constants, and $\epsilon_{i}$ a general strain deformation (the latter two quantities are 
expressed in Voigt notation and the subscripts indicate Cartesian directions). The symmetry 
of the crystal determines the number of elastic constants that are inequivalent and non-zero. 
For a crystal to be dynamically stable, its change in free energy due to an arbitrary strain 
deformation must be always positive; this requirement leads to a number of mechanical stability 
conditions that need to be fulfilled at any stable or metastable state, and which depend on the 
particular symmetry of the crystal (Born and Huang, 1954; Grimvall \emph{et al.}, 2012). 

The elastic constants of a solid can be measured with ultrasonic techniques since the velocity of density 
waves depends on the elastic properties of the medium in which they propagate. Brillouin scattering spectroscopy 
and synchrotron x-ray diffraction techniques can be employed also to this end. Likewise, the calculation 
of elastic constants with quantum simulation methods is a well established technique. At zero temperature, 
one can calculate the energy of the solid as a function of strain, using for instance the DMC or PIGS method (see 
Sec.~\ref{subsec:zeroT}), and then simply compute the value of its second derivative numerically (Cazorla, 
Lutsyshyn, and Boronat, 2012; Cazorla and Boronat, 2013). At $T \neq 0$, the calculation of $C_{ij}$'s 
is not so straightforward since one has to consider also the effects of thermal excitations. Sch\"offel and M\"user 
(2001) were to first to undertake such a type of calculation by using the path-integral Monte Carlo method. They estimated 
the elastic constants in solid Ar and $^{3}$He through direct derivation of the partition function with respect 
to the strain components. More recently, Pe${\rm \tilde{n}}$a-Ardila, Vitiello, and de Koning (2011) have proposed 
an alternative path-integral approach in which a suitable expression for the estimation of the stress-tensor is 
worked out. 

In this section, we review the elastic properties of \emph{perfect} quantum solids, that is, free of crystalline 
defects. Crystalline defects can affect considerably the elastic behavior of quantum (and also classical) crystals, 
so that we leave those aspects for Sec.~\ref{sec:defects}. Our analysis here is divided into low and high
pressures because the fundamental character of elasticity in quantum crystals changes when moving from one regime
to the other.    

\begin{figure}
\centerline
        {\includegraphics[width=1.0\linewidth]{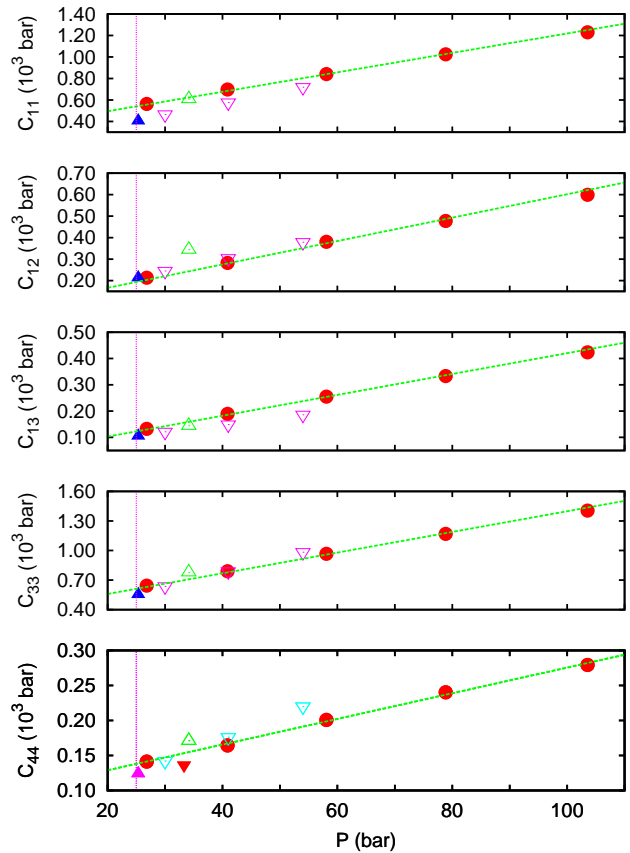}}
\caption{(Color online) Elastic constants in solid $^{4}$He at moderate pressures.  
         Experimental data from Crepeau \emph{et al.} (1971) and Greywall (1977) are represented with 
         $\blacktriangle$ and $\triangledown$, respectively; $C_{44}$ measurements from Syshchenko, Day, and Beamish 
         (2009) with $\blacktriangledown$; VMC calculations from Pessoa, Vitiello, and de Koning (2010) with $\vartriangle$; 
         DMC ground-state calculations from Cazorla, Lutsyshyn, and Boronat (2012) with $\bullet$. The dashed 
         lines (green) in the plots represent linear fits to the DMC results. The freezing pressure in the crystal 
         is marked with a vertical (magenta) line. Adapted from Cazorla, Lutsyshyn, and Boronat (2012).}
\label{fig:cij4he}
\end{figure}

\subsection{Low-pressure regime}
\label{subsec:lowpelas}
The elastic properties of traditional quantum solids like helium (Crepeau \emph{et al.}, 1971; Greywall, 1977) and hydrogen 
(Nielsen and M\o ller, 1971; Wanner and Meyer, 1973; Nielsen, 1973), have been measured extensively. In experiments, 
however, it is difficult to determine the exact contribution of quantum nuclear effects (QNE) to elasticity. 
In this context, the outcomes of first-principles studies can be very valuable. For instance, Sch\"offel and M\"user (2001) 
performed a thorough PIMC study on the elastic properties of solid $^{3}$He in the hexagonal hcp and cubic bcc and fcc 
phases, considering low temperatures and pressures. Their results were in good agreement with the reported 
experimental data, and they concluded that QNE accounted for about $30$~\% of the $C_{ij}$ values. A similar quantum 
influence on the elasticity of solid Ar was also reported (namely, $\sim 20$~\%), a crystal that is considered to behave 
much more classically than helium.           

More recently, the elastic properties of solid $^{4}$He have been studied in detail by several authors using different 
QMC techniques. Cazorla, Lutsyshyn, and Boronat (2012) have employed the DMC method to calculate the 
zero-temperature elastic constants, Gr\"uneisen parameters, sound velocities, and Debye temperature over a wide pressure 
interval of $\sim 100$~bar. The computed $C_{ij}$ values are in overall good agreement (i. e., discrepancies to less than $5$~\% 
in most cases) with the reported experimental data and results obtained by other authors using the VMC (Pessoa, Vitiello, 
and de Koning, 2010; Pessoa, de Koning, and Vitiello, 2013) and PIMC (Pe${\rm \tilde{n}}$a-Ardila, Vitiello, and de 
Koning, 2011) methods (see Fig.~\ref{fig:cij4he}). It has been found that the pressure dependence of all five elastic 
costants close to equilibrium is practically linear (see Fig.~\ref{fig:cij4he}). Interestingly, the contribution of QNE to 
the elastic constants in hcp $^{4}$He has been shown to be $\sim 30$~\%, which roughly coincides with the results obtained 
by Sch\"offel and M\"user (2001) in solid $^{3}$He. In essence, all these theoretical studies conclude that QNE profoundly 
affect the elastic properties of quantum crystals at low pressures (that is, $P \le 0.01$~GPa). 

A fundamental question that can be easily addressed with simulations but not with experiments is: what is the limit of 
mechanical stability in a quantum crystal? When the density in a system is reduced progressively, eventually this becomes 
unstable against long wavelength density fluctuations. This limit, also known as the spinodal point, has been analysed 
comprehensively in liquid $^{4}$He and $^{3}$He (Boronat, Casulleras, and Navarro, 1994; Maris, 1995; Maris and Edwards, 
2002); however, it has not been until recently that has been estimated directly in the crystal phase (Cazorla and Boronat, 
2015c). Theoretically, the spinodal point in a crystal is identified with the thermodynamic state at which any of the mechanical 
stability conditions involving the elastic constants is not satisfied. One can expect that, due to the presence of QNE 
and inherent structural softness, the limit of mechanical stability in quantum crystals lies very low in density. 

Based on $C_{ij}$ calculations performed with the DMC method and a semi-empirical pairwise potential, Cazorla and Boronat 
(2015c) have estimated that the ground-state spinodal pressure in solid $^{4}$He is $P_{s} = -33.8(1)$~bar, which corresponds 
to a volume of $V_{s} = 50.81(5)$~\AA$^{3}$. In particular, it has been found that the mechanical stability condition 
$(C_{33} - P)(C_{11} + C_{12}) - 2(C_{13} + P)^{2} > 0$ is violated at $P_{s}$. Regarding the propagation of 
density waves, previous calculations based on phenomenological models (Maris, 2009; Maris, 2010) had suggested that, in 
analogy to the liquid phase, the sound velocities in hcp $^{4}$He near the spinodal density could follow a power law of 
the form $\propto (P - P_{s})^{1/3}$. However, Cazorla and Boronat (2015c) have shown that quantum solids and liquids 
behave radically different in the vicinity of their mechanical stability limits; in particular, none of the sound velocity 
components, either propagating along the $c$-axis or in the basal plane, follow the previously proposed ``1/3'' power law.

\subsection{High-pressure regime}
\label{subsec:highpelas}
The elastic properties of archetypal quantum solids under high pressure (i. e., $P > 1$~GPa) have been thoroughly investigated 
with experiments (Zha \emph{et al.}, 1993; Zha, Mao, and Hemley, 2004). Surprisingly, the results of first-principles DFT 
studies in which QNE are completely or partially neglected, show very good agreement with the measured $C_{ij}$ and sound 
velocity data (Nabi \emph{et al.}, 2005; Freiman \emph{et al.}, 2013; Grechnev \emph{et al.}, 2015). 
In view of the importance of QNE on the elastic properties of quantum crystals at low pressures (see previous section), 
such a good agreement could be explained in terms of (i)~a systematic error cancellation involving the disregard of QNE,
on one hand, and an inaccurate description of the system provided by standard DFT functionals, on the other, or (ii)~a steady 
diminishing of the importance of QNE on elasticity as pressure is increased. 

To the best of our knowledge, there are not fully \emph{ab initio} studies (i. e., works in which both the electronic and 
ionic degrees of freedom are described with quantum mechanical methods) on the elastic properties of highly compressed 
quantum crystals. The reason for this is likely to be the large computational expense involved in the calculation of  
partition function derivatives or the stress-tensor with sufficient accuracy (Sch\"offel and M\"user, 2001;
Pe${\rm \tilde{n}}$a-Ardila, Vitiello, and de Koning, 2011). On the other hand, the semi-empirical two-body 
potentials that at low pressures describe the interactions between atoms in quantum crystals correctly, become unreliable at 
high pressures (see Sec.~\ref{sec:modeling}). In addition to this, pairwise interaction models are in general not well suited 
for the study of elasticity in very dense crystals since they inevitably lead to zero values of the Cauchy relations (Wallace, 
1972; Pechenik, Kelson, and Makov, 2008), which is in contrast to the observations (Zha, Mao, and Hemley, 2004). An estimation
of the importance of QNE in the elasticity of quantum crystals using such unrealistic interatomic potentials, therefore, could 
be misleading. 

Cazorla and Boronat (2015b) have recently introduced a set of effective three-body potentials based on Eq.~(\ref{eq:bmpot}),
to simulate solid $^{4}$He at high pressures realistically and with affordable computational effort (see Sec.~\ref{subsec:helium}). 
The new parametrisations have been obtained from fits to \emph{ab initio} energies and atomic forces calculated with a 
dispersion-corrected DFT functional (see Sec.~\ref{subsubsec:dft}). It has been shown that an overall improvement 
in the description of $^{4}$He elasticity at zero temperature and pressures $0 \le P \le 25$~GPa can be achieved 
with some of the proposed three-body interaction models. 
Interestingly, Cazorla and Boronat (2015b) have found that the role of QNE on helium elastic constants becomes secondary 
at very large densities. For instance, the inclusion of QNE makes the value of the shear modulus, $C_{44}$, to decrease by 
less than $4$~\% at a pressure of $\sim 50$~GPa (to be compared with $\sim 30$~\% found near equilibrium conditions). This 
conclusion appears to be consistent with the results of a recent PIMC study performed by Landinez-Borda, Cai, and de Koning 
(2014), in which the ideal shear strength on the basal plane of solid helium (that is, the maximum stress that the crystal
can resist without yielding irreversibly) has been found to behave quite similarly to that in classical hcp solids. 
Consequently, it can be expected that by treating QNE with approximate methods (e. g., the Debye model or quasi-harmonic 
approximations) one can obtain a reasonably good description of elasticity in quantum solids at high pressures. It will 
be very interesting to test whether this is also valid in quantum molecular crystals, the elastic properties of which 
remain largely unexplored at high densities.

%===============================================================
\section{Crystalline defects}
\label{sec:defects}
\begin{figure}
\centerline
        {\includegraphics[width=0.8\linewidth]{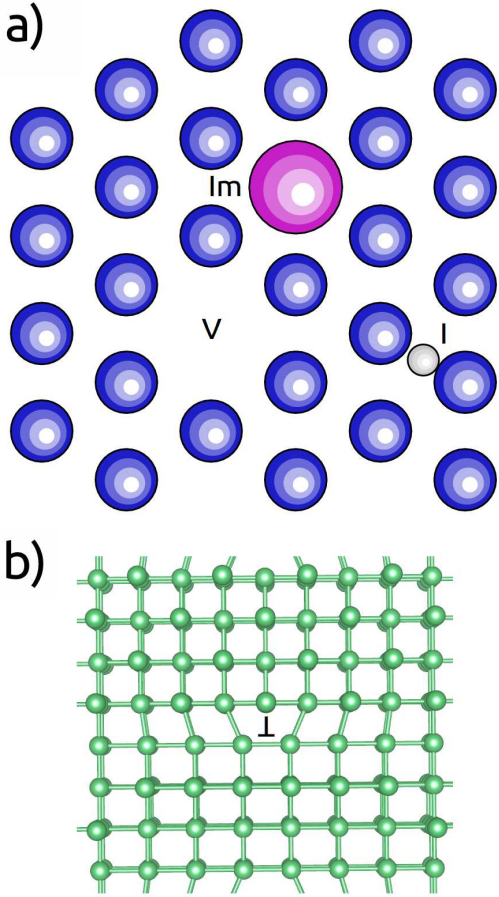}}
\caption{(Color online) Common types of defects in crystals. (a)~Representation of
         point defects; ``V'', ``Im'', and ``I'' respectively stand for vacancy, 
         impurity, and interstitial. (b)~Representation of an edge dislocation, a 
         class of line defect, in a solid with cubic symmetry.} 
\label{fig:crystal-defects}
\end{figure}

%===============================================================
A crystal is characterised by a periodic arrangement of atoms or molecules defined by an unit cell. 
The regular pattern in a solid, however, normally is interrupted by crystalline defects, which can be classified
into point, line, and planar types. Point defects occur only at or around one lattice site, and typical examples 
include vacancies, impurities, and interstitials (Kittel, 2005; see Fig.~\ref{fig:crystal-defects}). Line defects 
entail entire rows of atoms that are misaligned with respect to the others; common examples of line defects are 
dislocations, which in turn are classified into ``edge'' and ``screw'' (Bulatov and Cai, 2006; Hull and Bacon, 2011). 
An edge dislocation, for instance, is created by introducing an extra half-plane of atoms in mid-way through the 
crystal (see Fig.~\ref{fig:crystal-defects}). Planar defects can occur in single crystals or in the boundaries 
between single crystals, and include grain and twin boundaries, steps, and stacking faults (Kittel, 2005). The study 
of crystalline defects is very important since these can affect considerably the mechanical, electrical, structural, 
and adsorption properties of materials. Dislocations, for example, are key to understand the microscopic origins of 
plasticity, that is, the regime in which mechanical deformations in a crystal become non-reversible (Kosevich, 2005).    

In $1969$, Andreev and Lifshitz proposed that a state of matter in which crystalline order and
Bose-Einstein condensation coexisted could occur in a quantum crystal, the so-called supersolid. 
For this supersolid to exist, the presence of crystal vacancies was a necessary condition. At that 
time some experimentalists got attracted by the possibility of realising such an exotic state of matter, 
and several mass flow and torsional oscillator experiments were carried out in solid $^{4}$He at ultra-low 
temperatures. In all those experiments, however, no evidence for a supersolid state was found. Several 
decades later, a renewed interest in supersolids blossomed after the torsional oscillator experiments 
performed by Kim and Chan (2004a, 2004b). In their experiments, Kim and Chan observed a shift in the period 
of the torsional oscillator in solid helium as the temperature was lowered below $\sim 0.1$~K. This
sign was interpreted as the mass decoupling between the normal and superfluid fractions in the crystal. 
Meanwhile, few years later Day and Beamish (2007) measured the shear modulus in solid helium and found 
a striking resemblance with the temperature dependence of the oscillation period reported by Kim and Chan:
the shear modulus increased as the temperature was lowered below $\sim 0.1$~K. Day and Beamish attributed 
that increase in stiffness to the temperature dependence of the mobility of dislocations in the solid, 
which could be pinned by static $^{3}$He impurities. Day and Beamish's findings motivated a series of subsequent 
theoretical and experimental studies which have demonstrated that a change in the moment of inertia of the 
experimental torsional cell can be correlated to a change in the structure of the solid inside of it (Reppy, 
2010; Maris, 2012; Shin \emph{et al.}, 2016). In 2012, Kim and Chan completely redesigned their torsional 
oscillator setup making it stiffer, and the original mass-decoupling signal disappeared to within the 
experimental errors (see also, Kim and Chan, 2014). Thus, any convincing evidence of the existence of a 
supersolid yet is to be found.  

As a by-product of the frustrated investigations on supersolids, an interest in the plastic
behavior of quantum solids has emerged. Recently, Haziot \emph{et al.} (2013c) have shown that 
in ultra-pure single crystals of hcp $^{4}$He the resistance to shear along one particular direction
nearly vanishes at around $T = 0.1$~K due to free gliding of dislocations within the basal 
planes. This intriguing effect has been termed as ``giant plasticity'' and disappears 
in the presence of numerous $^{3}$He impurities or when the temperature is raised. 

In this section, we review the current understanding of crystalline defects in quantum crystals. 
Our analysis is focused on vacancies and dislocations since these are the two types of defects that
have been studied in more detail in solid $^{4}$He. Special emphasis is put on identifying those 
aspects that remain unknown or controversial.   

\subsection{Vacancies}
\label{subsec:vacancies}
Both experiments and theory agree in that the vacancy formation enthalpy, $\Delta H_{\rm v}$ , in solid 
$^{4}$He at ultra-low temperatures amounts to $\sim 15$~K (Fraass, Granfors, and Simmons, 1989; Galli and 
Reatto, 2004; Lutsyshyn \emph{et al.}, 2010). The general understanding then is that vacancies cannot 
be activated thermally in this crystal at temperatures as low as $0.1-1.0$~K. In fact, the classical 
equilibrium concentration of vacancies in a crystal is given by the expression 
$x_{\rm v}^{\rm class} = \exp{\left( -\Delta G_{\rm v} / T \right)}$, 
where $\Delta G_{\rm v}$ is the Gibbs free energy difference between the perfect and incommensurate (that is, 
defective) system. $\Delta G_{\rm v}$ is equal to $\Delta H_{\rm v} - T \Delta S_{\rm v}$, where $\Delta S_{\rm v}$
is the entropy change induced by the presence of vacancies. In turn, $\Delta S_{\rm v}$ can be estimated 
as the sum of a vibrational and a configurational contribution. The vibrational contribution corresponds to the 
variation of the lattice phonon frequencies as a result of the local relaxation occurring 
around the vacancy; in the limit of very small $x_{\rm v}$, this contribution can be safely neglected. 
The configurational entropy stems from the equivalency between lattice sites when creating a vacancy;
this contribution is $\Delta S_{\rm v}^{\rm conf} = -\ln{\left( x_{\rm v} \right)}$ and cannot be disregarded 
in the $x_{\rm v} \ll 1$ limit. By neglecting vibrational contributions to $\Delta G_{\rm v}$ and substituting
the value of $\Delta S_{\rm v}^{\rm conf}$ in $x_{\rm v}^{\rm class}$, one has that the classical equilibrium 
concentration of vacancies in a crystal is:
\begin{equation}
x_{\rm v}^{\rm class} = \exp{\left( -\frac{\Delta H_{\rm v}}{2T} \right)}~. 
\label{eq:xvclass}
\end{equation}       
In the case of solid $^{4}$He at $T = 0.1$~K, for instance, it follows that $x_{\rm v}^{\rm class} \sim 10^{-22}$
when considering $\Delta H_{\rm v} \sim 10$~K. In fact, such a classical equilibrium concentration of vacancies
is so extremely small that in principle it is physically irrelevant. 

Interestingly, Rossi \emph{et al.} (2008) and Pessoa, de Koning, and Vitiello (2009a, 2009b) have recently 
estimated, by using a reversible-work approach that exploits a quantum-classical isomorphism, that the zero-point 
vacancy concentration in solid $^{4}$He is $x_{v} \sim 10^{-3}$. Actually, this result is many orders of 
magnitude larger than the classical result obtained with Eq.~(\ref{eq:xvclass}), and it follows from assuming 
that the crystal is correctly described with a shadow wave function (Vitiello, Runge, and Kalos, 1988; MacFarland
 \emph{et al.}, 1994). An hypothetical equilibrium vacancy concentration of $\sim 10^{-3}$, although probably 
still is not experimentally detectable, would start being relevant to understand the physical behavior of defective 
solid $^{4}$He. Nevertheless, since Rossi and Vitiello's estimations ultimately rely on a variational model 
the large $x_{\rm v}^{\rm class} - x_{v}$ difference cannot be rigorously ascribed to the quantum nature of 
the crystal.    

Even when assuming that the equilibrium concentration of vacancies in solid helium is practically null, it cannot 
be discarded that in the process of growing a crystal from the liquid phase a small concentration of point defects 
is created. A possible question to answer next then is: do vacancies in a quantum crystal clusterise or keep dispersed? 
If vacancies clusterised, then they would segregate from the perfect system and become irrelevant. On the contrary, 
if vacancies remained separated, they could affect the general properties of the quantum crystal quite noticeably (Rota
 \emph{et al.}, 2012). Unfortunately, there is not a general consensus between theorists about how vacancies interact 
and distribute in solid $^{4}$He. Pollet \emph{et al.} (2008) have estimated from thermodynamic arguments that the 
binding energy of a divacancy is $E_{\rm div}^{\rm bind} = 1.4(5)$~K; we note that this result is about two times 
larger than the energy found by Clark and Ceperley (2008) using the PIMC method and a semi-empirical pairwise potential. 
It has been argued then that if vacancies existed they would separate into a vacancy-rich region and segregate from the 
perfect crystal. However, as we have noted earlier, at finite temperatures is crucial to consider also the entropic 
contributions to the Gibbs free energy, which cannot be obtained directly from the simulations. Actually, as we show 
next, it turns out to be much more favorable for the configurational entropy of the crystal to have two independent 
vacancies rather than a bound divacancy state. By completely ignoring vibrational effects, the resulting 
entropy gain can be estimated as:
\begin{equation}
\delta S_{\rm 2v-div}^{\rm conf} \approx 2 \cdot \Delta S_{\rm v}^{\rm conf} - \Delta S_{\rm div}^{\rm conf} = -\ln{\left( 2 \cdot x_{\rm v} \right)}~,                
\label{eq:div}
\end{equation}
where the constraint $x_{\rm div} = x_{\rm v}/2$ is employed. By considering the temperature and concentration 
of vacancies employed in PIMC simulations (Clark and Ceperley, 2008), namely $0.2$~K and $\sim 10^{-2}$, one 
obtains that $T \cdot \delta S_{\rm 2v-div}^{\rm conf} \sim 1$~K, which actually is of the same order of magnitude 
than the estimated $E_{\rm div}^{\rm bind}$. We note that the same conclusion also holds when considering smaller 
$T$'s and $x_{\rm v}$'s. Therefore, an attractive interaction between vacancies does not necessarily implies the 
existence of vacancy clusters or vacancy segregation. 

An alternative analysis to discern whether $^{4}$He vacancies coalescence or not, consists in monitoring their 
spatial correlations in quantum Monte Carlo simulations. For instance, if a multiple-vacancy bound state was to 
exist then an exponential decay in the corresponding vacancy-vacancy pair-correlation function should appear 
at separations beyond a specific interaction distance. Following this approach, Lutsyshyn, Cazorla, and Boronat 
(2010) and Lutsyshyn, Rota, and Boronat (2011) have not found any evidence for the existence of a multiple-vacancy 
bound state at zero temperature. In particular, at the freezing point and also at higher densities the tail in the 
vacancy-vacancy pair-correlation function always exhibits an asymptotic plateau. Pessoa, de Koning, and Vitiello 
(2009b) have arrived at a similar conclusion by means of VMC calculations performed with a shadow wave function 
model. Contrarily, Rossi \emph{et al.} (2010) have reported, based on the results of PIGS simulations, that 
when vacancies are present in large concentrations ($x_{\rm v} \sim 1$~\%) they tend to form bound states. 

We note that other possible processes involving vacancies, appart from clusterising or dispersing in bulk, have 
been also suggested; these include nucleation of dislocations (Rossi \emph{et al.}, 2010) and annealing towards 
the interface regions with the system container (Rossi, Reatto, and Galli, 2012). Test of these hypotheses in 
realistic crystals with first-principles methods, however, appears to be challenging due to the large system-size 
and relaxation-time scales involved in the simulations. 

In spite of the ongoing controversy about the possible existence of vacancies, the effects that hypothetically 
dispersed vacancies would have on the physical properties of solid helium have already been investigated 
thoroughly. For instance, the elastic properties of the incommensurate crystal in the limit of zero temperature
have been analysed by Cazorla, Lutsyshyn, and Boronat (2013); it has been shown that when considering 
large vacancy concentrations ($x_{\rm v} \sim 1$~\%) the shear modulus of the solid undergoes a small 
reduction of just few percent with respect to the perfect crystal case. 

Based on PIGS simulations and fundamental arguments, Galli and Reatto have demonstrated (2006) that Bose-Einstein 
condensation (BEC) occurs in the ground-state of incommensurate solid $^{4}$He, that is, $n_{0} \neq 0$ (see 
Sec.~\ref{subsub:pimc}), and that the corresponding critical temperature follows the relation $T_{0} \propto 
x_{\rm v}^{2/3}$. Recently, Rota and Boronat (2012) have corroborated the occurrence of vacancy-induced BEC in solid 
helium at low temperatures by means of PIMC simulations. It has been shown that below $T_{0}$ vacancies become quantum 
entities that completely delocalise in space; they have found also that the dependence of the critical temperature 
on $x_{\rm v}$ is best represented by a power law with coefficient $1.57$ (rather than of $2/3$), suggesting that the 
correlations between vacancies are stronger than previously thought. Interestingly, recent experiments performed by 
Benedek \emph{et al.} (2016) appear to support the possibility of a vacancy-induced BEC scenario in solid helium under 
non-equilibrium conditions.

\subsection{Dislocations}
\label{subsec:linedefects}
Since the seminal work by Day and Beamish (2007), there is little doubt that dislocations play a pivotal 
role on interpreting the mechanical behavior of solid $^{4}$He. If in the case of point defects it 
can be said that theory has led the way to their (partial) understanding, in the case of dislocations is the 
other way around. At present, most of what we know about dislocations in quantum solids comes from recent 
experiments performed by the groups of Beamish, in the University of Alberta, and Balibar, in the Ecole Normale 
Sup\'erieure de Paris (see, for instance, Haziot \emph{et al.}, 2013a; Haziot \emph{et al.}, 2013b; Fefferman 
\emph{et al.}, 2014; Souris \emph{et al.}, 2014a). Such a gap between theory and experiments is due to several 
reasons. First, in order to simulate dislocations reliably, large simulation cells containing at least several 
thousands of atoms need to be considered (Bulatov and Cai, 2006; Proville, Rodney, and Marinica, 2012); this system 
size is actually too large to be handled efficiently in quantum simulations. And second, dislocations are complex 
topological objects that until recently were not studied in depth in the context of low temperature physics, 
as a preponderant interest in ground-state properties leads to consider perfect crystals by default. However, 
as we describe next, quantum simulation of dislocations is critical for advancing our understanding of quantum 
crystals.  

\begin{figure}
\centerline
        {\includegraphics[width=1.0\linewidth]{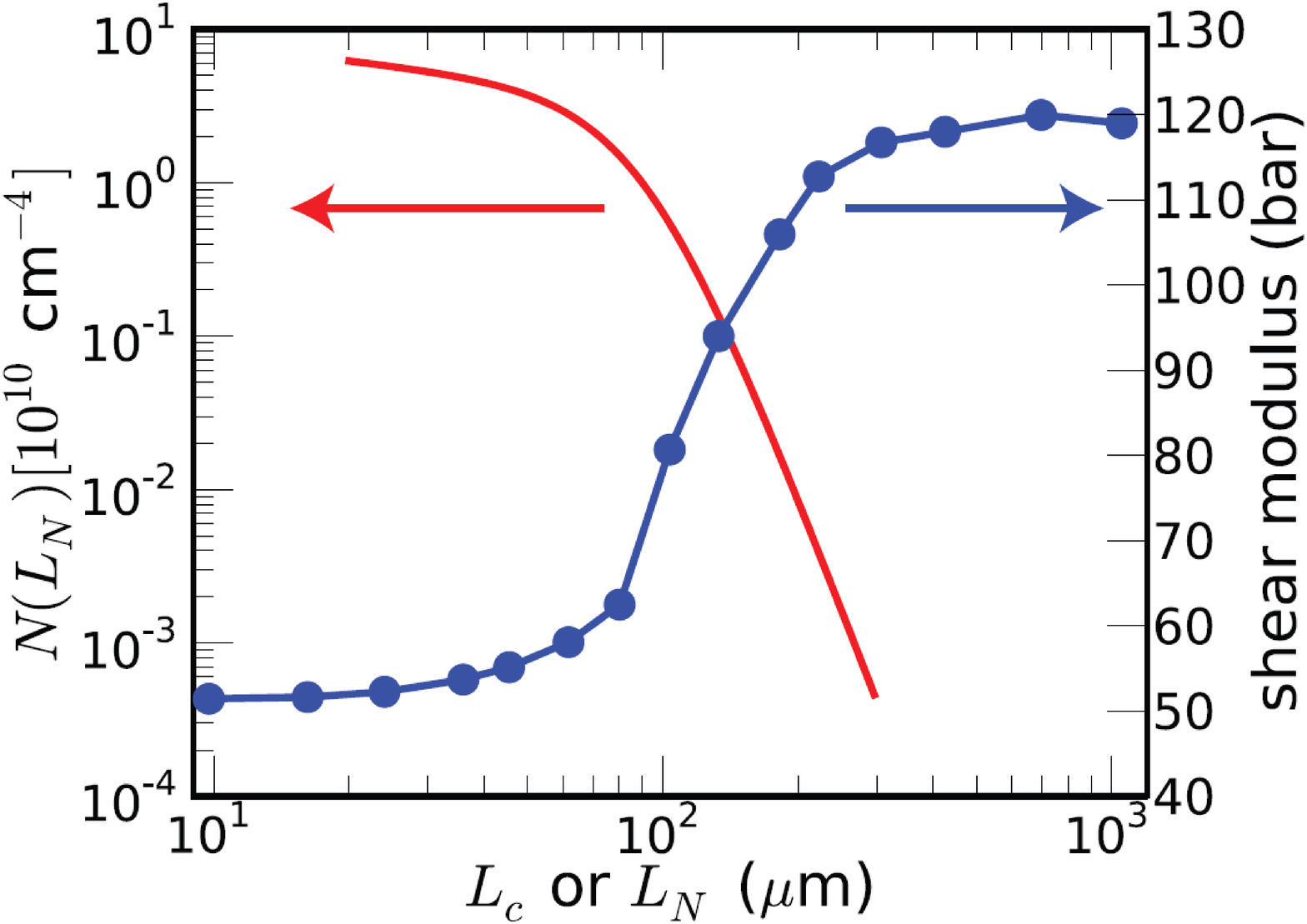}}
\caption{(Color online) The measured distribution, $N(L_{N})$, of lengths, $L_{N}$, between dislocation 
         network nodes in a $^{4}$He crystal at $T = 27$~mK; the contribution to the shear modulus from 
         each dislocation length is also indicated. From Fefferman \emph{et al.}, 2014.}
\label{fig:disldens}
\end{figure}

Dislocations are created during the growth process of solid helium (e. g., due to thermal contraction 
of the samples during cooling) as rough estimations of their formation energy amount to several thousands 
of K, hence they cannot be thermally activated at low $T$. Consider the classical elastic contribution to 
the formation energy per unit length of an edge dislocation (Cotterill and Doyama, 1966):   
\begin{equation}
E_{\rm disl}^{\rm elast}/L = \frac{\mu b^{2}}{4 \pi \left( 1 - \nu \right)} \ln{\left( \frac{r_{\rm d}}{r_{\rm c}} \right)} + E_{\rm core}^{\rm elast}~, 
\label{eq:elasdislo}
\end{equation}
where $\mu$ is the shear modulus of the crystal, $\nu$ its Poisson ratio, $b$ the length of the 
Burgers vector describing the dislocation, $r_{\rm d}$ the dislocation radius, $r_{\rm c}$ the 
dislocation core radius, and $E_{\rm core}^{\rm elast}$ the elastic energy of the dislocation 
core. Since we are interested in obtaining an approximate order of magnitude for $E_{\rm disl}^{\rm elast}$, 
we can neglect the second term in Eq.~(\ref{eq:elasdislo}), which is always positive, and assume 
that $\ln{\left( r_{\rm d} / r_{\rm c} \right)} \sim 1$. Using the elastic data reported for 
perfect solid $^{4}$He by Pessoa, Vitiello and de Koning (2010), that is, $\mu = 17.1$~MPa and $\nu = 0.15$, 
and adopting an usual Burgers vector of modulus $b = a / \sqrt{3} = 2.1$~\AA, one obtains 
that $E_{\rm disl}^{\rm elast}/L \sim 1$~K/\AA~. Considering now that dislocations in solid $^{4}$He 
typically are several $\mu$m long (see Fig~\ref{fig:disldens}), one finally obtains that, at least,  
$E_{\rm disl}^{\rm elast} \sim 10^{4}$~K. We note that although this rough estimation of the 
elastic formation energy of line defects is several orders of magnitude larger than the cost of 
creating, for instance, a vacancy (see Sec.~\ref{subsec:vacancies}), in principle it is not possible 
to grow $^{4}$He crystals free of dislocations with current state-of-the-art synthesis methods (Souris 
\emph{et al.}, 2014b). The apparently inevitable presence of dislocations in solid helium near the 
zero temperature limit already poses a puzzle to the theorist's mind.  

In a series of compelling experimental works, Balibar, Beamish and collaborators have characterised 
the energy, structural, and dynamic properties of dislocations in solid $^{4}$He (for a recent 
review, see Balibar \emph{et al.}, 2016). The usual experimental setup in those studies consists 
in a measurement cell supplied with two piezoelectric shear plates that are placed facing each to 
the other with a separation of few millimeters; the narrow gap that is formed between the transducers 
then is filled with a crystal that is oriented in a particular direction. By applying a voltage to one 
of the piezoelectric plates a shear strain is induced in the crystal, and the resulting stress is 
measured by the opposite shear plate. This process is done repeatedly by using alternating currents. 

The theory underlying most of Balibar and Beamish's results is that due to Granato and L\"ucke 
(1956), in which an analogy is made between the vibration of a dislocation pinned down by impurity 
particles under an alternating stress field and the classical (that is, not quantum mechanical) problem 
of the forced damped vibration of a string. In Granato and L\"ucke's classical theory it is assumed 
that at high temperatures dislocations interact with thermal lattice phonons, and that as a 
consequence a maximum shear modulus change of:
\begin{equation}
\delta_{\mu} \equiv \frac{\Delta C_{44}}{C_{44}} = \frac{\alpha \Lambda L^{2}}{1 + \alpha \Lambda L^{2}}
\label{eq:lukesm}
\end{equation}   
and a dissipation (that is, the phase difference between the applied strain and resulting stress) of:
\begin{equation}
\frac{1}{Q} = \delta_{\mu} B L^{2} \omega T^{3}~,
\label{eq:lukedis}
\end{equation}
occur in the crystal. In the context of solid $^{4}$He, ``high temperatures'' are considered to be 
$T \ge 0.3$~K (Balibar \emph{et al.}, 2016). In the equations above $\alpha$ and $B$ represent two thermal phonon 
damping parameters (which in solid $^{4}$He are equal to $0.019$ and $905$~s$\cdot$m$^{-2}$~K$^{-3}$, respectively, 
see Souris \emph{et al.}, 2014a), $\Lambda$ the density of dislocation lines per surface unit, $L$ a typical 
length between nodes in the dislocation network, and $\omega$ the frequency of the alternating strain field. 

Using Eqs.~(\ref{eq:lukesm}) and~(\ref{eq:lukedis}) and from direct measurements of $\delta_{\mu}$ and $1/Q$, 
Haziot \emph{et al.} (2013a) have found that typical values of $\Lambda$ and $L$ in solid helium are 
$10^{4}-10^{6}$~cm$^{-2}$ and $100-230$~$\mu$m (see Fig.~\ref{fig:disldens}), which in the latter case turn 
out to be macroscopic. In very high quality crystals (Souris \emph{et al.}, 2014b), it has been observed 
that dislocations avoid crossing each other by forming two-dimensional arrays of parallel lines called 
``sub-boundaries'', and that they glide together parallel to the basal planes. Remarkably, in the limit of 
zero temperature the dissipation associated to the gliding of dislocations in the basal plane vanishes 
(Fefferman \emph{et al.}, 2014), an effect that has been interpreted as evidence of quantum behavior. 
Nevertheless, whether such an observation implies that the formation energy of dislocation kinks and jogs 
(that is, defect perturbations that affect the straightness of the dislocation line) also vanishes at ultra-low 
temperatures, or that dislocation kinks and jogs are able to quantum tunneling through small energy barriers, 
yet needs to be clarified (Kuklov \emph{et al.}, 2014). In this context, the outcomes of quantum simulations 
could be highly valuable. 

At temperatures below $0.2$~K, it is found that the dynamics of dislocations is greatly influenced by the 
presence of isotopic $^{3}$He impurities. When the concentration of $^{3}$He atoms, $x_{3}$, is large enough
(i. e., $\sim 10^{-7}$ or larger) and $T$ is progressively reduced, the impurities start to bind to the 
dislocations with an energy that, according to Souris \emph{et al.}'s (2014a) measurements, is of $0.7~(0.1)$~K.
At those conditions, the mobility of the dislocations depends also on the frequency of the applied strain. 
At high frequencies, that is, at high dislocation speeds of $> 45$~$\mu$m/s, the impurities cannot move fast 
enough to follow the line defects so that they end up anchoring them. However, at lower frequencies, and always 
considering Souris \emph{et al.}'s (2014a) arguments, dislocations can actually move dressed with $^{3}$He 
atoms. 

A pertinent comment has to be made here. Several nuclear magnetic resonance studies have shown that at low 
temperatures $^{3}$He atoms in solid $^{4}$He behave as quantum quasi-particles that can move through the 
lattice at velocities as high as $\sim 1$~mm/s (Allen, Richards, and Schratter, 1982; Kim \emph{et al.}, 
2013), that is, significantly higher than $45$~$\mu$m/s. It has been argued then that near the dislocation 
the mobility of isotopic impurities could be reduced considerably by the existing local strain (Balibar 
\emph{et al.}, 2016); however, there is not quantitative evidence showing that such a huge variation of about 
three orders of magnitude in the mobility of $^{3}$He impurities could be actually possible. Clearly, a 
microscopic understanding of what are the interactions between dislocations and quantum isotopic impurities, 
and the factors that can affect the mobility of the latter, is necessary. The outcomes of quantum simulation 
studies again could be very useful in clarifying these issues.

With regard to theory, Boninsegni \emph{et al.} (2007) have shown using PIMC simulations that the core of 
screw dislocations with Burgers vectors oriented perpendicular to the basal plane in solid $^{4}$He are 
superfluid. Boninsegni \emph{et al.}'s predictions have led to a number of hypotheses about possible new 
phenomena involving quantum dislocations like, for instance, ``superclimb'' (S\"{o}yler \emph{et al.}, 2009; 
Aleinikava, Dedits, and Kuklov, 2011) and superfluidity in dislocation networks (Boninsegni \emph{et al.}, 
2007). In a recent PIMC study by Landinez-Borda, Cai, and de Koning (2016) on solid helium, it has been 
reported that either screw or edge dislocations with Burgers vectors along the basal plane are not superfluid. 
In particular, both types of dislocations are predicted to dissociate into non-superfluid Shockley partial 
dislocations separated by ribbons of stacking fault, as it normally occurs in classical hcp crystals (Bulatov 
and Cai, 2006; Hull and Bacon, 2011). Landinez-Borda, Cai, and de Koning (2016) have also concluded that in 
solid helium the resistance to flow of partial dislocations is negligible (that is, the corresponding Peierls 
stress is nominally zero) mostly due to zero-point fluctuations. The results presented in this latter simulation 
work provide valuable insight into the physical origins of the observed ``giant plasticity'' effect (Haziot 
\emph{et al.}, 2013c; Zhou \emph{et al.}, 2013; Haziot \emph{et al.}, 2013d). 

Apparently, there seems to be some inconsistencies among the conclusions presented by Boninsegni \emph{et al.} 
(2007) and Landinez-Borda, Cai, and de Koning (2016) as to what concerns the superfluid properties of dislocation 
cores. Nevertheless, we must note that the linear defects analysed in those two studies are different as their 
respective Burgers vectors are either oriented along the $c$-axis or contained in the basal plane. Further 
quantum simulation studies indeed appear to be necessary for clarifying the role of Burgers vector orientation 
on the transport properties of dislocation cores in solid $^{4}$He.       

Finally, recent experiments done in the group of Hallock in the University of Massachusetts have shown unexpected 
mass flow through $^{4}$He crystals at low temperatures ($T < 600$~mK) when sandwiched between two regions of 
superfluid liquid in which a pressure gradient is applied (Ray and Hallock, 2008; Ray and Hallock, 2010; Vekhov 
and Hallock, 2012). This phenomenon has been dubbed as ``giant isochoric compressibility'' or the ``syringe effect''.
The observed mass flow has been interpreted in terms of two possible scenarios (Hallock, 2015), namely the climbing 
(i. e., the passing of an obstacle to start moving again) of either superfluid dislocations (S\"{o}yler \emph{et 
al.}, 2009) or grain boundaries (Burovski \emph{et al.}, 2005; Sasaki \emph{et al.}, 2006; Pollet \emph{et al.}, 
2007; Cheng and Beamish, 2016). Recently, analogous mass flow phenomena have been observed also in an inverted 
solid-superfluid-solid setup by Cheng \emph{et al.} (2015), in which the effects of $^{3}$He impurities concentration 
and distribution have been analysed in detail. The exact atomistic mechanisms underlying the inverted and direct 
syringe effects, however, still remain an open question. New systematic experiments and quantum simulation studies 
certainly are necessary to achieve a more accomplished knowledge of mass transport along quantum linear and planar 
defects (i. e., dislocations and grain boundaries).

%===============================================================
\section{The role of dimensionality}
\label{sec:dimensionality}
%===============================================================
Quantum crystals at reduced dimensionality have been the object of numerous experimental
and theoretical studies. The interplay between quantum correlations and structural 
confinement opens a series of interesting new prospects that since the beginning of 
the quantum Monte Carlo era have been investigated meticulously with theory. The 
search for novel phases and physical phenomena in quantum gases adsorbed on graphitic 
and metallic substrates or on the surface of carbon nanostructures and the interior of 
narrow silica pores, represent well-known examples.

\subsection{Quantum films}
\label{subsec:films}
We focus here on helium and hydrogen since in these species QNE are most pronounced. 
In very thin films one can expect that two-dimensional effects become dominant, and for 
this reason many works have concentrated in studying the thermodynamic, structural, and 
dynamical properties of purely 2D quantum many-body systems.  

At zero temperature and zero pressure 2D $^4$He is a liquid with an estimated
equilibrium density of $\sigma_{0} = 0.043$~\AA$^{-2}$ and binding energy of 
$E/N = -0.90$~K (Giorgini, Boronat, and Casulleras, 1996). By increasing the 
density, the liquid solidifies into a triangular lattice (Whitlock, Chester, and 
Kalos, 1988). The liquid and solid are in equilibrium at densities $\rho_{f} = 0.068$~\AA$^{-2}$ 
(freezing) and $\rho_{m} = 0.072$~\AA$^{-2}$ (melting), respectively. On the other
hand, $^3$He at low densities remains in the gas phase due to its lower mass, and more 
importantly, fermionic character (Grau, Boronat, and Casulleras, 2002). Upon steady 
increase in density the gas eventually transforms into a triangular solid, although  
the critical point associated to this transition has not been characterised yet with 
precision.

\begin{figure}
\centerline
        {\includegraphics[width=1.00\linewidth,angle=0]{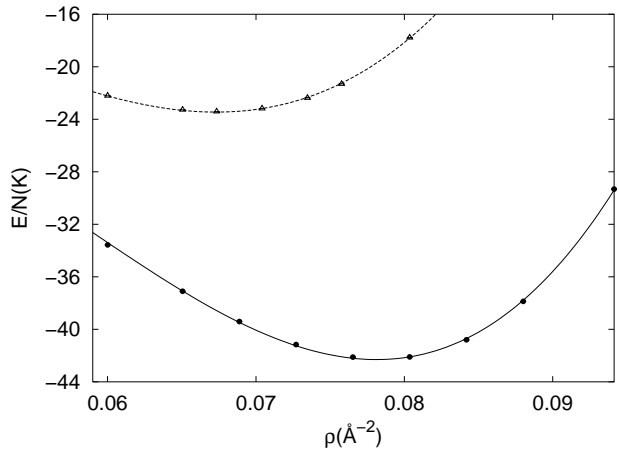}}
        \caption{Energy per particle in two-dimensional solid ${\rm D_{2}}$ (solid 
        line and filled circles), and two-dimensional solid ${\rm H_{2}}$ (dotted 
        line and empty triangles). From Cazorla and Boronat (2008c).}
\label{figd22d}	
\end{figure}

The ground-state of two-dimensional molecular hydrogen and deuterium have been  
investigated also with QMC methods (Cazorla and Boronat, 2008c; Boninsegni, 2004). The 
primary interest of these studies was to discern whether by reducing the dimensionality 
it was possible to stabilise the liquid phase. Those theoretical works, however, have 
shown that the fluid is never stable, neither when considering negative pressures 
(Cazorla and Boronat, 2008c; Boninsegni, 2004). In Fig.~\ref{figd22d}, we enclose the 
density dependence of the energy calculated in 2D H$_{2}$ and D$_{2}$ with the DMC 
method and Silvera-Goldman potential (Cazorla and Boronat, 2008c). At zero pressure, 
both crystals stabilise in a triangular lattice of density $\sigma_0=0.0673$~\AA$^{-2}$ 
and $0.0785$~\AA$^{-2}$, respectively. The corresponding energy per particle at those 
conditions are $-23.45$~K in H$_2$ and $-42.30$~K in D$_2$. This large energy difference 
indicates that the presence of quantum isotopic effects is also significant when considering 
only two dimensions.
 
With regard to the possibility of realising H$_{2}$ superfluidity (see Sec.~\ref{subsec:hydrogen}), 
several strategies have been explored also in reduced dimensionality. It was first proposed by 
Gordillo and Ceperley (1997) that the intercalation of alkali atoms could frustrate the formation 
of the solid due to the weaker interaction between impurities and H$_2$ than between hydrogen
molecules. K and Cs were considered as the likely candidates to induce H$_{2}$ melting in a PIMC 
study by Gordillo and Ceperley (1997). Large superfluid fractions of $\rho_{s}/\rho \sim 0.2-0.5$ 
were reported in the resulting hydrogen-alkali thin films. However, subsequent quantum simulation 
studies performed with a larger number of particles have found very small values of $\rho_{s}/\rho$ 
in equivalent systems (Boninsegni, 2005; Cazorla and Boronat, 2004). More recently, Cazorla and 
Boronat (2013) have predicted by using the DMC method and semi-empirical pairwise potentials that 
frustration of 2D solid H$_{2}$ could be achieved with sodium atoms arranged in a triangular lattice 
of constant $10$~\AA~. The main reason for this is that the forces between Na atoms and hydrogen 
molecules are weaker than those considered in previous studies, hence a significant reduction of 
the equilibrium density is induced that favors stabilization of the liquid phase. We note, however, 
that in a posterior PIMC study by Boninsegni (2016) this conclusion has been disputed by arguing 
that the system remains in the solid phase independently of its density and type of alkali impurity 
that is considered. 

Experimental realisation of quasi-two dimensional quantum solids is achieved through adsorption of 
quantum gases on attractive substrates. In this context, one of the most extensively investigated 
substrates is graphite. The physics of gas-adsorption phenomena in graphite is very rich (Bruch, 
Cole, and Zaremba, 1997) as a large sequence of transitions have been experimentally observed and 
described with microscopic theory (Clements \emph{et al.}, 1993; Clements, Krotscheck, and Saarela, 
1997). We concentrate here on describing the properties of the first adsorbed layer and other related 
crystalline phases. It is worth noticing that when corrugation effects are included in the simulations 
(that is, the spatial distribution of carbon atoms in the underlying substrate is explicitly taken 
into consideration), denser commensurate phases are normally obtained. 

According to recent quantum simulation studies performed with semi-empirical pairwise potentials, the 
ground state of $^4$He adsorbed on graphite (and graphene) is a $\sqrt{3}\times\sqrt{3}$ commensurate 
phase with a surface density of $0.0636$~\AA$^{-2}$ (Gordillo, Cazorla, and Boronat, 2011). The liquid 
phase is metastable with respect to the crystal. As the density is increased, the commensurate crystal 
transforms into a triangular incommensurate solid of density $\sim 0.08$~\AA$^{-2}$ (Gordillo and Boronat, 
2009b; Pierce and Manousakis, 2000; Corboz \emph{et al.}, 2008). This description is in excellent agreement 
with the available experimental data (Bruch, Cole, and Zaremba, 1997). By increasing further the density, 
a second layer develops on top of the first with an equilibrium density of $0.12$~\AA$^{-2}$.

Recent QMC studies of the registered phases of H$_2$ adsorbed on graphite and graphene provide a 
description that is identical to that obtained in $^4$He, and which is in very good agreement with 
the experiments (Gordillo and Boronat, 2010). In particular, the ground state is a commensurate 
$\sqrt{3}\times\sqrt{3}$ phase that undergoes a first-order transition towards an incommensurate 
triangular crystal at $\rho =0.077$~\AA$^{-2}$. The phase diagram of D$_2$ on graphite has been 
investigated thoroughly in experiments (Bruch, Cole, and Zaremba, 1997) but not yet with theory. 
It is known that this is richer than its H$_2$ counterpart since at least two additional commensurate 
phases appear in the first adsorbed layer: the $\varepsilon$ phase, which is a $4\times 4$ structure 
($0.0835$~\AA$^{-2}$), and the $\delta$ one, corresponding to a $5\sqrt{3}\times 5\sqrt{3}$ lattice 
($0.0789$~\AA$^{-2}$). According to DMC calculations none of these latter commensurate phases are 
stable in H$_2$ (Gordillo and Boronat, 2010).

\subsection{One-dimensional systems}
\label{subsec:onedim}
Carbon nanotubes and nanopores embedded in solid matrices have opened the possibility of 
studying quantum systems in quasi-one dimensional geometries (Calbi \emph{et al.}, 2001). 
Investigating individual carbon nanotubes, however, has proved challenging due to the fact 
that they are normally capped and adsorption of atoms on their interior is energetically unfavored. 
Alternatively, adsorption on the intersites and grooves formed between adjacent nanotubes have been 
observed. A topic of interest in these systems is the study of the crossover between three-dimensional 
and one-dimensional physics (Gordillo and Boronat, 2009a). For instance, changes on the superfluid 
fraction and condensate fraction of liquid $^{4}$He upon variation of the nanopore radius have 
been systematically studied by Vranje\v{s} Marki\'c and Glyde (2015) with PIMC simulations.

Strictly one-dimensional quantum systems possess unique features as compared to the rest of 
low-dimensional systems. One of the most relevant aspects is that the presence of a hard core 
in the interatomic interactions makes quantum statistics irrelevant. This means that 
simulation of a Fermi system (e. g., $^3$He) can be made exactly because the nodes of the 
corresponding wave function are known \emph{a priori}, hence one can get rid of the sign problem 
in practice (see Sec.~\ref{subsubsec:qmc}). Another important characteristic is the absence of 
continuous phase transitions (that is, as described by Landau's theory), although the limiting 
$T=0$ case may be an exception. Finally, if the excitation spectrum of the system is gapless, 
i. e., $E_{k} = \hbar |k| c$, Luttinger theory applies and consequently the behavior of correlation 
functions at large-distance (or small-momenta) is known analytically (Giamarchi, 2004; Imambekov, 
Schmidt, and Glazman, 2012). In this latter case, the behavior of the system is universal in terms 
of the Luttinger parameter, $K$, which in a homogeneous system is directly related to the Fermi 
velocity, $v_{F}$, and speed of sound, $c$, as $K = v_{F}/c$~. The Fermi velocity $v_{F}$ is 
completely fixed by the linear density, $\rho$, whereas the speed of sound $c$ depends on the 
many-body interactions.  

In Luttinger's theory, the height of the $l$-th peak in the static structure factor, $S(k)$, is 
given by (Haldane, 1981):
\begin{equation}
S(k = 2l k_{F}) = A_{l} N^{1-2l^{2}K}~,
\label{Eq:Sk:peak}
\end{equation}
which diverges when $K < 1/2l^{2}$. If the first peak in $S (k)$ diverges, that is, $K<1/2$, 
the system is called a ``quasi-crystal''. There is a number of important differences between 
quasi and real crystals. A three-dimensional crystal possesses diagonal long-range order 
since the density oscillations in the two-body distribution function remain in phase over 
long distances. In one dimension, on the contrary, that order is lost according to a power-law 
decay. The height of the first peak diverges in both cases, however in a true crystal the 
Bragg peak grows linearly with the number of particles, $S(k_{peak}) \propto N$, whereas in a 
quasi-crystal the corresponding exponent is smaller than unity [see Eq.~(\ref{Eq:Sk:peak})].

\begin{figure}
\centerline
{\includegraphics[width=0.8\linewidth,angle=90]{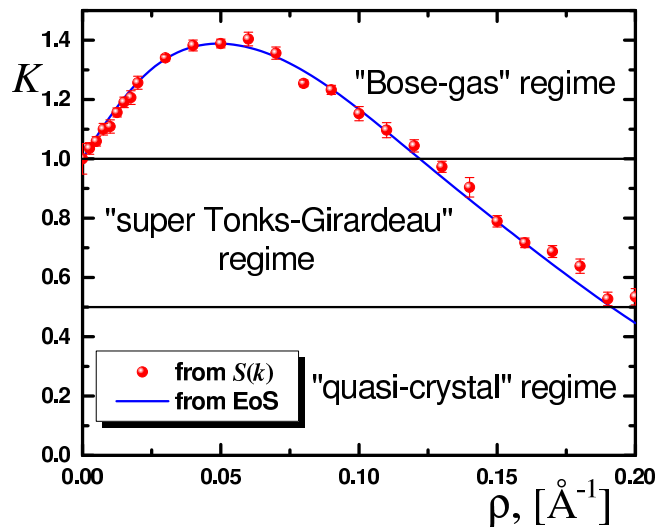}}
\caption{(Color online) Luttinger parameter, $K$, in one-dimensional $^3$He expressed as a 
function of the linear density, $\rho$. The corresponding speed of sound, as extracted from 
the phononic part of the static structure factor (symbols) and thermodynamic compressibility 
(line), is also shown. Adapted from Astrakharchik and Boronat (2014).}
\label{fig:luttinger}
\end{figure}

QMC calculations of 1D $^4$He at equilibrium have shown that this is a self-bound system with 
a tiny binding energy of $\sim -4$~mK (Gordillo, Boronat, and Casulleras, 2000a; Boninsegni 
and Moroni, 2000). When the density is increased, the system eventually becomes a quasi-crystal. 
Recently, the ground state of one-dimensional $^3$He has been studied thoroughly with the DMC 
method (Astrakharchik and Boronat, 2014). The lower mass of the isotope makes the system to be 
non self-bound, and thus it remains in the gas phase down to zero pressure. Through calculation 
of the corresponding Luttinger parameter one can appreciate the richness of its behavior as a 
function of density (see Fig.~\ref{fig:luttinger}). 

As the interaction between hydrogen molecules is more attractive than between helium atoms, 
H$_{2}$ is also self bound in the one-dimensional limit. When H$_2$ molecules, or helium atoms, 
are adsorbed inside of a nanopore the resulting phases depend strongly on the amount of space 
that is available. In very narrow nanotubes, for instance, one observes the existence of real 
quasi-1D systems, that is, in the Luttinger sense (Gordillo and Boronat, 2009a). On the contrary, 
if the nanopore diameter is wide enough, particles migrate towards the nanopore walls due to the 
strong attractive interactions. Eventually, if the nanopore interior is further enlarged, nucleation 
of a narrow channel containing a liquid may occur (Rossi, Galli,and Reatto, 2005). In the case of 
molecular hydrogen, however, the possible stabilisation of a 1D fluid remains controversial 
(Gordillo, Boronat, and Casulleras, 2000b; Boninsegni, 2013b; Omiyinka and Boninsegni, 2016; 
Rossi and Ancilotto, 2016).

Recently, the adsorption of quantum gases on the external surface of a single nanotube has 
drawn some attention. State-of-the-art resonance experiments on a single suspended carbon 
nanotube have been able to determine the phase diagram of the deposited rare gases with high 
precision (Wang \emph{et al.}, 2010; Tavernarakis \emph{et al.}, 2014). For instance, in the 
$T = 0$ limit one can identify either a registered $\sqrt{3}\times \sqrt{3}$ phase, already 
known from adsorption on planar substrates, or incommensurate phases, depending on the chemical 
species. Theoretical predictions on these systems (Gordillo and Boronat, 2011) agree well with 
the experimental findings.

\subsection{Clusters}
\label{subsec:clusters}
Helium and hydrogen drops can be generated in the laboratory by means of  
free jet expansions from a stagnation source chamber that go through a thin walled nozzle 
(Grebenev, Toennies, and Vilesov, 1998). Helium drops are the most clean example of 
inhomogeneous quantum liquids with either boson ($^4$He) or fermion ($^3$He) quantum 
statistics. In recent years, the relevance on He drops has been reinforced by the increasing 
interest in studying the behavior of small molecules placed in their interior. In fact, 
quantum clusters can act as ideal matrices in which to carry out accurate spectroscopy 
analysis of the embedded molecules. When the guest molecule is surrounded by $^4$He atoms, 
the corresponding rotational spectrum presents a peaked structure that has been attributed 
to the superfluid nature of helium. By contrast, in $^3$He drops a broad peak is recorded. 
This phenomena, termed as microscopic superfluidity, has been the object of many QMC 
studies in the last years (Sindzingre, Klein, and Ceperley, 1989; Sola, Boronat, and 
Casulleras, 2006). 

H$_2$ clusters have been produced also in the laboratory with jet expansion techniques 
(Tejeda \emph{et al.}, 2004). The behavior of H$_{2}$ drops is richer than that of $^4$He 
since they can be either liquid or solid depending on the number of constituent particles. 
The first PIMC study on H$_2$ clusters was carried out by Sindzingre, Ceperley, and Klein 
(1991), and it was found that clusters comprising a number of molecules up to $N \simeq 18$ 
were superfluid at temperatures below $T=2$~K. In a subsequent PIMC work (Khairallah \emph{et al.}, 
2007) the limiting number of molecules exhibiting superfluid behavior has been raised to 
$N \simeq 26$. The results reported by Khairallah \emph{et al.} (2007) appear to show 
superfluidity mostly localised in the surface of the cluster, which points to an inhomogeneous 
structure formed by a solid core surrounded by a liquid \emph{skin}, that at low temperatures 
becomes superfluid. This interpretation, however, has been challenged in a posterior PIMC 
work in which it has been argued that, in spite of the local variation in molecular order, 
superfluidity remains a global property of the entire cluster (Mezzacapo and Boninsegni, 
2008). 

\begin{figure}
\centerline
{\includegraphics[width=1.0\linewidth,angle=0]{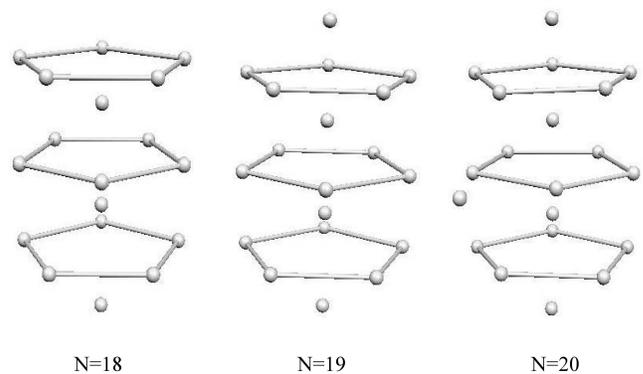}}
\caption{(Color online) Optimal distribution of equilibrium sites in solid H$_2$ 
clusters with $N=18$, $19$, and $20$ molecules at zero temperature. From Sola and 
Boronat (2011).}
\label{gotesh2}
\end{figure}

The structure and energy of small H$_2$ clusters in the limit of zero temperature have 
been studied accurately with both the DMC (Guardiola and Navarro, 2008) and PIGS (Cuervo 
and Roy, 2006) methods. The presence of \emph{magic}-cluster sizes, identified with a 
kink in the chemical potential, have been reported in those studies. The number of 
molecules contained in the smallest and energetically most stable clusters, appear to 
coincide with the results of Raman spectroscopy measurements (Tejeda \emph{et al.}, 2004).
A combination of the DMC and stochastic optimization (i. e., simulated annealing) techniques 
has allowed to determine the equilibrium structure in most stable solid H$_2$ clusters 
(Sola and Boronat, 2011). Examples of optimal molecular arrangements obtained in those 
clusters are shown in Fig.~\ref{gotesh2}.

%===============================================================
\section{Molecular crystals}
\label{sec:molsol}
%===============================================================
Molecular systems are of critical importance in astronomy, biology, and environmental science. 
Hydrogen is the most abundant element in the universe and over wide thermodynamic conditions 
is most stable in molecular form (see Sec.~\ref{subsec:hydrogen}). Water is vital to all known 
forms of life and it covers around three quarter parts of the Earth's surface. Nitrogen and methane 
are found in the interior and crust of many celestial bodies and also in organic substances. 
When all these species are compressed eventually they become crystals in which, due to 
the light weight of their atoms and weak interparticle interactions, QNE play a pivotal role at low 
temperatures (see Fig.~\ref{fig:deboer}). Next, we briefly review the knowledge of the phase diagram 
of these compounds and highlight the aspects that remain contentious. Due to their intrinsically rich 
and complex nature, it is not possible to provide here a detailed description of H$_{2}$, H$_{2}$O, 
N$_{2}$, and CH$_{4}$, hence we address the interested reader to other recent and more specialised 
articles (see, for instance, McMahon \emph{et al.}, 2012; Goncharov, Howie, and Gregoryanz, 2013; 
Herrero and Ram\'irez, 2014). For the sake of focus, only those aspects related to the crystalline 
phases are considered in this section. 

\subsection{H$_{2}$ at extreme $P-T$ conditions}
\label{subsec:h2extreme}

\begin{figure}
\centerline
        {\includegraphics[width=1.0\linewidth]{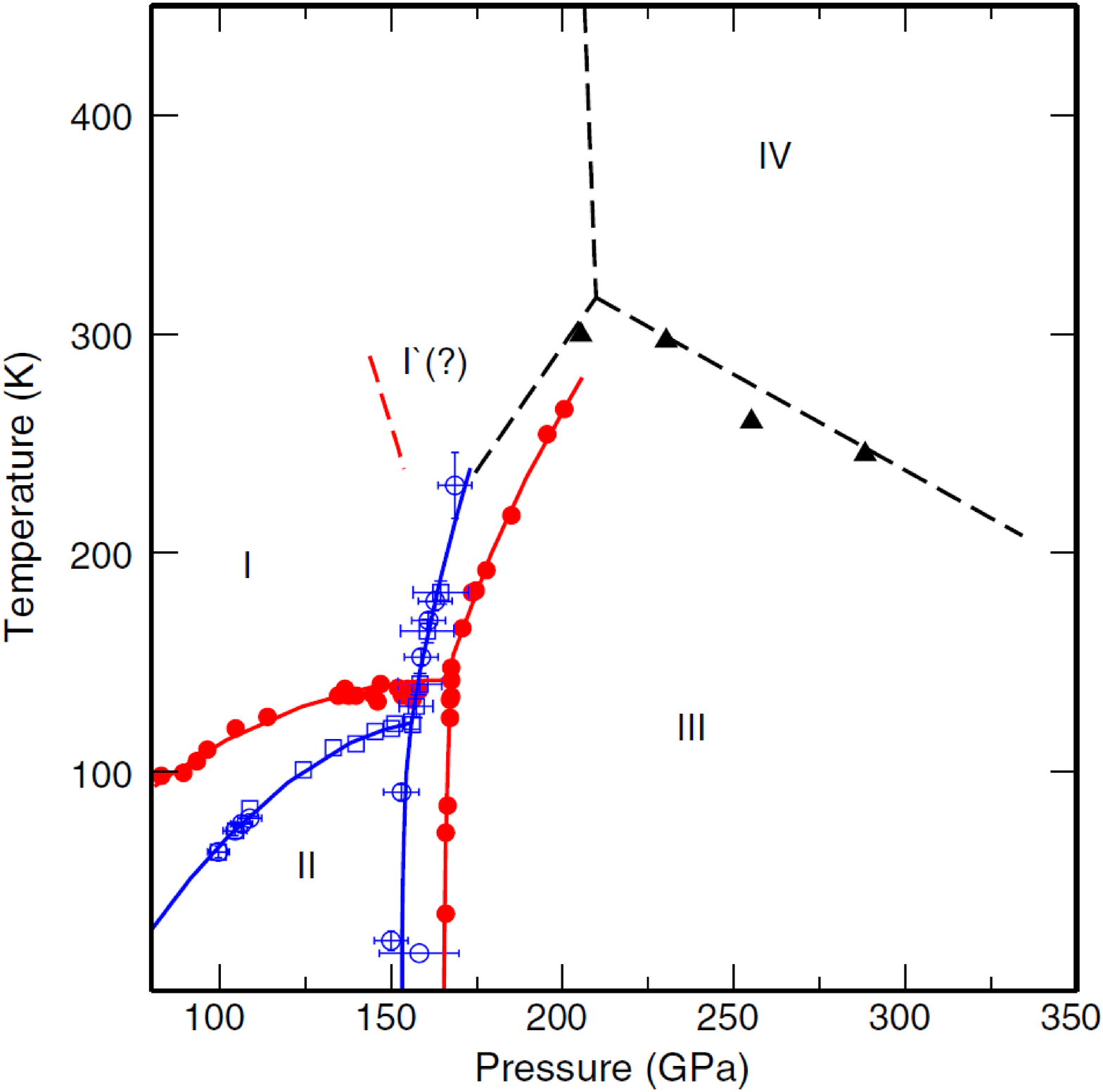}}
\caption{(Color online) High-$P$ phase diagram of solid molecular hydrogen and deuterium. Open and 
filled circles are Raman measurements for hydrogen and deuterium, respectively, and open squares 
are IR data for the hydrogen (Goncharov, Hemley, and Mao, 2011). Triangles and the  
dashed lines indicate inferred boundaries for the recently proposed phase IV (Howie \emph{et al.}, 
2012b). The dashed line and the existence of phase I' are still matters of debate (see text). From 
McMahon \emph{et al.}, 2012.} 
\label{fig:h2solpd}
\end{figure}

Due to its low $Z$ number, hydrogen's x-ray scattering cross section is very low. This means that it 
is extremely challenging to determine with accuracy its atomic structure under extreme thermodynamic 
conditions in the laboratory. Infrared (IR) and Raman spectroscopy techniques have been used to monitor 
the changes in the vibrational properties of the crystal that can be ascribed to a phase transition.
However, due to the high reactivity, mobility and diffusion of the molecules already at moderate 
temperatures (i. e., $\ge 250$~K), this type of measurements turn out to be very difficult (Goncharov,
Howie,and Gregoryanz, 2013). Here is where the inputs of theory and computer simulations become critical. 
By comparing the vibrational phonon spectra of low-energy structures obtained in first-principles searches 
with experimental data, candidate atomic structures can be identified for each of the detected transformations. 
Unfortunately, for the reasons highlighted in Sec.~\ref{subsec:challengesquantumsolids}, the theoretical 
study of hot and dense solid hydrogen is technically difficult and very sensitive to the employed method 
(that is, the free energy differences between phases normally are very small, of the order of few meV, 
which coincide with the typical accuracy threshold in first-principles calculations). As a consequence, the 
description of hydrogen-based systems obtained with various levels of theory may differ (Morales \emph{et al.}, 
2013; Drummond \emph{et al.}, 2015), complicating even further the characterisation of solid hydrogen.     

The H$_{2}$ crystal phases that are experimentally well established are denoted by I, II, III, 
and IV (see Fig.~\ref{fig:h2solpd}). Phase I corresponds to the close-packed hcp structure, in which 
para-H$_{2}$ molecules have zero angular momentum and spherically symmetric wave functions (Silvera, 1980). 
At low temperatures and as pressure is increased, breaking of rotational symmetry eventually occurs and 
the crystal stabilises in phase II (Lorenzana, Silvera, and Goettel, 1990); the boundary between phases I 
and II is strongly dependent on isotope type (see Fig.~\ref{fig:h2solpd}), which indicates the presence 
of important QNE. Around $150$~GPa, molecular hydrogen undergoes another phase transformation into phase III 
(Hemley and Mao, 1988; Lorenzana, Silvera, and Goettel, 1989), which has been shown to extend up to 
pressures of $\sim 300$~GPa and temperatures of $\sim 300$~K (Zha, Liu, and Hemley, 2012). 
Experiments have been able also to provide constraints on the molecular orientation in phases II and III 
although, for the reasons specified above, not full structural characterisations (Goncharov \emph{et al.}, 
1998). 

There have been many attempts to identify the structure of phases II and III using theoretical methods. Due 
to the technical difficulties encountered in the treatment of weak dispersive intermolecular interactions and 
of the indispensable consideration of QNE, however, there is not yet general agreement on this matter. For 
phase II, there is a number of candidate structures including the orthorhombic $Cmc2_{1}$ (Kitamura 
\emph{et al.}, 2000), monoclinic $P2_{1}/c$ (Zhang \emph{et al.}, 2007), and orthorhombic $Pca2_{1}$ (Kohanoff 
\emph{et al.}, 1997; St\"{a}dele and Martin, 2000). From all these structures, $Pca2_{1}$ emerges as one of 
the most likely molecular models (Moraldi, 2009; McMahon \emph{et al.}, 2012). Raman experiments, however, 
indicate that phase II possesses one vibrational mode whereas the $Pca2_{1}$ phase has four (Cui, Chen, and Silvera, 
1995). More recently, a new monoclinic $P2_{1}/c$ phase containing $24$ atoms in the primitive cell has been 
proposed also as a likely candidate for phase II (Pickard and Needs, 2009). This structure has been obtained 
through the \emph{ab initio} random structure searching method (Pickard and Needs, 2006) and its vibrational 
phonon features appear to be consistent with Raman experiments (Drummond \emph{et al.}, 2015). 

With regard to phase III, it was initially proposed that a hcp lattice with molecules tilted roughly $60^{\circ}$ 
with respect to the $c$ axis could be a strong candidate (Natoli, Martin, and Ceperley, 1995). This suggestion is 
consistent with the reported spectroscopy data, in which intense IR activity is appreciated (Cui, Chen, and Silvera, 
1995), and with a recent x-ray diffraction study by Akahama \emph{et al.} (2010). Subsequently, Pickard and Needs 
(2007) proposed, again relying on the outcomes of DFT-based random structure searches, a different candidate  
structure consisting of $12$ molecules per unit cell with the centers close to those in a distorted hcp lattice. 
The symmetry of this phase is $C2/c$ (monoclinic) and its vibrational phonon features are also consistent with 
the available experimental data. More recently, a hexagonal structure with $P6_{1}22$ symmetry has been introduced
as another possible candidate for phase III (Monserrat \emph{et al.}, 2016).   

In 1995, Goncharov \emph{et al.} found in deuterium a small discontinuity in the vibron mode (that is, the intramolecular 
stretching mode) and a change in the slope of the corresponding I-III phase boundary at pressures around $150$~GPa and 
temperatures above $175$~K. This small vibron discontinuity practically disappeared at $T \ge 250$~K. These observations 
suggest the possible existence of a new phase denoted by I' (see Fig.~\ref{fig:h2solpd}), that is isostructural to  
phase III, and of a critical I-I'-III point. PIMC calculations by Surh \emph{et al.} (1997) on a system of quantum 
rotors interacting through an effective LDA model, provide some support to this hypothesis. In subsequent spectroscopy 
experiments, Baer, Evans, and Yoo (2007, 2009) have found vibron signatures that are also consistent with the existence 
of phase I'. However, for these latter observations to be consistent with those by Goncharov \emph{et al.} (1995), the 
slope of the I-I' phase boundary needs to be negative, a feature that was not reported in the earliest work. 
In a recent study Goncharov, Hemley and Mao (2011) have performed a refined vibrational spectroscopy analysis and 
concluded that the new data do not support the existence of phase I'. Further systematic investigations appear  
to be necessary to clarify these issues.   

Recently, room-temperature static diamond-anvil-cell (DAC) experiments have been performed in which pressures of
up to $300$~GPa have been reached (Eremets and Troyan, 2011; Howie \emph{et al.}, 2012a; Howie \emph{et al.}, 
2012b; Loubeyre, Occelli, and Dumas, 2013). Eremets and Troyan (2011) have reported that solid hydrogen becomes 
metallic at a pressure of $265$~GPa. Subsequent experimental studies (Howie \emph{et al.}, 2012a; Howie \emph{et al.},
2012b; Loubeyre, Occelli, and Dumas, 2013), however, do not appear to support the validity of this result. The 
pressure threshold for the insulator to metal transition in hydrogen still is believed to lie between $325$ (Goncharov 
\emph{et al.}, 2001) and $450$~GPa (Loubeyre, Occelli, and LeToullec, 2002). These recent room-temperature DAC studies, 
on the other hand, agree all in that hydrogen transforms to a new phase, denoted by IV (see Fig.~\ref{fig:h2solpd}), 
at a pressure near $220$~GPa. During the III-IV transformation, a large vibron Raman frequency discontinuity
and the appearance of two IR and two Raman vibron modes are observed. The existence of phase IV, therefore,  
now is regarded as well established. 

Again, several candidate structures have been proposed for phase IV. Howie \emph{et al.} (2012a, 2012b) have 
tentatively indexed it as $Pbcn$, based on the results of the DFT-based random structure searches carried out by 
Pickard and Needs (2007) and their experimental spectroscopy analysis. This new orthorhombic structure presents 
a quite peculiar molecular arrangement in which consecutive graphene-like layers alternate between ordered 
and disordered structures. Chiefly, proton tunneling occurs within the graphene-like disordered layers and the 
corresponding frequency increases under pressure (Howie \emph{et al.}, 2012a). Pickard, Martinez-Canales, and Needs 
(2012a, 2012b), however, have shown using DFT-based methods that the $Pbcn$ phase is vibrationally unstable at zero 
temperature. The same authors have proposed a monoclinic $Pc$ structure to represent phase IV. This monoclinic phase 
is dynamically stable and contains $96$ atoms in its unit cell; it consists of alternating layers of weakly coupled 
molecules with short intra-molecular bonds, and strongly coupled molecules forming graphene-like sheets with long
intra-molecular bonds. Recent synchrotron infrared measurements by Loubeyre, Occelli, and Dumas (2013) appear to 
support the validity of this structural layered model. By relying also on the results of first-principles simulations, 
Liu \emph{et al.} (2012) have proposed a monoclinic $Cc$ structure as a new possible candidate for phase IV; this 
phase is vibrationally stable and structurally very similar to the $Pc$ structure proposed by Pickard, Martinez-Canales, 
and Needs (2012a, 2012b), although is thermodynamically less stable and has no orientational order. Further systematic 
investigations appear to be necessary to determine with precision the molecular structure of phase IV.              

Several other phases have been predicted to exist in solid hydrogen at low temperatures and pressures beyond 
$250$~GPa. Most of those phases have been predicted based on the results of first-principles crystal structure 
searches that incorporate QNE through the quasi-harmonic approximation (see Sec.~\ref{subsub:approxzeroT}). 
Pickard, Martinez-Canales, and Needs (2012a, 2012b) have proposed that their candidate structure for phase III, 
that is, monoclinic $C2/c$, transforms into an orthorhombic $Cmca-12$ phase containing $12$ atoms per unit cell 
at $P = 285$~GPa, and that this subsequently transforms into another $Cmca$ phase with a smaller number of atoms 
at $P = 385$~GPa. Liu, Wang, and Ma (2012), have also predicted that at pressures higher than $\sim 500$~GPa 
hydrogen transforms into a new monoclinic $C2/c$ phase that possesses two types of intramolecular bonds with 
different lengths. In fact, new crystal phases (e. g., IV' and V) have been observed in DAC experiments performed 
at pressures beyond $\sim 300$~GPa (Howie \emph{et al.}, 2012a; Dalladay-Simpson, Howie, and Gregoryanz, 2016); 
however, their precise molecular arrangements still remain unknown.    

The possibility of stabilising an atomic, rather than a molecular, phase in solid hydrogen by means of pressure 
has been also explored by several authors with theory. This possibility is very interesting from a fundamental 
point of view as it could render a metallic system (Wigner and Huntington, 1935). Considering the orthorhombic 
$Cmca$ phase originally proposed by Pickard and Needs (2007) and relying on \emph{ab initio} random structure 
searches, McMahon and Ceperley (2011) have proposed that molecular hydrogen dissociates into a monoatomic body-centered 
tetragonal structure near $500$~GPa. Labet \emph{et al.} (2012) and Labet, Hoffmann, and Ashcroft (2012a, 2012b, 
2012c) have also analysed in the detail the process of molecular dissocation by focusing on the structures predicted 
by Pickard and Needs (2007); they have found a discontinuous shift in the distances between protons when transitioning 
from the orthorhombic $Cmca$ to the atomic phase, and have proposed an intermediate phase that would allow for a 
continuous dissociation. More recently, Azadi \emph{et al.} (2014) have concluded, based on electronic QMC methods 
(see Sec.~\ref{subsubsec:qmc}) and considering nuclear anharmonic contributions to the enthalpy through DFT, that 
a transition from the orthorhombic $Cmca-12$ to an atomic $I4_{1}/amd$ phase could occur at $P = 374$~GPa. Interestingly, 
Dalladay-Simpson, Howie, and Gregoryanz (2016) have just reported experimental evidence for a new phase in hydrogen, 
denoted by V, which at room temperature is stabilised at a pressure of $325$~GPa. The experimental evidence consist 
of a substantial weakening of the vibrational Raman activity, a change in the pressure dependence of the vibron, 
and a partial loss of the low-frequency excitations. Whether this new phase could be identified with the debated 
$P$-induced atomisation of solid hydrogen, is still not well established. 

As it has been explained in this section, many complex and controversial aspects still need to be solved in solid 
hydrogen under extreme thermodynamic conditions. On the theoretical side, most of the predictions on phases II, III, 
and IV rely on techniques that incorporate QNE only approximately (e. g., quasi-harmonic approaches) and on standard 
DFT methods. Using such approaches to reproduce the thermodynamic stability of highly compressed hydrogen, however, 
seems to be inadequate. For instance, Chen \emph{et al.} (2014) have recently shown in a thorough PIMC benchmark study 
on H$_{2}$ that those cases in which good agreement between standard DFT calculations and experiments is obtained, large 
error cancellations are likely to be affecting the simulations. Similar conclusions have been attained also by Geneste 
\emph{et al.} (2012), Morales \emph{et al.} (2013), and Drummond \emph{et al.} (2015) by using non-standard computational 
approaches (e. g., non-harmonic simulation methods in combination with electronic QMC). In order to provide more conclusive 
estimations in solid H$_{2}$, therefore, is necessary to employ quantum simulation methods that simultaneously describe 
QNE (e. g., PIMD, PIMC and PIGS, see Sec.~\ref{sec:simulation}) and long-range intermolecular forces (e. g., non-standard 
DFT functionals and eQMC, see Sec.~\ref{sec:modeling}) accurately.

\subsection{Solid water}
\label{subsec:h2o}

\begin{figure}
\centerline
        {\includegraphics[width=1.0\linewidth]{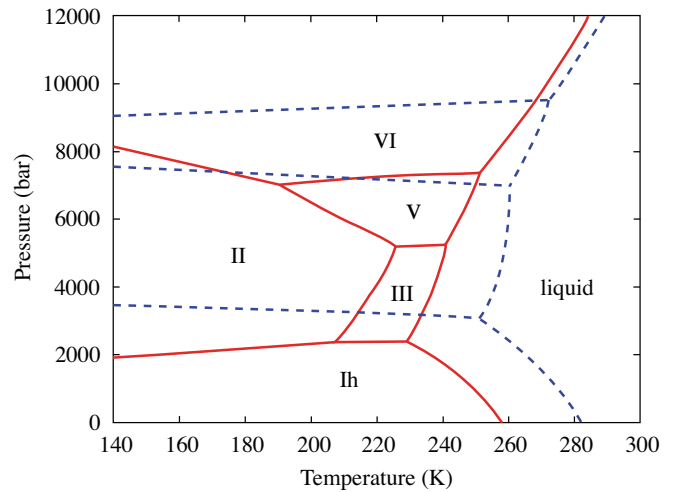}}
\caption{(Color online) The classical phase diagram of water based on the TIP4PQ/2005 force field (dashed 
         blue lines) as compared to the quantum phase diagram obtained with path-integral simulations 
         (solid red lines). Roman numerals label different crystal structures with hexagonal symmetry (I$_{h}$),
         rhombohedral (II), tetragonal (III and IV), and monoclinic (V). Adapted from McBride \emph{et al.}, 
         2012.}
\label{fig:icepd}
\end{figure}

QNE are unquestionably important for understanding the physical properties of ice. Due to the small moment of 
inertia of the H$_{2}$O molecule and relatively low strength of the intermolecular hydrogen bonds, QNE persist 
in ice up to temperatures of $\sim 100$~K (Gai, Schenter, and Garrett, 1996a; Ceriotti, Bussi, and Parrinello, 
2009; Vega \emph{et al.}, 2010; Moreira and de Koning, 2015). Numerous examples of these effects can be found 
in the literature. For example, incoherent single-particle tunneling has been disclosed in cubic ice at Mbar 
pressures, explaining so the origins of the measured H/D isotopic effects on the antiferroelectric ice 
VIII~$\to$~paraelectric VII phase transformation (Benoit, Marx, and Parrinello, 1998; Benoit, Romero, and Marx, 
2002). A recent neutron scattering study has also revealed an anomalous $T$-dependent dynamic effect in normal 
(hexagonal) ice I$_{h}$ (Bove \emph{et al.}, 2009), that has been explained in terms of collective tunneling of 
protons (up to six) within locally ordered rings (Drechsel-Grau and Marx, 2014). These findings suggest that 
quantum many-body tunneling could be important also in a variety of related H-bonded systems, including other 
phases of ice and cyclic water clusters on metal surfaces (Drechsel-Grau and Marx, 2014). 

In analogy to solid helium and hydrogen, an interest has developed in understanding the features of the momentum 
distribution in ice. Both inelastic neutron scattering experiments (Reiter \emph{et al.}, 2004; Flammini \emph{et 
al.}, 2012) and advanced path-integral calculations (Morrone and Car, 2008; Lin \emph{et al.}, 2010; Lin \emph{et al.}, 
2011) agree in describing the corresponding $n({\bf k})$ with an anisotropic Gaussian. This result implies that 
the protons experience an anisotropic quasi-harmonic effective potential with distinct principal frequencies that 
reflect the possible molecular orientation. According to both neutron scattering experiments and path-integral 
calculations (Flammini \emph{et al.}, 2012) the excess kinetic energy in ice I$_{h}$ at low temperatures 
amounts to $\sim 150$~meV, which evidences a marked quantum character (see Sec.~\ref{subsec:quantumvsclassical}).  

The presence of quantum isotopic effects is also notable in solid water. The effects of hydrogen isotope 
substitution on the structural, kinetic energy and atomic delocalization properties of ice, have been investigated 
in detail with experiments and path-integral calculations. For example, quantum simulations of D$_{2}$O in the 
I$_{h}$ phase at $T = 100$~K have found a decrease in the crystal volume and intramolecular O-D distance of $0.6$~\% 
and $0.4$~\%, respectively, as compared to H$_{2}$O (Herrero and Ram\'irez, 2011a). An increase of $\sim 6$~\% 
in the melting temperature of D$_{2}$O at ambient pressure has been also predicted with path-integral
simulations (Ram\'irez and Herrero, 2010). Similarly, the presence of quantum isotopic effects in highly 
compressed amorphous ice have been reported by several groups (Gai, Schenter, and Garret, 1996b; Herrero and
Ram\'irez, 2012). Interestingly, an anomalous thermal expansion isotopic effect has been observed in ice; 
the volume of solid D$_{2}$O is larger than that of solid H$_{2}$O (R\"{o}ttger \emph{et al.}, 1994), in contrast 
to what occurs in other crystals upon substitution with heavier species. This quantum nuclear effect has been 
rationalised recently by Pamuk \emph{et al.} (2012) with \emph{ab initio} calculations based on the quasi-harmonic 
approximation. 

The importance of QNE on the phase diagram of ice has been determined quantitatively with path-integral 
Monte Carlo simulations based on the TIP4PQ/2005 force field by McBride \emph{et al.} (2012) [see Fig.~\ref{fig:icepd}].
It is worth noting that although the intermolecular potential model employed by McBride \emph{et al.} (2012) 
is non-flexible and non-polarisable, the agreement obtained with the experiments is fairly good. In particular, 
quantum simulations provide phase boundaries that are shifted $\sim 20$~K to lower temperatures as compared
to the observations (see Fig.~$3$ in McBride \emph{et al.}, 2012). As it is shown in Fig.~\ref{fig:icepd}, QNE play 
a significant role on the thermodynamic stability of the different phases of ice: the melting lines are shifted to higher 
temperatures and the solid-solid transitions to higher pressures. Another important difference is that the region of 
thermodynamic stability of phase II is significantly reduced in the classical phase diagram, as this phase appears there 
only at temperatures below $80$~K (that is, in the classical phase diagram shown in Fig.~\ref{fig:icepd} phase II is missing). 
The origins of these quantum effects have been rationalised in terms of the tetrahedral angular order ascribed to 
each polymorph and the volume change involved in the phase transformations. The pressure dependence of the crystal 
volume, bulk modulus, interatomic distances, atomic delocalisation, and kinetic energy in hexagonal ice (I$_{h}$) 
under pressure, have been also analysed thoroughly with similar computational techniques by Herrero and Ram\'irez 
(2011b). 

Despite of the mounting experimental and theoretical evidence showing the importance of QNE, these effects are 
normally disregarded in most computational studies of water and ice at $T \neq 0$ conditions. This is due in part 
to the difficulties encountered in the description of molecular interactions in H$_{2}$O. Different types of 
computationally inexpensive empirical potentials, which either assume rigid or flexible molecules and polarisable 
or non-polarisable ions (e. g., the so-called SCP, TIP4P, q-TIP4P/F and TIP4PQ/2005 force fields), have been 
employed in most simulation studies of ice at finite temperatures. Some of those force fields have been fitted to 
reproduce experimental data, to be used subsequently in classical simulation studies, hence they already incorporate 
QNE effectively. Quantum calculations based on those interaction models, therefore, may provide in some cases worse 
agreement with the experiments than classical simulation studies due to double counting of quantum nuclear effects 
(Herrero and Ram\'irez, 2014). In other words, the inaccurracies affecting common empirical interaction models may 
disguise to some extent the real influence of QNE by providing reasonably good agreement with the experiments. In 
some cases it has been actually demonstrated that the combined description of molecular interactions and ionic effects 
at the quantum level is necessary for reproducing correctly the experimental findings in ice. Examples include the 
anomalous volume expansion observed in ice isotopes (Pamuk \emph{et al.}, 2012) and the interpretation of measured 
x-ray absorption spectra (Kong, Wu, and Car, 2012; Kang \emph{et al.}, 2013).  
 
\emph{Ab initio} treatment of the molecular interactions in ice has been mostly done with DFT methods. 
However, the presence of hydrogen bonds and dispersive long-range forces makes the description of this 
crystal difficult, demanding the use of computational methods going beyond standard DFT (see Sec.~\ref{subsubsec:dft}). 
For a detailed description of the strengths and weaknesses of different exchange-correlation DFT  
approximations in describing H$_{2}$O-based systems, we refer the interested reader to Morales \emph{et al.} 
(2014) and the recent review by Gillan, Alf\`e, and Michaelides (2016). We note in passing that application 
of electronic quantum Monte Carlo methods to the study of bulk ice is very rare (Santra \emph{et al.}, 2011), 
mainly due to computational affordability issues and the likely bias associated to the use of pseudopotentials 
(Driver and Militzer, 2012).       
 
An advantageous aspect in the simulation of ice, as compared to that of solid helium or hydrogen, 
is that QNE in principle can be described correctly with the quasi-harmonic approximation (QHA, see 
Sec.~\ref{subsub:approxzeroT}). This conclusion has been attained by several authors based on  
comparisons provided between \emph{ab initio} QHA results and neutron scattering experiments (see, 
for instance, Senesi \emph{et al.}, 2013) and, more convincingly, between QHA and path-integral 
simulations performed with a same effective interaction model (Pamuk \emph{et al.}, 2012; Ram\'irez 
\emph{et al.}, 2012). Consequently, the study of the low-temperature phase diagram and thermodynamic 
properties of H$_{2}$O in principle is affordable with first-principles methods. 
Nevertheless, a note of caution must be added here. Very recently, Engel, Monserrat, and Needs (2015) 
have shown in a DFT-based study that anharmonic contributions to the free energy can turn out 
to be decisive for describing correctly the thermodynamic stability of different ice polymorph with 
very close quasi-harmonic free energies. In particular, it has been shown that anharmonic quantum 
nuclear effects are decisive in stabilizing the hexagonal I$_{h}$ phase with respect to cubic I$_{c}$, 
the latter being a rare form of ice that presents a different stacking of layers of tetrahedrally 
coordinated water molecules. As noted by the authors of that study, treatment of anharmonicity in 
general could be crucial for correctly describing the energy differences between similar polymorph 
in hydrogen-bonded molecular crystals (which can be relevant, for instance, to the pharmaceutical 
sciences).

\subsection{Nitrogen and Methane}
\label{subsec:n2-ch4}
QNE are more pronounced in molecular nitrogen (N$_{2}$) and methane (CH$_{4}$) than in H$_{2}$O 
(see Fig.~\ref{fig:deboer}). This is due to the fact that the intermolecular interactions in the two 
former systems are dominated by long-range dipole-dipole (CH$_{4}$), dipole-quadrupole (CH$_{4}$), and 
quadrupole-quadrupole (N$_{2}$ and CH$_{4}$) forces, which are weaker than hydrogen bonds (Cazorla, 2015). 
Certainly, under normal thermodynamic conditions H$_{2}$O is a liquid whereas N$_{2}$ and CH$_{4}$ are gases. 
However, the study of QNE in solid nitrogen and methane is very marginal in comparison to that in ice (or 
hydrogen). Improving our quantitative understanding of solid N$_{2}$ and CH$_{4}$ is actually important for 
planetary and energy materials sciences. For example, these species are believed to abound in the surface and 
interior of Uranus, Neptune, and Pluto (Hubbard \emph{et al.}, 1991; Protopapa \emph{et al.}, 2008). Meanwhile, 
under high pressures ($\ge 110$~GPa) molecular nitrogen dissociates into singly bonded polymeric nitrogen, the 
so-called cubic gauge phase, that is being considered as a potential high-energy-density material because 
can exist in a metastable form at ambient pressure (Eremets \emph{et al.}, 2004).  

The $P-T$ phase diagrams of compressed nitrogen and methane are very complex, as it occurs in most molecular 
systems, due to the prominence of the orientational degrees of freedom. N$_{2}$ exhibits five solid molecular 
phases at pressures below $\sim 10$~GPa and temperatures $T \le 300$~K (Gregoryanz \emph{et al.}, 2007; Tomasino 
\emph{et al.}, 2014). The low-temperature phases in molecular nitrogen are governed by quadrupole-quadrupole 
interactions and in moving from zero to higher pressures the crystal first transforms from an orientationally 
disordered cubic ($\alpha$) to an ordered tetragonal ($\gamma$) phase, and then to an ordered rhombohedral 
phase ($\epsilon$); when increasing $T$, a disordered hexagonal phase ($\beta$) first appears at $2.4$~GPa 
that subsequently transforms into a cubic phase ($\delta$) with orientational disorder by effect of pressure. 
It is worth noting that large isotopic effects have been observed in the $P$-induced $\alpha \to \gamma$  
transformation occurring at low temperatures (Scott, 1976), which indicates the presence of important QNE. 
Some other phases have been observed to stabilise at higher pressures in the experiments, the structures 
of which are unknown in most cases. This lack of knowledge has motivated an intense theoretical activity 
in solid nitrogen. Over a dozen of different structures have been predicted to be stable in the pressure 
range $0 \le P \le 400$~GPa; among these we highlight the layered $Pba2$ or $Iba2$ ($188-320$~GPa) and 
helical tunnel $P2_{1}2_{1}2_{1}$ structures ($>320$~GPa)[Ma \emph{et al.}, 2009], and the cluster form 
of nitrogen diamondoid ($>350$~GPa) [Wang \emph{et al.}, 2012], which have been obtained through systematic 
crystal structure searches based on DFT methods.     

Unquestionably, the results of DFT-based studies on molecular nitrogen are invaluable for advancing in the 
knowledge of its phase diagram; however, we must note that most of the first-principles investigations presented 
thus far systematically neglect two basic aspects in N$_{2}$ crystals: long-range dispersion interactions and 
QNE (i. e., they have been performed with standard LDA and GGA DFT exchange-correlation functionals and 
disregarding likely zero-point motion effects even through the quasi-harmonic approximation). It could be argued 
that the importance of these two elements become secondary at high pressures or that somehow they cancel out when 
comparing the enthalpy of different phases. However, by taking into consideration all the similarities between N$_{2}$ 
and H$_{2}$ in terms of intermolecular interactions and degree of quantumness, one can suspect that this is not 
the case (i. e., as it has been explicitly shown in solid hydrogen, see Sec.~\ref{subsec:h2extreme}). Therefore, it is 
reasonable to think that the transition pressures and phase boundaries reported in standard DFT studies of N$_{2}$
are likely to be inaccurate. With regard to this last point, it was first predicted from standard DFT calculations that 
molecular nitrogen transforms into a polymeric phase (i. e., cubic gauche $cg$-N) prior to metallization at a 
pressure of $\sim 50$~GPa (Mailhiot, Yang, and McMahan, 1992); this transformation has been observed subsequently 
in experiments, however, at thermodynamic conditions much higher than the predicted ones, namely, $P \ge 110$~GPa 
and $T \ge 2000$~K (Eremets \emph{et al.}, 2004). Whether the causes of the discrepancies between theory and 
experiments lie on the use of inaccurate DFT functionals and neglecting of QNE, or the use of incorrect molecular 
structures in the calculations, or the existence of large kinetic barriers for the dissociation of N$_{2}$ molecules 
that complicate the measurements, or a combination of all these factors, is not clear yet. Systematic 
computational studies analysing the importance of QNE and benchmarking the description of intermolecular 
interactions in highly compressed nitrogen, are very desirable for clarifying these issues.   

To the best of our knowledge, there is only one computational study by Presber \emph{et al.} (1998) in which 
the importance of QNE on the orientational phase transitions in bulk solid N$_{2}$ at low $P$ and low $T$
has been assessed. By using the PIMC method and a classical N$_{2}$-N$_{2}$ interaction potential, Presber 
\emph{et al.} (1998) found that the transition temperature corresponding to the $\alpha \to \gamma$ transformation 
is reduced by about $11$~\% with respect to the result obtained with classical methods. We note that, in spite 
of the simplicity of the employed interaction model, Presber \emph{et al.}'s quantum predictions show reasonably 
good agreement with the experiments. 

Similarly, the impact of quantum nuclear effects on the orientational ordering of N$_{2}$ molecules adsorbed 
on graphite has been investigated with PIMC methods by Marx and co-workers in a series of works (Marx \emph{et al.}, 
1993; Marx, Sengupta, and Nielaba, 1993; Marx and M\"{u}ser, 1999). To this end, rigid rotors with their centers of 
mass pinned on a triangular lattice commensurate with the graphite basal plane, and molecule-molecule and 
molecule-surface interactions treated with atomistic models and point charges, were analysed. The main conclusions 
from those studies can be summarised as that quantum fluctuations lead to ``10~\% effects'' on the physical properties 
of N$_{2}$ films (Marx \emph{et al.}, 1993; Marx, Sengupta, and Nielaba, 1993). For example, the temperature 
corresponding to the so-called ``2-in'' herringbone orientational transition that occurs at low temperatures and low 
densities is shifted down by about 10~\% as a result of zero-point motion, in good agreement with the experiments. 
These results imply that in order to make quantitatively correct predictions in N$_{2}$ crystals QNE must be taken 
into account.     

Regarding methane, i. e., CH$_{4}$, the ground-state phase at low pressures is a cubic structure that can be thought 
of two molecular sublattices, one of which is orientationally ordered and the other disordered (James and Keenan, 1959). 
A first-order phase transition between this cubic and a partially ordered phase occurs at a temperature of $20.4$~K 
(Press and Kollmar, 1975); in CD$_{4}$, a similar transition occurs but at a higher temperature of $27.4$~K (Press, 1972). 
This large isotopic effect, again, marks the presence of significant QNE. In fact, by using the PIMC method and a model 
potential based on \emph{ab initio} results, M\"{u}ser and Berne (1996) were able to replicate such a large isotopic shift 
in the transition temperature, otherwise not reproducible with classical methods.   

The phase diagram of methane at high pressures, on the other hand, remains contentious. Up to nine different phases
have been observed in CH$_{4}$ at pressures below $\sim 10$~GPa and temperatures $0 \le T \le 300$~K (Bini and Pratesi, 
1997; Maynard-Casely \emph{et al.}, 2010), and only three of them have been determined. For instance, based on neutron 
scattering measurements Maynard-Casely \emph{et al.} (2010) have proposed that the so-called phase A, which appears at 
pressures about $1$~GPa and temperatures above $\sim 100$~K, consists of $21$ molecules in a rhombohedral unit cell that 
is strongly distorted with respect to the cubic ground-state. Using systematic crystal structure searches based on a 
genetic algorithm and dispersion-corrected DFT methods, Zhu \emph{et al.} (2012) have predicted a similar candidate 
structure for phase A that, in contrast to the experimentally determined one, presents orientationally disordered 
molecules. At pressures beyond $\sim 100$~GPa, CH$_{4}$ is expected to become chemically unstable and to decompose 
(Gao \emph{et al.}, 2010). Unfortunately, possibly due in part to the lack of knowledge on the molecular phases that 
appear below that pressure limit, the impact of QNE on the high-$P$ and low-$T$ phase diagram of solid methane remains 
largely unexplored.  

Recent simulation studies by Goldman, Reed, and Fried (2009) and Qi and Reed (2012), have shown that quantum 
nuclear effects in fact are crucial for understanding the behavior of solid CH$_{4}$ at high-$P$ and high-$T$ 
conditions. By adopting a quantum thermal bath scheme to treat QNE (see Sec.~\ref{subsub:qtherbath}) and a multi-scale 
simulation approach to model the molecular interactions, Qi and Reed (2012) have quantified the impact of QNE on the 
Hugoniot of compressed methane. It has been found that quantum nuclear effects are responsible for a huge shift of 
$\sim 40$~\% to lower pressures in the onset of decomposition. The primary factor behind such a tremendous effect 
has been ascribed to the large variation in the heat capacity that occurs when QNE are considered. In a previous 
work, some of those authors had already shown that quantum temperature corrections to classical DFT calculations 
on the Hugoniot of methane were as large as $20-30$~\%, and that these improved the agreement with the experiments 
(Goldman, Reed, and Fried, 2009). 

Analogously to the situation explained for solid N$_{2}$, there is a pressing need for unravelling the influence 
of QNE on the thermodynamic and structural properties of methane at low and high pressures. From this knowledge, 
our description and understanding of quantum molecular crystals could be improved significantly.

%===============================================================
\section{Quantum materials science}
\label{sec:matscience}
%===============================================================

\begin{figure}
\centerline
        {\includegraphics[width=1.0\linewidth]{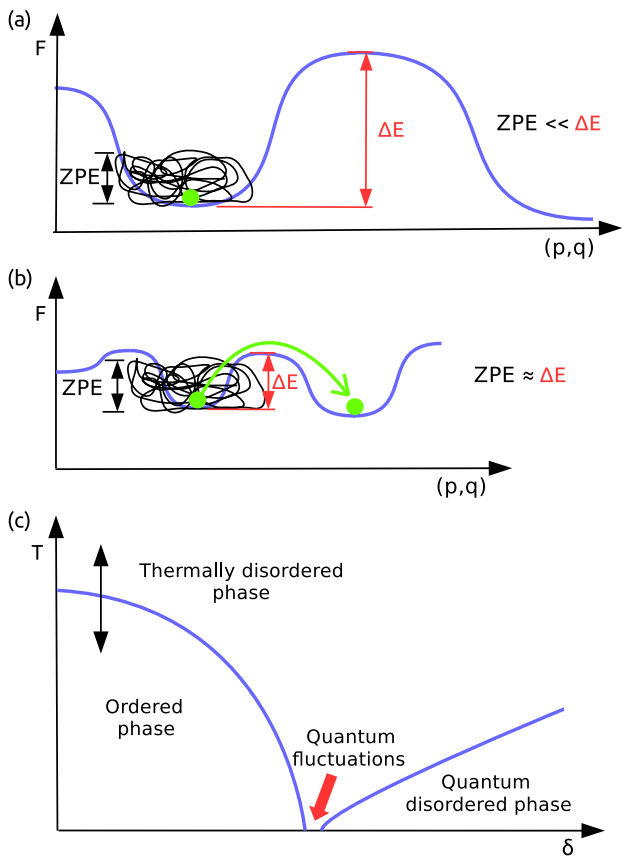}}
\vspace{-0.75cm}
\caption{(Color online) The influence of QNE in normal systems and materials with a shallow 
         multi-well potential energy surface (PES). (a)~The free energy barrier separating two 
         local PES minima is too large as compared to the atomic zero-point motion (ZPE), hence 
         the system remains indifferent. (b)~The free energy barrier separating two different 
         phases is similar in magnitude to the ZPE, hence the system transits from one to 
         the other. (c)~Sketch of the phase diagram in a quantum paraelectric; $\delta$ represents 
         a particular tuning parameter (e. g., pressure).}
\label{fig:qmatsc}
\end{figure}

Here we explain the physical properties of crystals that are technologically relevant and in
which at the same time the influence of QNE is significant. In this category we include from 
light-weight and metallic crystals (e. g., Li) to heavy-weight and insulator compounds (e. g., 
BaTiO$_{3}$). The former systems respond to the traditional definition of a quantum crystal, 
that is, solids composed of low-$Z$ atoms that interact through relatively weak forces. The 
latter systems are better described as highly anharmonic crystals with a shallow multi-well 
potential energy surface (PES). In this case, disparate phases, identified with local minima 
in the PES, are energetically very competitive and thus QNE play a crucial role on stabilising 
one or another (see Fig.~\ref{fig:qmatsc}a-b).

\subsection{Perovskite oxides}
\label{subsec:oxides}

\begin{figure}
\centerline
        {\includegraphics[width=1.0\linewidth]{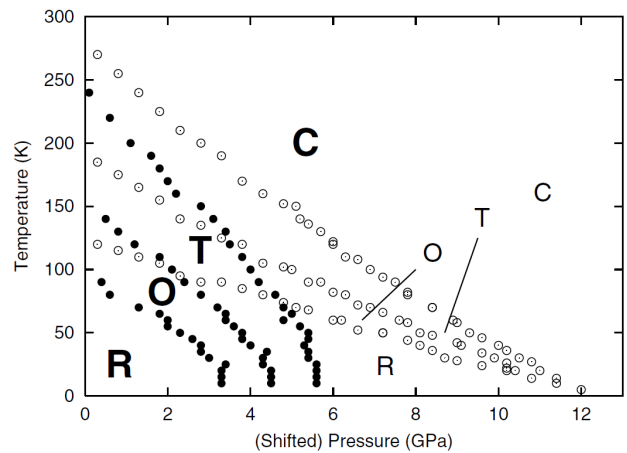}}
\caption{(Color online) Phase diagram of BaTiO$_{3}$ calculated with classical (open circles 
          and small labels) and quantum (solid circles and large labels) simulation methods. 
          From ${\rm \acute{I}}$${\rm \tilde{n}}$iguez and Vanderbilt, 2002.}
\label{fig:btopd}
\end{figure}

Perovskite oxides have the general formula ABO$_{3}$ in which A and B are cations, the latter  
being a transition metal element with a smaller radius than A. The ideal perovskite structure 
is cubic with space group $Pm\overline{3}m$, where the B cation is $6$-fold coordinated with the
oxygen anions and A $12$-fold coordinated. Perovskite oxides display many interesting physical 
properties like, for instance, ferroelectricity (e. g., BaTiO$_{3}$), ferromagnetism (e. g., 
SrRuO$_{3}$), multiple coupled ferroic orders (e. g., BiFeO$_{3}$), and insulator-to-metallic 
transitions (e. g., LaCoO$_{3}$). All these properties are very sensitive to the chemistry, crystalline 
defects, electrical boundary conditions, and applied stress, so that they can be tuned externally. 
For this reason, pervoskite oxides normally are referred to as functional materials in the literature 
(Ohtomo and Hwang, 2004; Schlom \emph{et al.}, 2007; Cazorla and Stengel, 2012).    

BaTiO$_{3}$ is an archetypal ferroelectric. At room temperature this material adopts a rhombohedral (R)
phase that displays a spontaneous and switchable ferroelectric polarisation. As temperature is raised
from zero to $\sim 300$~K, BaTiO$_{3}$ goes through the series of phase transformations R~$\to$~O~$\to$~T~$\to$~C,  
where ``O'' stands for an orthorhombic phase, ``T'' for a tetragonal, and ``C'' for a cubic. The same
sequence of phases is observed under pressure (see Fig.~\ref{fig:btopd}).  The high-$T$ 
cubic phase corresponds to the ideal perovskite structure, which is non-polar (that is, has null ferroelectric 
polarisation). The appearance of ferroelectricity in perovskite oxides is originated by a delicate balance 
between long-range Coulomb interactions, that favor the ferroelectric state, and short-range repulsive forces, 
that favor the cubic non-polar state; the hybridisation between B cation $d$ electronic orbitals and oxygen 
$2p$ plays an essential role on that equilibrium (Cohen, 1992). The ferroelectric phase transition in BaTiO$_{3}$ 
is considered to be an example of displacive transition, in which a zone-center vibrational mode, called 
``soft'', has a vanishing frequency at the phase transition and its eigenvector is similar to the atomic 
displacements observed in the ferroelectric state. Therefore, BaTiO$_{3}$ is a highly anharmonic crystal 
in which several phases are energetically very competitive.  

In 2002, ${\rm \acute{I}}$${\rm \tilde{n}}$iguez and Vanderbilt estimated the impact of QNE on the $P-T$ phase
diagram and ferroelectric properties of BaTiO$_{3}$. Using the PIMC approach based on an effective Hamiltonian 
model fitted to DFT results, the authors of that study found that the $P-T$ boundaries separating the regions
of stability of different phases varied considerably when considering QNE. As it is shown in Fig~\ref{fig:btopd}, 
the phase boundaries are noticeably shifted towards lower pressures and temperatures as compared to those found 
in the classical phase diagram. As a result, the agreement between theory and experiments was improved significantly. 
${\rm \acute{I}}$${\rm \tilde{n}}$iguez and Vanderbilt (2002) also found that the electrical polarisation in BaTiO$_{3}$ 
shrinks by about $10$~\% when considering quantum fluctuations, and that the same quantity exhibits null variation 
in the $T \to 0$ limit [as it is expected from quantum arguments (Hayward and Salje, 1998)]. 

More recently, Geneste \emph{et al.} (2013) have analysed the influence of QNE on the dielectric permittivity 
and piezoelectric constants of rhombohedral BaTiO$_{3}$. Using a path-integral molecular dynamic approach 
based on the same effective Hamiltonian model than employed by ${\rm \acute{I}}$${\rm \tilde{n}}$iguez and Vanderbilt 
(2002), Geneste \emph{et al.} (2013) have found that inclusion of quantum nuclear effects systematically enhances 
the dielectric response and piezoelectric constants of the crystal by approximately a factor of $2$. This huge effect 
has been explained in terms of the strong anharmonicity of BaTiO$_{3}$, which is retained by the crystal down 
to zero temperature. Geneste \emph{et al.}'s findings suggest that quasi-harmonic approaches are not adequate for 
describing the behavior of displacive ferroelectrics at low temperatures.

QNE can influence strongly the low-$T$ response of a system when is near to a structural phase transition. In 
most ferroelectrics, the stability limit of the polar phase, $T_{C}$, falls within a region governed by classical
Boltzmann fluctuations. In few crystals known as ``quantum paraelectrics'', however, $T_{C}$ is very close to 
the zero-temperature limit and thus quantum statistical fluctuations play a dominant role in the transformation 
(M\"{u}ller and Burkard, 1979; Rytz, H\"{o}chli, and Bilz, 1980; Conduit and Simons, 2010). Examples of 
quantum paraelectrics include the perovskite oxides SrTiO$_{3}$, KTaO$_{3}$, and KTaO$_{3}$-NaTaO$_{3}$ and 
KTaO$_{3}$-KNbO$_{3}$ solid solutions (H\"{o}chli and Boatner, 1979; Samara, 1988). At low temperatures, the 
dielectric properties of a quantum paraelectric are appreciably different from those of a classical material.
For instance, the Curie-Weiss law describing the variation of the static dielectric constant, $\epsilon$, near 
$T_{C}$, namely, $ \propto (T - T_{C})^{-1}$, is not longer fulfilled; instead $\epsilon$ follows a $\propto 
T^{-2}$ relation (H\"{o}chli and Boatner, 1979). 

As the temperature is raised, or as a specific tuning parameter that induces atomic displacements is varied 
(e. g., pressure), the dielectric behavior in quantum and classical polar materials eventually become analogous. 
Therefore, a classical-quantum crossover regime exists in quantum paraelectrics in which intriguing quantum 
phenomena can be expected to occur (see sketch in Fig~\ref{fig:qmatsc}c). This is the case, for instance, of  
ferroelectric quantum criticality, which have been recently observed in SrTiO$_{3}$ and KTaO$_{3}$ crystals (Rowley 
\emph{et al.}, 2014). In particular, the inverse of the dielectric constant in these materials, which below 
$50$~K follow the non-classical $\propto T^{2}$ dependence, experiences an anomalous upturn at very low temperatures 
that extends into the millikelvin range. This unexpected effect has been rationalised in terms of quantum criticality 
theory, after considering the influence of long-range dipolar interactions and of the coupling of the electrical 
polarisation with acoustic phonons (Rowley \emph{et al.}, 2014). The quantum critical regime associated to quantum 
paraelectrics is significantly different (e. g., in terms of the collective dynamics and tuning parameter) from 
the better known quantum regime occurring in quantum ferromagnetic materials (e. g., Ni$_{3}$Ga and ZrZn$_{2}$); 
interesting new prospects in the field of quantum phase transitions, therefore, appear to be opened.      

Quantum paraelectrics are also important from a technological point of view. Currently, there is a great interest 
in exploiting magnetoelectric (ME) effects, which are responsible for the coupling between the electrical 
and magnetic degrees of freedom in multiferroic crystals, for nanoelectronics applications. ME effects could be used 
to induce the reversal of the magnetisation in a material with an electric field, making it possible to store 
information in advanced electronic devices with minimal power consumption (i. e., creating magnetic fields generally 
involves higher energy expenses than electric fields). For ME effects to be practical, the value of the magnetic and 
electrical susceptibilities need to be large around a same transition temperature. Unfortunately, this rarely occurs 
in any material. Recently, Shvartsman \emph{et al.} (2010) have measured a large ME effect in EuTiO$_{3}$ near $T_{N} 
= 5$~K, a quantum paraelectric that undergoes an anti-ferromagnetic to paramagnetic phase transition at very low 
temperatures. The magnetoelectric moments revealed at the magnetic phase transition are comparable to those found in 
benchmark multiferroic crystals such as TbPO$_{4}$. Shvartsman \emph{et al.}'s findings suggest that quantum paraelectrics 
could be promising for nanoelectronic applications. 

Nonetheless, for the realisation of practical devices based on quantum paraelectrics the observed magnetoelectric 
activity should be brought closer to room temperature. High compression could represent a solution to this problem 
as it can extend the regime in which QNE remain important while simultaneously shifting $T_{N}$ towards higher 
temperatures. In this last regard, the outcomes of quantum simulation studies based on first-principles methods 
could be very insightful. Actually, recent phonon calculations by Evarestov \emph{et al.} (2011) performed with hybrid 
DFT methods (see Sec.~\ref{subsec:firstprinciples}) have accurately reproduced the experimental $T$-dependence of 
the heat capacity in SrTiO$_{3}$. On the other hand, theoretical approaches that allow to estimate $T$-renormalised 
phonon modes and frequencies are already well established (e. g., velocity auto-correlation and self-consistent 
harmonic methods, see: Teweldeberhan, Dubois, and Bonev, 2010; Errea, Rousseau, and Bergara, 2011; Errea, Calandra, 
and Mauri, 2014). Nevertheless, to the best of our knowledge, full \emph{ab initio} studies of quantum paraelectrics 
under pressure are absent in the literature.

\subsection{H-bond ferroelectrics}
\label{subsec:hferro}
H-bond ferroelectrics normally consist of polar stacks of sheets of hydrogen-bonded molecules. Hydrogen bonding 
can create electrical dipoles in crystals among hydrogen-donating molecules, which become partially negative, and 
hydrogen-accepting molecules, which become partially positive. Upon application of an electric field, protons 
associated with one molecule shift cooperatively towards a hydrogen-bonded neighbor, switching the molecular 
dipole and thus producing a large electrical polarisation (Horiuchi and Tokura, 2008). Examples of H-bond 
ferroelectrics include the molecular compounds KH$_{2}$PO$_{4}$ (KDP) and C$_{4}$H$_{2}$O$_{4}$ (H2SQ), which 
present some well-characterised crystal structures and can be synthesised both in standard and 
deuterated forms. A key aspect in the ferroelectric behavior of most H-bonded ferroelectrics is the motion of H
atoms in correlated double well potentials. This correlation consists of H atoms in neighboring H-bonds being strongly 
coupled due to the energetic requirement for satisfying the ``ice  rules'' (Singer \emph{et al.}, 2005); in the 
KDP and H2SQ systems this condition implies that each molecule participates in $4$ different H-bonds, two of 
which have donating character and the rest accepting. H-bond ferroelectric materials currently are attracting 
a lot of attention because polar order near room-temperature has been revealed in some organic species (Horiuchi 
\emph{et al.}, 2005; Horiuchi \emph{et al.}, 2010). This finding opens the possibility for manufacturing cheaper 
and more environment friendly nanoelectronic components and devices. 

Interestingly, the Curie transition temperature, $T_{C}$, in H-bond ferroelectrics can increase by about 
$100$~K upon deuteration. The origins of such an enormous isotopic effect however, remain contentious. 
Originally, a simple quantum model consisting of proton quantum-tunneling in a double-well potential was 
proposed to rationalise the $T_{C}$ observations (Blinc and Svetina, 1966); however, this simple model failed 
to explain the so-called Ubbelohde effect, which relates the experimentally observed elongation of H bonds 
upon deuteration to a purely geometric origin (Ubbelohde and Gallagher, 1955). Subsequently, models involving 
coupled vibrational lattice modes and proton dynamics were proposed (Dalal, Klymachyov, and Bussmann-Holder, 
1998) that led to the conviction that quantum-tunneling effects were not necessary for explaining the isotopic
influence on $T_{C}$ (McMahon \emph{et al.}, 1990). More recently, however, neutron Compton scattering 
experiments performed on KDP have found evidence for coherent proton quantum tunneling occurring at temperatures
above the ferroelectric transition ($T \sim 125$~K), whereas no such evidence is found in the analogous deuterated 
system (Reiter, Mayers, and Platzman, 2002; Reiter \emph{et al.}, 2008). On the theoretical side, some authors 
have attempted to reconcile the differing interpretations by arguing that a mechanism behind the Ubbelohde 
effect itself might be collective quantum tunneling in atomic clusters (Koval \emph{et al.}, 2002). 

Quantum \emph{ab initio} studies of H-bond ferroelectrics are very desirable to help in clarifying the 
controversy about the relevance of QNE on the observed $T_{C}$ isotope dependence. Nevertheless, due to 
the large size of the unit cells involved and complex collective dynamics of hydrogen/deuterium 
atoms, quantum simulation of H-bond ferroelectrics turns out to be very challenging; to the best of our 
knowledge, the number of published works on this topic can be counted with one hand. 

Srinivasan and Sebastiani (2011) have performed \emph{ab initio} PIMD simulations in KDP (i. e., KH$_{2}$PO$_{4}$) 
and DKDP crystals, in order to estimate the degree of quantum-mechanical localisation of hydrogen and 
deuterium atoms in the paraelectric phase. In both systems, they have found that proton quantum delocalisation 
in the OH$\cdots$O hydrogen bond is necessary for stabilisation of the disordered state. The only  
difference between KDP and DKDP is that quantum tunneling occurs coherently in the former system whereas 
incoherently in the latter, an effect that has been linked by the authors to the observed $T_{C}$ dependence
on the isotope.    
        
Recently, Wikfeldt and Michaelides (2014) have employed \emph{ab initio} PIMD  simulations based on DFT to 
investigate the importance of QNE on the atomic ordering and structure of H2SQ (i. e., C$_{4}$H$_{2}$O$_{4}$).
We note that in this case the authors have explicitly considered long-range van der Waals interactions 
by using a dispersion-corrected DFT functional (see Sec.~\ref{subsec:firstprinciples}). It has
been found that concerted proton jumps along H-bond chains are facilitated dramatically by quantum tunneling 
of several protons occurring at the same time. According to Wikfeldt and Michaelides' results, QNE are 
crucial in this order-disorder phase transition (that is, ferroelectric to paraelectric). The same 
phenomenology has been observed also in the analogous deuterated crystal but in a smaller magnitude, leading 
to an Ubbelohde effect that is in good agreement with the experiments (i. e., elongation of the oxygen-oxygen 
distances by $\sim 0.02$~\AA). Subsequently, Wikfeldt (2014) has introduced a simple model for a coupled 
one-dimensional H-bond chain that has been parametrised to DFT calculations performed in H2SQ. Such an 
effective model allows for an efficient exploration of QNE in larger systems over longer simulation times. 
The PIMC results obtained with Wikfeldt's H2SQ model in fact appear to be consistent with the conclusions 
presented in a previous full \emph{ab initio} work (Wikfeldt and Michaelides, 2014). 

Further systematic studies are necessary to determine exactly how important QNE are for the understanding of 
proton dynamics and proton order in H-bond ferroelectrics. The computational evidence gathered to date appears 
to indicate that quantum nuclear effects certainly are crucial. Reassuringly, in a recent \emph{ab initio} PIMC 
study Li, Walker, and Michaelides (2011) have shown that the quantum nature of the H-bond manifests appreciably
in most hydrogen-bonded materials.

\subsection{Lithium and related compounds}
\label{subsec:metals}
Lithium (Li) is the lightest metallic element; at ambient conditions it is most stable in a cubic bcc 
crystal. Li represents the prototype of a simple metal with a Fermi surface that is nearly spherical. 
As pressure is increased, however, this material undergoes a series of symmetry-breaking structural transformations 
that provoke an increase in complexity on its electronic band-structure (Guillaume \emph{et al.}, 2011). 
The presence of QNE in solid Li under pressure is notable. Experimentally, large isotopic effects have been 
observed in the equation of state and elastic properties of the bulk crystal at low temperatures ($T \le 77$~K) and 
pressures up to $\sim 2$~GPa (Gromnitskaya, Stal'gorova, and S. M. Stishov, 1999). For instance, differences of 
about $7$~\% have been reported for the transversal and longitudinal sound-wave velocities in solid $^{6}$Li and 
$^{7}$Li at low and high pressures. On the theoretical side, Filippi and Ceperley (1998) have analysed the 
influence of quantum nuclear effects on the kinetic energy of the crystal with PIMC simulations based on a pairwise 
interaction potential. It has been found that the excess kinetic energy in Li decreases from about $10.4$~\% of the 
classical value at $300$~K to $4.5$~\% at $450$~K, hence QNE are important all the way up to melting. The role of 
QNE in the structural and electronic properties of small Li clusters has been estimated also as very influential 
by Rousseau and Marx (1998).      

One of the effects of applying pressure in a crystal is to increase the kinetic energy of the atoms. It has been 
argued theoretically that if the increment in zero-point motion due to compression is higher than that attained in 
the potential energy, eventually the crystal could melt at low temperatures. This possibility has attracted a lot 
of attention in hydrogen since according to some theoretical arguments and effective models a metallic liquid with 
exotic properties could be stabilised in the regime of Mbar pressures (Babaev, Sudb\o, and Ashcroft, 2004). In solid 
lithium it has been experimentally observed (Lazicki, Fei, and Hemley, 2010) and calculated with first-principles 
methods (Hern\'andez \emph{et al.}, 2010) that at a pressure of $\sim 10$~GPa the corresponding melting line develops 
a negative slope (in analogy to what occurs in sodium at $P \sim 30$~GPa). This finding appears to open an alternative 
for the possible realisation of a ground-state metallic liquid at high pressures (although possibly in the Mbar 
regime, or even at much higher pressures). 

The role of QNE on the sudden drop observed in the melting line of Li, however, remains controversial. Guillaume 
\emph{et al.} (2011) have measured a melting temperature of $\sim 190$~K at a pressure of $\sim 40$~GPa, which represents 
by far the lowest melting temperature observed for any material at such pressures. The authors of this work have suggested 
that QNE play an important role in shaping the phase diagram of Li. This suggestion seems to be consistent with the fact 
that classical first-principles simulations (Hern\'andez \emph{et al.}, 2010) provide melting temperatures which are about 
$100$~K higher than the experimental points obtained by Guillaume \emph{et al.} (2011). A more recent experiment by Schaeffer 
\emph{et al.} (2012), however, has revealed a totally different scenario in which excellent agreement with the classical 
\emph{ab initio} results by Hern\'andez \emph{et al.} (2010) is obtained. Then, what is the real extent of QNE on 
the melting properties of Li? Very recently, Feng \emph{et al.} have performed a systematic first-principles PIMD study aimed
at answering this question. These authors have found that the net effect of considering QNE on the melting temperature of Li 
is minimal (e. g., a small shift of $15$~K towards lower temperatures at $P = 45$~GPa). Interestingly, QNE influence 
noticeably the free energy of the solid and liquid phases by separate, however there is a strong QNE compensation effect 
between the two phases at melting (Feng \emph{et al.}, 2015).  

Light-weight materials based on lithium are important from a technological point of view, and thus so are QNE. 
Two interesting examples are lithium hydride (LiH) and lithium imide (Li$_{2}$NH). LiH is used in the nuclear 
industry either as a shielding agent or fuel in energy reactors (Welch, 1974; Veleckis, 1977). LiH is an ionic crystal 
that stabilises in the rocksalt structure at ambient conditions; Li is the cation (positively charged ion) 
and H the anion (negatively charged ion). The presence of large QNE in LiH has been reported both in experiments 
and quantum simulations (Boronat \emph{et al.}, 2004). At room temperature the experimental Lindemann ratio of the hydrogen 
ion amounts to $0.12$ (Vidal and Vidal-Valat, 1986), which lies in between those measured for solid H$_{2}$ and Ne (i. 
e., $0.18$ and $0.09$, respectively), thus indicating very strong quantum character. 

Large quantum isotopic effects have been reported in lithium hydride for numerous quantities including the kinetic energy, 
Lindemann ratio, and lattice parameter. For example, Cazorla and Boronat (2005) have estimated by means of VMC calculations 
based on classical interatomic potentials that the kinetic energy of the hydrogen anion at zero temperature changes from $84(1)$ 
in LiH to $67(1)$~meV in LiD. Meanwhile, the corresponding Lindemann ratio is reduced by about $24$~\% in LiD as compared to 
that in LiH. More recently, Dammak \emph{et al.} (2012) have found by using a quantum thermostat approach in combination with  
first-principles methods, that the lattice parameter difference between LiH and LiD amounts to $0.019$ and $0.016$~\AA~ at 
$0$ and $300$~K, respectively, in fairly good agreement with the experiments (namely, $0.016$ and $0.014$~\AA). 

QNE can also affect considerably the electronic properties of a quantum solid, in particular the electronic energy band gap, 
$E_{g}$, due to the presence of electron-phonon couplings. By using a first-principles approach that consistently takes into 
account anharmonic and zero-point motion effects, Monserrat, Drummond and Needs (2013) have calculated the quantum-mechanical 
expectation value of $E_{g}$ in LiH and LiD over the temperature interval $0 \le T \le 800$~K. They have found that the isotopic 
effect in $E_{g}$ roughly amounts to $4-7$~\%, with LiD exhibiting always the largest energy band gap. Interestingly, QNE at 
zero temperature account for a $E_{g}$ variation of $\sim 2$~\% as compared to the value calculated with classical methods, 
which is $\sim 3.00$~eV.         

Lithium imide (Li$_{2}$NH) is a very promising hydrogen storage material due to its low molecular weight and central 
role played on the decomposition reaction (Shevlin and Guo, 2009):
\begin{eqnarray}
{\rm LiNH_{2} + 2LiH} && \leftrightarrow  {\rm Li_{2}NH + LiH + H_{2}} \nonumber \\ 
                      && \leftrightarrow {\rm Li_{3}N + 2H_{2}}~,
\label{eq:hydstor}
\end{eqnarray}
where in the first stage a total of $5.5$~wt\% H$_{2}$ is released and of $5.2$~wt\% H$_{2}$ in the second. A lot of research,
both of computational and experimental nature, has been devoted to understand the atomic structure and phase transitions 
occurring in Li$_{2}$NH. Due to the light mass of the atoms and relatively weak interactions between particles, QNE are likely 
to affect the fundamental properties of this material. Zhang, Dyer, and Alavi (2005) have solved the Schr\"{o}dinger 
equation of a proton in the potential energy surface of Li$_{2}$NH calculated with DFT methods, to analyse the influence of QNE 
on its dynamics. It has been found that the quantum character of H atoms is very strong, leading to partial delocalisation 
of the proton around certain N centers through quantum tunneling. The origin of this effect has been traced back to the 
relatively flat potential energy landscape of the system. The results of a more recent computational study by Ludue${\rm \tilde{n}}$a 
and Sebastiani (2010) based on \emph{ab initio} PIMD simulations, appear to support the validity of these results. The proton 
momentum distribution in Li$_{2}$NH has been experimentally measured with inelastic neutron scattering techniques, and calculated 
with quantum thermostatted \emph{ab initio} molecular dynamics (Ceriotti \emph{et al.}, 2010b). The reported experimental and 
computational $n({\bf k})$ results are in good agreement, providing a large average kinetic energy of $415$~K for Li atoms, of 
$410$~K for N, and of $858$~K for H.    

The presence of proton quantum tunneling and large zero-point motion in Li$_{2}$NH has several implications. First, the conventional 
treatment of quantum nuclear effects through the quasi-harmonic approximation should not be adequate in this system. And second, 
the real energy barrier for hydrogen diffusion, a key parameter for understanding and designing new H-storage materials,  
must be lower than predicted with classical simulation methods. In fact, Zhang, Dyer, and Alavi (2005) have estimated that at room 
temperature the H diffusion coefficient in Li$_{2}$NH is about $4$ orders of magnitude higher than the one expected from classical 
theory. In the light of these results, full quantum treatment of hydrogen atoms in crystals similar to Li$_{2}$NH (e. g., LiBH$_{4}$) 
may allow for an improved rational engineering of H-storage materials.

\subsection{Carbon-based crystals and nanomaterials}
\label{subsec:c6}
Carbon atoms are found in a large number of technologically relevant materials, including diamond and the prolific family of
carbon nanostructures (e. g., graphene, nanotubes, and nanohorns). Diamonds, which due to their strong covalent atomic bonds 
possess superlative hardness and thermal conductivity, are used as anvil cells to study condensed matter systems 
over wide $P-T$ ranges, and also have a major industrial application as cutting and polishing tools. At low temperatures, 
diamond is an archetypal quasi-harmonic crystal (Ceriotti, Bussi, and Parrinello, 2009) in which the presence of QNE has a 
profound impact on its structural, elastic, and electronic band-structure features. Herrero and Ram\'irez (2000) have studied 
the influence of zero-point motion on the thermodynamic properties of this solid with PIMC simulations based on an empirical 
interatomic potential. They have found that QNE account for an increase of $0.5$~\% in the lattice parameter and a decrease of 
$5$~\% in the bulk modulus with respect to the values obtained with classical simulation methods. More recently, Monserrat, Drummond, 
and Needs (2013) have estimated with a fully anharmonic first-principles approach that the zero-point motion renormalisation 
of the electronic energy band gap in diamond amounts to $-461$~meV. The origins of this large effect have been discussed in 
detail by Monserrat and Needs (2014), in terms of important electron-phonon interactions that affect differently valence and 
conduction band electrons (e. g., the latter being specially sensitive to the size of the Lindemann ratio in the crystal).   

Diamondoids, namely nanocages with formula C$_{x}$H$_{y}$ in which the carbon atoms are $sp^{3}$ bonded like in diamond, are 
biocompatible and superhard molecules. These organic nanoparticles are found in large concentrations in petroleum fluids and 
currently are attracting a lof of attention due to their potential use in drug delivery and nanotechnology applications (Mochalin 
\emph{et al.}, 2012). In analogy to diamond, a strong electron-phonon coupling is expected to occur in diamondoids. Recently, 
Patrick and Giustino (2013) have demonstrated by means of first-principles simulations combined with Monte Carlo sampling techniques 
that the role of QNE on the ``photophysics'' of these molecules is pivotal. In particular, for the theoretically calculated 
optical absorption spectra of diamondoids to be in quantitative agreement with the experiments, the zero-point motion 
of the atoms must be taken into account. Also, the accompanying renormalisation of the electronic energy band gaps amounts to 
$0.4-0.6$~eV, depending on the selected C$_{x}$H$_{y}$ species, which coincides with the results obtained by Monserrat, 
Drummond, and Needs (2013) in diamond.    

QNE can affect significantly the gas adsorption and transport properties of carbon-based nanostructures from zero up to room 
temperature. A case study that has been thoroughly investigated both with theory and experiments is the adsorption and diffusion of 
hydrogen molecules and atoms on graphene and other related nanomaterials. To understand this problem correctly is critical from a 
technological point of view [e. g., for the design of improved hydrogen storage materials (Cazorla, 2015)] and also for fundamental 
reasons [e. g., to rationalise the formation of molecular hydrogen in the interstellar medium and improve the astrophysical models 
of star evolution (Bromley \emph{et al.}, 2014)]. Experimental evidence of the importance of QNE in hydrogenated carbon-based
surfaces and cavities is abundant. Tanaka \emph{et al.} (2005) have measured the adsorption of H$_{2}$ and D$_{2}$ on single-wall 
carbon nanohorns at $T = 77$~K and reported appreciably different behaviors in the two cases. In particular, around $6-7$~\% 
more of deuterium molecules are adsorbed on the interior of the nanoparticles. The observed kinetic isotope effect has been ascribed, 
on basis to the results of path-integral grand canonical MC simulations performed with semi-empirical potentials, to the presence of 
QNE that favor the localisation of D$_{2}$ in the cone part of the nanohorns. A similar adsorption isotope effect has been 
reported also for graphene, which has been interpreted in terms of similar quantum-mechanically nuclear arguments (Paris \emph{et al.}, 
2013). Lovell \emph{et al.} (2009) have studied the room-temperature adsorption of H$_{2}$ in the graphite intercalation 
compound KC$_{24}$ with inelastic neutron scattering techniques. By comparing their experimental data to the results of quantum 
first-principles simulations, they have concluded that QNE are responsible for a tremendous reduction of $\sim 60$~\% in the 
amount of taken gas. 

On the purely computational side, Kowalczyk \emph{et al.} (2007) have described the physical adsorption of molecular hydrogen in 
slit-like carbon nanopores at low temperatures and high gas densities, using classical and path-integral grand canonical Monte 
Carlo simulations based on semi-empirical interatomic potentials. It has been found that classical simulations overestimate the 
amount of hydrogen in carbon nanopores due to neglecting of QNE (although the differences between the classical and quantum 
predictions are ameliorated when the size of the slit-carbon pore diameter is wider than $\sim 6$~\AA~). Herrero and Ram\'irez 
(2010) have studied with PIMD simulations and a tight-binding potential fitted to DFT calculations, the finite-temperature 
properties of H$_{2}$ molecules adsorbed in graphite. It has been shown that H$_{2}$ molecules are disposed parallel to the 
graphite-layer plane and that they can rotate freely about their center of mass in that plane. The stretching mode of the hydrogen 
molecule is found to change considerably under graphitic confinement by reducing its frequency $\sim 3.5$~\% with respect to the 
isolated molecule. Herrero and Ram\'irez (2010) have also reported strong quantum isotopic effects in this system; for instance, at 
room temperature the ratio between the kinetic energy of H$_{2}$ and D$_{2}$ amounts to $1.31$, where $E_{\rm k}$(H$_{2}$) 
is equal to $0.238$~eV. Kowalczyk \emph{et al.} (2015) have also investigated the structural and dynamical properties of hydrogen 
and deuterium molecules adsorbed in the interior of carbon-based nanotubes at low temperatures, using PIMD techniques and classical 
force fields. A large isotope effect caused by QNE has been revealed that consists of H$_{2}$ molecules diffusing seven to eight 
times faster than D$_{2}$ on the inner H$_{2}$/D$_{2}$ monolayer that coats the carbon atoms. This effect, which is quantum in 
nature, could be exploited in light-weight isotope separation processes employing nanoporous molecular sieves.

Several quantum studies involving a variety of simulation techniques have been performed to investigate also the chemisorption 
and diffusivity of H atoms on graphene (see, for instance, Herrero and Ram\'irez, 2009; Garashchuk \emph{et al.}, 2013; 
Karlick\'y, Lepetit, and Lemoine, 2014; Bonfanti \emph{et al.}, 2015). The general picture deriving from all these works 
is that QNE appreciably facilitate both the adsorption and posterior diffusion of hydrogen atoms on the carbon surface. 
Consequently, hydrogenation of large areas of graphene could be achieved more easily in practice than previously inferred 
from classical simulation studies. Interestingly, Davidson \emph{et al.} (2014) have pointed to the need of explicitly 
considering van der Waals forces in this type of quantum simulation studies; the estimated energetic barriers for the
chemisorption and diffusion of H atoms then are reduced further, in some cases as much as $\sim 25$~\% (depending on the
employed DFT functional). In view of the results presented in the last part of this section, we can conclude that inclusion 
of QNE and long-range dispersive interactions in modeling of hydrogenated carbon-based nanomaterials is necessary for 
providing a realistic estimation of gas-adsorption capacities and transition states at low temperatures.

%===============================================================
%===============================================================
\section{Summary and Outlook}
\label{sec:conclusions}
%===============================================================
%===============================================================
We have presented an overview of the current understanding of quantum crystals formed by atoms and small 
molecules over wide thermodynamic intervals, focusing on the insights provided by quantum simulations. We have 
described the fundamentals of the computational methods that are used to study QNE in quantum solids including 
variational, projector and path-integral Monte Carlo techniques, among others. Also, we have explained the basic 
notions of popular first-principles electronic band-structure methods (e. g., DFT and eQMC) as applied to the 
description of atomic interactions in crystals. 

Our analysis shows that consideration of QNE in computer simulation studies of rare-gases, molecular solids, 
H-bond ferroelectrics, light-weight ionic compounds, carbon-based nanomaterials, and even some perovskite 
oxides, is crucial for understanding the origins of their energy, structural and functional properties at low 
temperatures. In most quantum crystals (e. g., $^{4}$He, H$_{2}$, Li$_{2}$NH, and BaTiO$_{3}$) quasi-harmonic 
approaches turn out to be inadequate for describing their thermodynamic stability and the energy differences 
between energetically competitive phases; one instead has to consider methods that fully take into account 
anharmonicity. Meanwhile, the interatomic interactions in quantum solids normally are not described correctly 
by standard first-principles (LDA and GGA DFT functionals) or semi-empirical approaches. Combination of these 
two factors makes the simulation of quantum solids very challenging. 

QNE are important in a large number of systems and processes that are relevant to materials science. These include, 
hydrogen storage (e. g., Li$_{2}$NH and LiH), perovskite oxides (e. g., BaTiO$_{3}$ and SrTiO$_{3}$), ferroelectricity 
(e. g., H-bonded polar compounds), solid plasticity (e. g., $^{4}$He), and high-energy density materials (e. g., N$_{2}$). 
It is also likely that QNE are more influential than previously assumed in systems that are relevant to the pharmaceutical 
industry (molecular crystal polymorph) and catalysis (diffusion and adsorption of small molecules on carbon-based and 
metallic surfaces). We hope that our review will motivate new investigations in the context of materials science that 
will take into consideration the quantum nature of atoms in particular systems and processes.  

In spite of all the insight gathered in quantum solids, there are still a few remaining aspects that need to be better 
understood. These are essentially related to comprehension of (i)~the behavior of different types of crystalline 
defects and the interactions between them, and (ii)~the energy, structural, and dynamical properties of quantum crystals 
under extreme $P-T$ conditions. Advancing in the first of these two challenges is crucial for substantiating the microscopic 
arguments that have been proposed to explain the intriguing plastic phenomena observed in solid $^{4}$He at ultra-low 
temperatures. In particular, a quantitative description of dislocations at the atomic scale and of their interactions 
with isotopic $^{3}$He impurities is still pending. Quantum simulations could contribute significantly to this endeavor. 
Nevertheless, due to the large size of the simulation cells involved ($\sim 10^{4}-10^{5}$ atoms) and inherent structural 
complexity of line defects, this progress is slow at the moment (see, for instance, Boninsegni \emph{et al.}, 2007
and Landinez-Borda, Cai, and de Koning, 2016). 

Meanwhile, the crystal structures appearing in the phase diagram of most molecular solids at high pressures either 
are vaguely characterised or unknown. The outcomes of systematic structural searches based on first-principles methods
in fact have been very useful to better identify them. Nevertheless, the influence of QNE on the thermodynamic 
stability of different high-$P$ polymorph generally are disregarded in computational studies (see, for instance, 
the case of N$_{2}$ and CH$_{4}$), or considered straightforwardly within the quasi-harmonic approximation. It is worth 
stressing once again that a consequence of applying pressure in a crystal is to extend the temperature range over which 
QNE are relevant; therefore, the presence of quantum nuclear effects like zero-point motion, quantum atomic exchanges 
and quantum tunneling, the majority of which are not reproduced correctly by harmonic-based approaches, are key aspects
for understanding the properties of molecular solids under extreme thermodynamic conditions. Such a comprehension is
crucial to advance our knowledge in planetary sciences.                        

Common to these challenges is the underlying problem about how to describe the interactions between atoms in quantum 
crystals correctly. As we have explained before, these interactions require to go beyond standard first-principles 
approaches which, in addition to the unavoidable task of treating QNE, sometimes makes the simulation of quantum solids 
prohibitive in terms of computational expense. To this regard, the outcomes of systematic benchmark studies involving 
non-standard DFT and eQMC methods are crucial for rigorously establishing acceptable balances between numerical accuracy 
and computational load. Further progress in current electronic-band structure algorithms, on one hand, and improvement 
on the availability of quantum computer packages that allow to simulate QNE, on the other, would facilitate enormously 
this task.    

As a final reflection, we would like to mention that in not a few situations QNE are ``put under the rug'' by arguing 
that they should play a minor role or somehow cancel out. This is normally supported by a reasoning of 
the type ``good agreement with the experiments'' obtained in classical studies. Nevertheless, several authors have 
demonstrated that the causes behind such a good agreement sometimes can be traced back to an inaccurate representation 
of the atomic forces, which can disguise the real magnitude of QNE [see, for instance, the case of the predicted 
atomisation transition in solid H$_{2}$ under pressure, Chen \emph{et al.} (2014)]. Therefore, tests on the influence of 
QNE in light-weight and highly anharmonic crystals should not be avoided but instead performed systematically. As expressed 
by Miller (2005), ``If one performs only classical simulations, one will never know whether quantum effects are important. 
One must have the ability to include quantum effects into a simulation, even if only approximately, to know when they 
are important and when they are not.''

\begin{acknowledgments}
This research was supported under the Australian Research Council's Future Fellowship funding 
scheme (project number FT140100135), and MICINN-Spain (Grants No. MAT2010-18113, CSD2007-00041, 
and FIS2014-56257-C2-1-P). Computational resources and technical assistance were provided by 
the Australian Government through Magnus under the National Computational Merit Allocation Scheme.
\end{acknowledgments}

\end{document}